\documentclass[twoside]{article}

\usepackage{url,intmacros,graphicx}
\usepackage{amssymb,amsmath}
\usepackage{capt-of}
\usepackage{float}

\def\BibTeX{{\rm B\kern-.05em{\sc i\kern-.025em b}\kern-.08em
    T\kern-.1667em\lower.7ex\hbox{E}\kern-.125emX}}

\numberwithin{equation}{section}

\newcommand{\email}[1]{\\ \small{\url{#1}} \\}
\newcommand{\institution}[1]{\\ \parbox{3.0in}{\small{#1}}}
\newcommand{\keywords}[1]{\small\textbf{Keywords: }#1}
\newcommand{\AMSsubj}[1]{\noindent\textbf{AMS subject classifications: }#1}
\newcommand\whenaccepted{}

\newcommand{\Verbose}{}

\title{Statistical Taylor Expansion: A New and Path-Independent Method for Uncertainty Analysis \footnote{\whenaccepted}}

\ifdefined\NoAuthor
\else
\author{Chengpu Wang
\institution{40 Grossman Street, Melville, NY 11747, USA}
\email{Chengpu@gmail.com}}

\fi

\begin{document}
\maketitle
\begin{abstract}

Statistical Taylor expansion is a rigorous extension of conventional Taylor expansion that replaces each precise input variable with a random variable of known distribution and sample count, then computes the mean, deviation, and a bounding reliability of every result.
By tracking the propagation of input uncertainties through all intermediate steps, it renders the final result path-independent, with precise quantification of the tracking quality.
This path-independence sets it fundamentally apart from conventional numerical approaches, which are path-dependent.
This study presents an implementation called variance arithmetic and demonstrates its performance across diverse mathematical applications.

This study also reveals the potentially substantial impact of numerical errors in library functions, the defect of applying input uncertainties as weights in conventional regression, and the implied assumption of discrete Fourier transformation.
The concept of statistical algebra is also introduced.
\end{abstract}


\AMSsubj{G.1.0}

Copyright \copyright{2024}

\section{Introduction}
\label{sec: introduction}

Let a signal value contain a value $x$ and an uncertainty deviation $\delta x$.
As input, $x$ and $\delta x$ are usually the mean and standard deviation of a measurement, respectively \cite{Statistical_Methods, Precisions_Physical_Measurements}.
If $\delta x = 0$, then $x$ is a \emph{precise value}; otherwise, the pair specifies an \emph{imprecise value} $x \pm \delta x$.
Let $P(x) \equiv \delta x / |x|$ be the \emph{statistical precision} (hereafter referred to as precision) of $x \pm \delta x$.
A smaller $P(x)$ indicates a higher measurement quality of $x \pm \delta x$.

Statistical Taylor expansion determines the result $f \pm \delta f$ of a general analytic expression $f(x, \dots)$ on the basis of inputs $x \pm \delta x, \dots$ and their corresponding distributions and sample counts. 
\begin{itemize}
\item 
Previous studies have examined the effect of input uncertainties on output values for specific cases \cite{Lower_Order_Variance_Expansion}.
Statistical Taylor expansion generalizes these effects as uncertainty bias, as shown in Formulas \eqref{eqn: Taylor 1d mean} and \eqref{eqn: Taylor 2d mean} of this study.

\item
The traditional variance-covariance framework accounts only for first-order interactions between random variables through an analytic function \cite{Lower_Order_Variance_Expansion, Probability_Statistics, Numerical_Recipes}, whereas statistical Taylor expansion extends this framework to include higher-order interactions as expressed in Formula \eqref{eqn: Taylor 2d variance}.
Carrying out the expansion in full also reveals when the variance series converges, which is the focus of Section \ref{sec: convergence}.

\item Calculating the reliability of $f \pm \delta f$ as a value between $0$ and $1$ from input uncertainty distributions and sample counts appears to be without precedent, as shown in Section \ref{sec: statistical bounding}.
\end{itemize}

Statistical Taylor expansion improves upon existing numerical arithmetic in several key respects.

Conventional floating-point arithmetic \cite{Computer_Architecture, Floating_Point_Arithmetic, Floating_Point_Standard} computes only the result value $f$.
Because $f$ contains an unknown amount of rounding error \cite{Rounding_Error, Precise_Numerical_Methods, Algorithms_Accuracy}, a 32-bit floating-point representation with $10^{-7}$ resolution may not be sufficient for non-trivial computations on inputs with $10^{-6}$ or coarser precision.
However, statistical Taylor expansion can account for rounding errors as part of the result deviation $\delta f$.

The bounding range in interval arithmetic \cite{Interval_Analysis, Worst_Case_Error_Bounds, Interval_Analysis_Theory_Applications, Interval_Arithmetic, Interval_Analysis_Notations} is inconsistent with the statistical nature of an $x \pm \delta x$ pair, and it answers a different question from statistical Taylor expansion.
The direct application of Taylor expansion in interval arithmetic over-estimates result intervals; consequently, it has to be treated as a polynomial \cite{Taylor_Models, Taylor_Forms_Limits}, and needs interval partition and other techniques to reduce but not eliminate such over-estimation \cite{Interval_Analysis, Interval_Analysis_Theory_Applications, Interval_Global_Optimization}.
The subdivision can cause branched results \cite{Interval_Analysis, Interval_Analysis_Theory_Applications}, such as when computing the square root of an input interval containing zero.
The over-estimation of bounding range is a well-known difficulty of interval arithmetic \cite{Interval_Global_Optimization}.
In contrast, statistical Taylor expansion is statistically precise, and it diverges when the input precision is not fine enough.

\begin{equation}
\label{eqn: dependency expression}
x^2 - x = x(x - 1) = (x - \frac{1}{2})^2 - \frac{1}{4};
\end{equation}
Both conventional floating-point arithmetic and interval arithmetic depend strongly on the specific algebraic form of an analytic function, a phenomenon known as the \emph{dependency problem} \cite{Precise_Numerical_Methods, Algorithms_Accuracy, Interval_Analysis, Interval_Analysis_Theory_Applications}, which can make conventional numerical computation less systematic and more practitioner-dependent.
For example, when the equivalent algebraic form in Formula \eqref{eqn: dependency expression} is executed by ostensibly independent arithmetic operations such as negation, addition, multiplication, division, square root, only $(x - \frac{1}{2})^2 - \frac{1}{4}$ yields the correct result in interval arithmetic, while the result difference using floating-point arithmetic is attributed to rounding errors.
For an analytic express, Because interval arithmetic has to partition input range into pair-wise monotonic range, and use objective functions such as $\min$ and $\max$ to stitch the outputs together, interval arithmetic produces less closed-form result than that from statistical Taylor expansion, e.g., Formula \eqref{eqn: multiplication variance} and its counterpart in interval arithmetic for the multiplication range show clearly such differences.

Taylor expansion can address the algebraic dependency problem because different algebraic forms result in the same Taylor expansion.
However, because the result of Taylor expansion using interval arithmetic is implementation dependent with various degrees of overestimation \cite{Taylor_Models, Taylor_Forms_Limits,Interval_Global_Optimization}, \emph{path dependency} extends the original algebraic dependency problem to more general cases of implementation dependency.
In contrast, statistical Taylor expansion yields compact closed-form analytic expressions for Taylor expansion, and it is path independent.

The uncorrelated uncertainty condition distinguishes statistical Taylor expansion from uncertainty propagation \cite{Uncertainty_Propagation}:
The correlation of two signals does not apply to the uncertainties of their individual signal values, with the latter usually arising from noise \cite{Statistical_Methods}.

To ensure mathematical and statistical rigor, statistical Taylor expansion abandons the significance-arithmetic representation used in its predecessor \cite{Prev_Precision_Arithmetic}, in which precision was tracked implicitly through the number of significant digits.

As a statistical sampling process, stochastic arithmetic \cite{Stochastic_Arithmetic, CADNA_library} is computationally expensive, whereas statistical Taylor expansion provides a direct characterization without sampling.

To demonstrate the wide applicability of statistical Taylor expansion, the remainder of this paper is organized as follows:
\begin{itemize}
\item Section \ref{sec: statistical Taylor expansion} develops the theoretical foundation of statistical Taylor expansion.

\item Section \ref{sec: variance arithmetic} describes variance arithmetic as a numerical implementation of statistical Taylor expansion.

\item Section \ref{sec: validation} presents standards for validating variance arithmetic.

\item Section \ref{sec: polynomial} illustrates variance arithmetic in polynomial computation, demonstrating its ability to trace floating-point rounding errors and its continuity in parameter space.

\item Section \ref{sec: matrix} describes the applications of variance arithmetic to matrix inversion, distinguishing between distribution tests and value tests.

\item Section \ref{sec: Math Library} discusses the evaluation of variance arithmetic on common mathematical library functions, showing the effect of a distributional pole.

\ifdefined\Verbose
\item Section \ref{sec: Moving-Window Linear Regression} reveals the accumulation of numerical errors by reusing an input multiple times in a moving-window progressive algorithm.
\fi

\item Section \ref{sec: FFT} examines the impact of numerical library errors and shows that these errors can be significant.
It also shows that statistical Taylor expansion is more suitable than interval arithmetic to characterize the result uncertainty when the input uncertainty is random in nature.

\item Section \ref{sec: recursion} showcases variance arithmetic in catching catastrophic cancellation in a recursive algorithm.

\item Section \ref{sec: regression} demonstrates imprecise analysis as a new concept in linear regression.

\item Section \ref{sec: equations} comments on solving equations.

\item Section \ref{sec: conclusion and discussion} concludes with a summary and a discussion of the findings.

\end{itemize}

\section{Statistical Taylor Expansion}
\label{sec: statistical Taylor expansion}

\subsection{Distributional Zero and Distributional Pole}

\begin{figure}[p]
\centering
\includegraphics[height=2.5in]{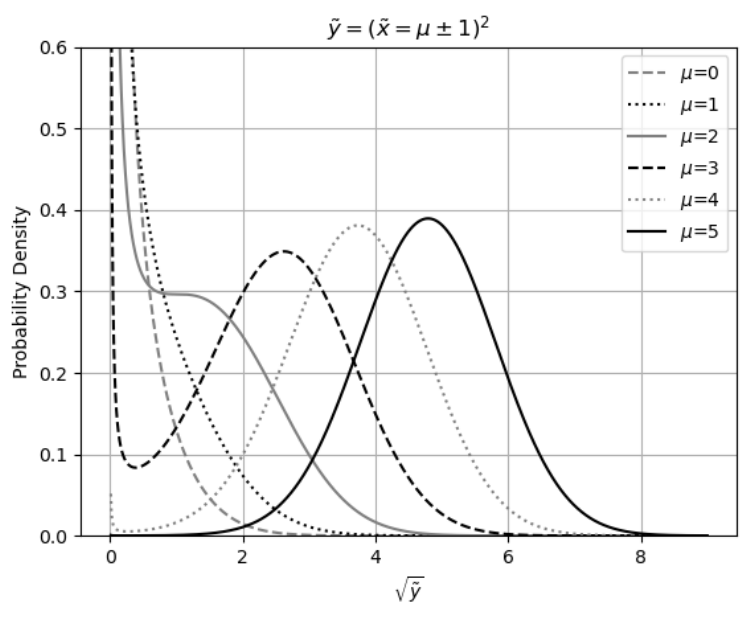} 
\captionof{figure}{
Probability density function of $\tilde{y} = \tilde{x}^2$, for various values of $\mu$ as indicated in the legend. 
The variable $\tilde{x}$ follows a Gaussian distribution with distribution mean $\mu$ and deviation $1$.
The horizontal axis is scaled as $\sqrt{\tilde{y}}$.
}
\label{fig: Square_Distribution}
\end{figure}

\begin{figure}[p]
\centering
\includegraphics[height=2.5in]{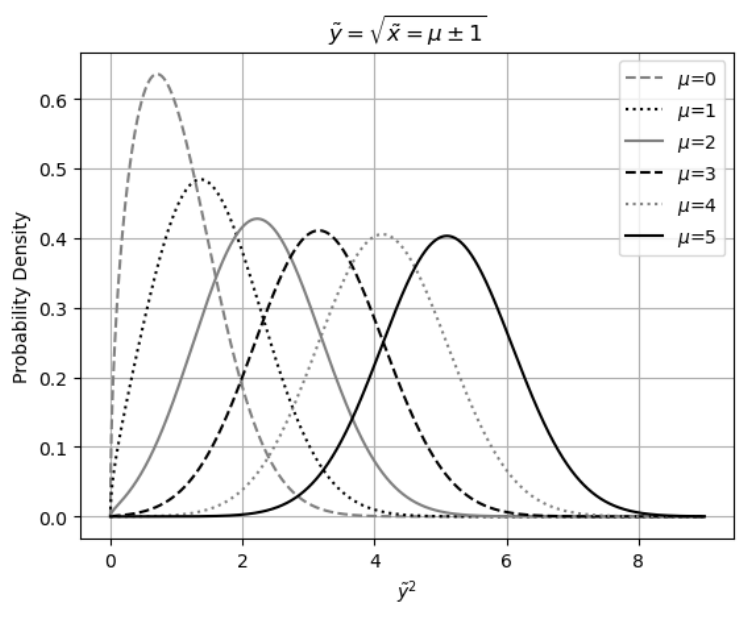} 
\captionof{figure}{
Probability density function for $\tilde{y} = \sqrt{\tilde{x}}$, for various values of $\mu$ as indicated in the legend. 
The variable $\tilde{x}$ follows a Gaussian distribution with distribution mean $\mu$ and deviation $1$.
The horizontal axis is scaled as $\tilde{y}^2$.
}
\label{fig: Square_Root_Distribution}
\end{figure}

Let $\rho(\tilde{x}, \mu, \sigma)$ denote the probability density function of a random variable $\tilde{x}$ with distribution mean $\mu$ and distribution deviation $\sigma$.
Let $f(x)$ be a strictly monotonic function of $x$ so that the inverse $f^{-1}$ exists, and let $\tilde{y} = f(\tilde{x})$.
Formula \eqref{eqn: function distribution} states that after the change of variables $\tilde{y} = f(\tilde{x})$, the probability density function $\rho(\tilde{y}, \mu_y, \sigma_y)$ with distribution mean $\mu_y$ and distribution deviation $\sigma_y$ describes the same underlying probability as $\rho(\tilde{x}, \mu, \sigma)$, viewed in a different coordinate.
\begin{align}
\label{eqn: function distribution}
\rho(\tilde{y}, \mu_y, \sigma_y) d\tilde{y} &\equiv \rho(f^{-1}(\tilde{y}), \mu, \sigma) \frac{d\tilde{x}}{d\tilde{y}} d\tilde{y} = \rho(\tilde{x}, \mu, \sigma) d\tilde{x};
\end{align}

Viewed in the $f^{-1}(\tilde{y})$ coordinate, $\rho(\tilde{y}, \mu_y, \sigma_y)$ is $\rho(\tilde{x}, \mu, \sigma)$ multiplied by $1/f^{(1)}_x$, where $f^{(1)}_x$ is the first derivative of $f(x)$ with respect to $x$.
\begin{itemize}
\item 
A \emph{distributional pole} occurs when $f^{(1)}_x=0 \rightarrow \rho(\tilde{y}, \mu_y, \sigma_y) = \infty$.
For example, $(\mu \pm 1)^2$ (which means $\tilde{x}^2$ for $\tilde{x} \sim N(\mu, 1)$) has a distributional pole at $\tilde{x} = 0$, as shown in Figure \ref{fig: Square_Distribution}.

\item 
A \emph{distributional zero} occurs when $f^{(1)}_x=\infty \rightarrow \rho(\tilde{y}, \mu_y, \sigma_y) = 0$.
For example, $\sqrt{\mu \pm 1}$ (which means $\sqrt{\tilde{x}}$ for $\tilde{x} \sim N(\mu, 1)$) has a distributional zero at $\tilde{x} = 0$, as shown in Figure \ref{fig: Square_Root_Distribution}.

\end{itemize}

In both Figures \ref{fig: Square_Distribution} and \ref{fig: Square_Root_Distribution}, $\rho(\tilde{y}, \mu_y, \sigma_y)$ closely resembles $\rho(\tilde{x}, \mu, \sigma)$ when the mode lies sufficiently far from any distributional pole or zero, as in $(5 \pm 1)^2$ and $\sqrt{5 \pm 1}$.
In such cases, the output $f(x)$ can be well characterized by its distribution mean $\overline{f(x)}$ and distribution deviation $\delta f(x)$.

\subsection{One-Dimensional Statistical Taylor Expansion}

Define $\tilde{z} \equiv (\tilde{x} - \mu)/\sigma$ and let $\rho(\tilde{z})$ be the normalized form of $\rho(\tilde{x}, \mu, \sigma)$ such that $\tilde{z}$ has distribution mean $0$ and deviation $1$.
The Normal distribution is the normalized form of the Gaussian distribution.
The requirement of the normalized form excludes the distributions without finite $\sigma$, such as the Student's $t$-distributions whose degrees of freedom is either 1 or 2.

\begin{align}
\label{eqn: bound moment}
\zeta(n, \kappa) &\equiv \frac{\int_{\varrho}^{\kappa} \tilde{z}^n \rho(\tilde{z}) d \tilde{z}}{\int_{\varrho}^{\kappa} \rho(\tilde{z}) d \tilde{z}};\\
\label{eqn: mean-reverting bounding}
\zeta(1, \kappa) &= 0;
\end{align}
Let $\tilde{z} \in [\varrho, \kappa]$ where $\varrho$ and $\kappa$ specify the \emph{bounding range}. 
Formula \eqref{eqn: bound moment} defines the corresponding \emph{bound moment} $\zeta(n, \kappa)$, which further satisfies the \emph{mean-reverting condition} of Formula \eqref{eqn: mean-reverting bounding} such that $\kappa$ determines $\varrho$.
For any symmetric probability distribution: $\rho(-\tilde{z}) = \rho(\tilde{z})$, $\varrho = -\kappa$, and $\zeta(2n+1, \kappa) = 0$.

In reality, instead of distribution mean $\mu$ and deviation $\sigma$, only sample mean $x$ and deviation $\delta x$ are known for $f(x)$.
Section \ref{sec: statistical bounding} will discuss the statistical foundation of replacing $\mu$ and $\sigma$ with $x$ and $\delta x$, respectively.

\begin{align}
\label{eqn: Taylor 1d}
f(x + \tilde{x}) &= f(x + \tilde{z} \delta x) = f(x) + \sum_{n=1}^{\infty} \frac{f^{(n)}_x}{n!} \tilde{z}^n (\delta x)^n; \\
\label{eqn: Taylor 1d mean}
\overline{f(x)} &= f(x) + \sum_{n=2}^{\infty}(\delta x)^n \frac{f^{(n)}_x}{n!} \zeta(n, \kappa); \\
\label{eqn: Taylor 1d variance}
\delta^2 f(x) &= \sum_{n=2}^{\infty} (\delta x)^n \sum_{j=1}^{n-1} \frac{f^{(j)}_x}{j!} \frac{f^{(n-j)}_x}{(n-j)!} \nu(j, n - j); \\
\label{eqn: Taylor 1d covariance}
\nu(j, n - j) &\equiv \zeta(n, \kappa) - \zeta(j, \kappa) \zeta(n-j, \kappa); 
\end{align}
An analytic function $f(x)$ can be accurately evaluated near $x$ using the Taylor series as shown in Formula \eqref{eqn: Taylor 1d}.
Formulas \eqref{eqn: Taylor 1d mean} and \eqref{eqn: Taylor 1d variance} yield the mean $\overline{f(x)}$ and variance $\delta^2 f(x)$ of $f(x)$, respectively.
The difference $\overline{f(x)} - f(x)$ is defined as the \emph{uncertainty bias}, representing the effect of input uncertainty on the resulting value.
$\delta^2 f$ at the $(\delta x)^n$ order is calculated as the sum of all convolution between $\frac{f^{(j)}_x}{j!} \tilde{z}^j$ and $\frac{f^{(n - j)}_x}{(n - j)!} \tilde{z}^{n -j}$ with $\nu(j, n -j)$ in Formula \eqref{eqn: Taylor 1d covariance} as the co-variance between $\tilde{z}^j$ and $\tilde{z}^{n -j}$.

Formulas \eqref{eqn: exp mean} and \eqref{eqn: exp precision} give the mean and precision for $e^x$, respectively:
\begin{align}
\label{eqn: exp mean}
\frac{\overline{e^x}}{e^x}  &= 1 + \sum_{n=2}^{\infty} (\delta x)^n \frac{\zeta(n, \kappa)}{n!}; \\
\label{eqn: exp precision}
\frac{\delta^2 e^x}{(e^x)^2} &= \sum_{n=2}^{\infty} (\delta x)^n \sum_{j=1}^{n-1} \frac{\nu(j, n -j)}{j!\;(n - j)!};
\end{align}

Formulas \eqref{eqn: log mean} and \eqref{eqn: log precision} give the mean and variance for $\log(x)$, respectively:
\begin{align}
\label{eqn: log mean}
\overline{\log(x)}  &= \log(x) + \sum_{n=2}^{+\infty} P(x)^{n} \frac{(-1)^{n+1} \zeta(n, \kappa)}{n}; \\
\label{eqn: log precision}
\delta^2 \log(x) &= \sum_{n=2}^{+\infty} P(x)^{n} (-1)^n \sum_{j=1}^{n-1} \frac{\nu(j, n -j)}{j (n-j)};
\end{align}

Formulas \eqref{eqn: sin mean} and \eqref{eqn: sin precision} give the mean and variance for $\sin(x)$, respectively:
\begin{align}
\label{eqn: sin mean}
\overline{\sin(x)} &= \sin(x) + \sum_{n=2}^{\infty} (\delta x)^n \sin(x)^{(n)} \frac{\zeta(n, \kappa)}{n!}; \\
\label{eqn: sin precision}
\delta^2 \sin(x) &= \sum_{n=2}^{\infty} (\delta x)^n \sum_{j=1}^{n-1} \sin(x)^{(j)}\sin(x)^{(n-j)} \frac{\nu(j, n -j)}{j! (n-j)!};
\end{align}

Formulas \eqref{eqn: power mean} and \eqref{eqn: power precision} give the mean and precision for $x^c$, respectively:
\begin{align}
\label{eqn: power mean}
\frac{\overline{x^c}}{x^c}  &= 1 + \sum_{n=2}^{\infty} P(x)^{n} \zeta(n, \kappa) \binom{c}{n}; \\
\label{eqn: power precision}
\frac{\delta^2 x^c}{(x^c)^2} &= \sum_{n=2}^{\infty} P(x)^{n} \sum_{j=1}^{n-1} \binom{c}{j} \binom{c}{n - j} \nu(j, n -j);
\end{align}

The input and output in statistical Taylor expansion reflect the inherent characteristics of the calculation, such as $\delta x \rightarrow P(e^x)$, $P(x) \rightarrow \delta \log(x)$, $\delta x \rightarrow \delta \sin(x)$, and $P(x) \rightarrow P(x^c)$.

\subsection{Convergence of Variance}
\label{sec: convergence}

\begin{figure}[p]
\centering
\includegraphics[height=2.5in]{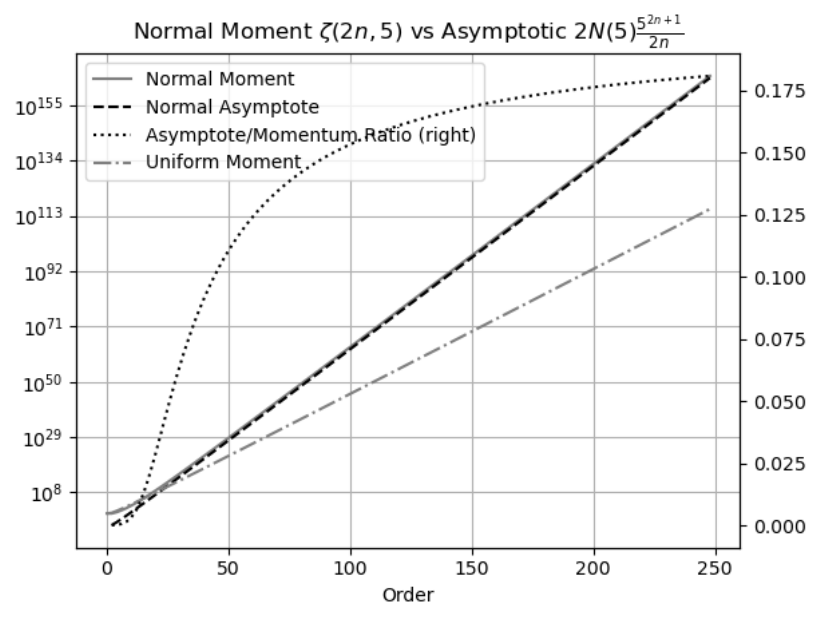} 
\captionof{figure}{
When $\kappa = 5$, the value of bound moment  $\zeta(2n, \kappa)$ for a Gaussian distribution with order $2n$ (x-axis) versus $2 \kappa \rho(\kappa) \frac{\kappa^{2n}}{2n}$ (left y-axis), and their ratio (right y-axis).
The value of moment  $\zeta(2n)$ for a Uniform distribution is also provided as a comparison.
}
\label{fig: Gaussian_5_Moment}
\end{figure}

\begin{align}
\label{eqn: uniform moment}
\rho(\tilde{z}) = \frac{1}{2\sqrt{3}}, \tilde{z} \in [-\sqrt{3}, \sqrt{3}] &: \zeta(2n, \kappa) = \frac{\kappa^{2n}}{2n + 1}; \\
\label{eqn: symmetric asymptotic moment}
\rho(-\tilde{z}) =\rho(\tilde{z})&: \lim_{n \rightarrow +\infty} \zeta(2n, \kappa) 
		= \frac{\kappa \rho(\kappa)}{\int_{0}^{\kappa} \rho(\tilde{z}) \;d \tilde{z}} \frac{\kappa^{2n}}{2n}; \\
\label{eqn: asymptotic moment}
|\varrho| < \infty&: \lim_{n \rightarrow +\infty} \zeta(n, \kappa) 
		= \frac{\kappa \rho(\kappa)}{\int_{\varrho}^{\kappa} \rho(\tilde{z}) \;d \tilde{z}} \frac{\kappa^{n}}{n}; \\
\label{eqn: variance asymptotic}
0 < j < n&: \lim_{n \rightarrow +\infty} \nu(j, n - j) \simeq \frac{\kappa^{n}}{n};
\end{align}
Formula \eqref{eqn: uniform moment} gives the closed form of $\zeta(2n, \kappa)$ for the Uniform distribution.
Formula \eqref{eqn: symmetric asymptotic moment} states the large-$n$ asymptote for any symmetric $\rho(\tilde{z})$.
Formula \eqref{eqn: asymptotic moment} states the corresponding asymptote for any asymmetric $\rho(\tilde{z})$ under the mean-reverting bounding condition of Formula \eqref{eqn: mean-reverting bounding} when both $\varrho$ and $\zeta(2, \kappa)$ are bounded.
Figure \ref{fig: Gaussian_5_Moment} shows $\zeta(2n, \kappa)$ for the Uniform distribution, $\zeta(2n, 5)$ for the Normal distribution, and the corresponding asymptote for the Normal distribution.
Formula \eqref{eqn: variance asymptotic} shows that the general asymptotic behavior of $\zeta(n, \kappa)$ determines whether Formula \eqref{eqn: Taylor 1d variance} converges:
\begin{itemize}
\item Formula \eqref{eqn: exp precision} for $e^{x \pm \delta x}$ and Formula \eqref{eqn: sin precision} for $\sin(x \pm \delta x)$ both converge unconditionally because the derivative coefficients decay factorially, dominating the $\kappa^{n+1}/(n+1)$ growth.

\item Formula \eqref{eqn: log precision} for $\log(x \pm \delta x)$ can be approximated by Formula \eqref{eqn: log convergence}, which converges when $P(x) \lesssim 1/\kappa$.

\item Formula \eqref{eqn: power precision} for $(x \pm \delta x)^c$ can be approximated by Formula \eqref{eqn: pow convergence}, which converges when $P(x) \lesssim 1/\kappa$; the precise upper bound for $P(x)$ varies with $c$.
\end{itemize}
\begin{align}
\label{eqn: log convergence}
\delta^2\log(x\pm\delta x) &\simeq \sum_{n = 2}^{+\infty} (-P(x) \kappa)^{n} \frac{2}{n^2} \sum_{j=1}^{n-1} \frac{1}{j}; \\
\label{eqn: pow convergence}
\frac{\delta^2 (x \pm \delta x)^c}{(x^c)^2} &\simeq \sum_{n=2}^{\infty} (P(x)\kappa)^n \frac{\binom{2c}{n} - 2\binom{c}{n}}{n};
\end{align}
Statistical Taylor expansion rejects the distributional zero of $\log(x \pm \delta x)$ or $(x \pm \delta x)^c$ in the range of $P(x) > 1/\kappa$ statistically because of the divergence of Formulas \eqref{eqn: log convergence} and \eqref{eqn: pow convergence} mathematically, with $\zeta(2n, \kappa)$ providing the connection between these two perspectives.

\subsection{Statistical Bounding}
\label{sec: statistical bounding}

\begin{figure}[p]
\centering
\includegraphics[height=2.5in]{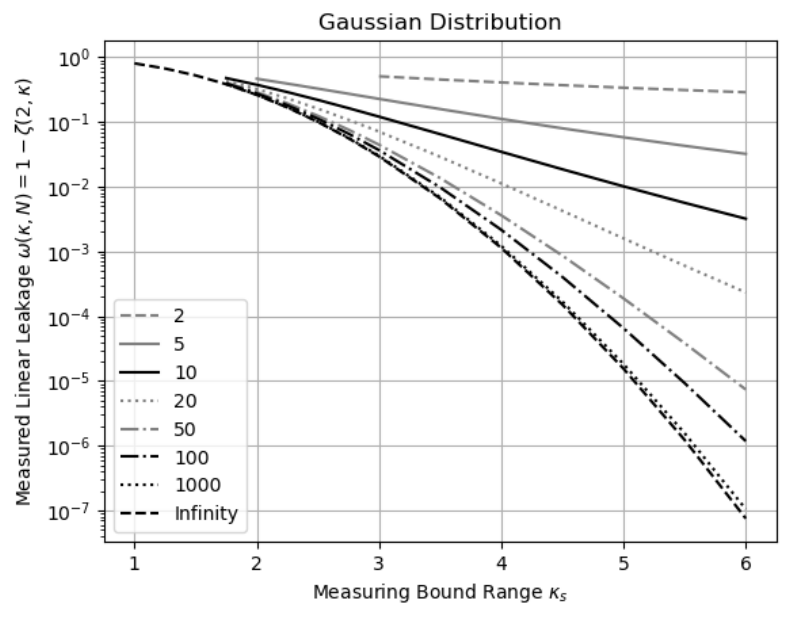} 
\captionof{figure}{
Measured linear leakage $\omega(\kappa_s, N)$ (y-axis) for varying measuring bound range $\kappa_s$ (x-axis) and sample count $N$ (legend).
}
\label{fig: Normal_Bounding_Leakage}
\end{figure}

\begin{figure}[p]
\centering
\includegraphics[height=2.5in]{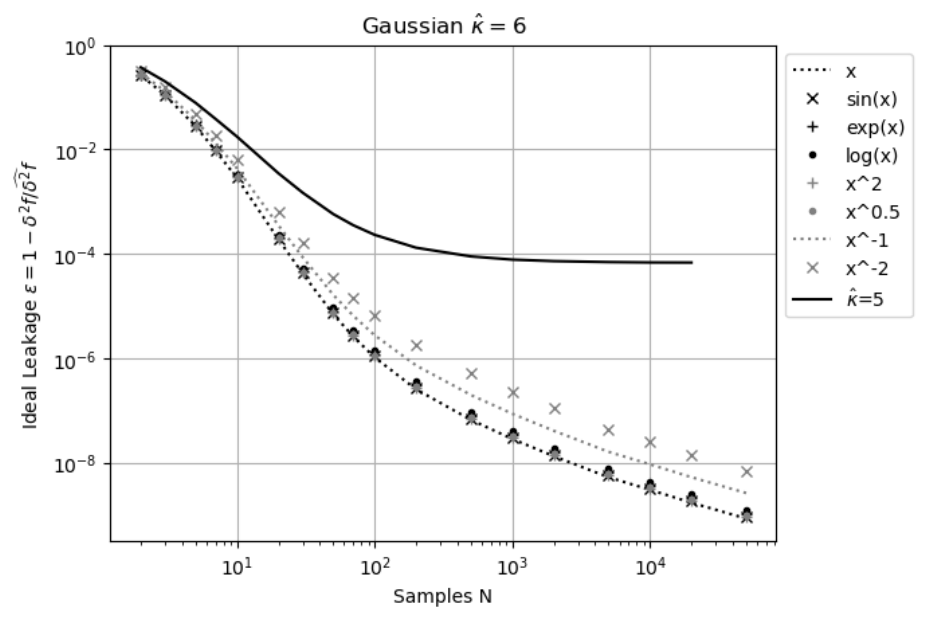} 
\captionof{figure}{
Ideal leakage $\varepsilon$ (y-axis) for varying sample count $N$ (x-axis) for the selected function $f(x=1 \pm 0.1)$ (legend) when $\hat{\kappa} = 5$ for Gaussian input uncertainties.
}
\label{fig: Normal_Function}
\end{figure}

\begin{figure}[p]
\centering
\includegraphics[height=2.5in]{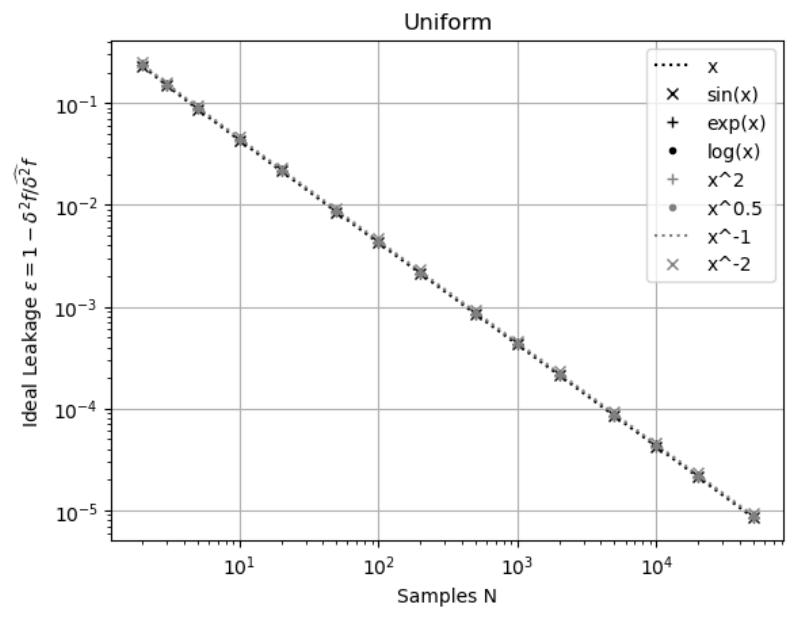} 
\captionof{figure}{
Ideal leakage $\varepsilon$ (y-axis) for varying sample count $N$ (x-axis) for the selected function $f(x=1 \pm 0.1)$ (legend) for Uniform input uncertainties.
}
\label{fig: Uniform_Function}
\end{figure}

A large $\kappa$ in Formula \eqref{eqn: bound moment} ensures that the calculated $\delta^2 f(x)$ is closer to the true statistical value, but it also means a smaller convergence range for $\delta^2 f(x)$, so that the choice of $\kappa$ is a compromise.
Let $\hat{\kappa}$ be the chosen \emph{ideal bounding range} with its corresponding \emph{ideal variance} $\hat{\delta^2} f(x)$.
The ideal bounding range limit the range of the distribution to where the sampling happens.
\begin{itemize}
\item Define \emph{ideal leakage} as $\varepsilon(\hat{\kappa}) \equiv 1 - \delta^2 f/\hat{\delta^2} f$. 

\item Define \emph{linear leakage} $\omega(\hat{\kappa}) \equiv 1 - \zeta(2, \hat{\kappa})$. 
$\omega(\hat{\kappa})$ the first order approximation of $\varepsilon(\hat{\kappa})$ according to Formula \eqref{eqn: Taylor 1d variance}.
For example, when $f(x)=x$, $\varepsilon = \frac{\omega(\hat\kappa,N) - \omega(\hat\kappa)}{1 - \omega(\hat\kappa)}$.
\end{itemize}

When sampling from a distribution, the sample mean $\overline{x}$ and sample deviation $\delta x$ approach the distribution mean $\mu$ and distribution deviation $\sigma$ respectively as the sample count $N$ increases \cite{Probability_Statistics}.
This yields the \emph{sample linear leakage} $\omega(\kappa, N)$ for the interval $[\overline{x} + \varrho \delta x, \overline{x} + \kappa \delta x]$, in contrast to the \emph{distributional linear leakage} $\omega(\kappa)$ for the interval $[\mu + \varrho \sigma, \mu + \kappa \sigma]$.
Let $\omega(\kappa) = \omega(\hat{\kappa}, N)$, to solve for $\kappa(\hat{\kappa}, N)$ which is used in calculating $\zeta(n, \kappa)$.

\begin{align}
\label{eqn: Gaussian linear leakage} 
\omega(\kappa) &= \frac{4 \rho(\kappa) \kappa}{\xi(2\kappa/\sqrt{2})}; \\
\label{eqn: Gaussian sample leakage}
\omega(\kappa, N) &= \frac{4 \rho(\kappa) \kappa}{\xi(\frac{\kappa \delta x - \overline{x}}{\sqrt{2}}) + \xi(\frac{\kappa \delta x + \overline{x}}{\sqrt{2}})};
\end{align}
For example, when the underlying distribution is Normal, Formula \eqref{eqn: Gaussian linear leakage} presents $\omega(\kappa)$, whereas Formula \eqref{eqn: Gaussian sample leakage} gives $\omega(\kappa, N)$, in which $\xi(\tilde{z})$ is the Normal error function \cite{Probability_Statistics}.
Figure \ref{fig: Normal_Bounding_Leakage} shows that as $N$ increases, $\omega(\hat{\kappa}, N)$ decreases monotonically to $\omega(\hat{\kappa})$, while $\kappa(\hat{\kappa}, N)$ increases monotonically to $\hat{\kappa}$. 
Figure \ref{fig: Normal_Function} displays that for a selected functions, $\varepsilon(6, N)$ decreases with $N$ following $\omega(6, N)$.
Figure \ref{fig: Normal_Function} also includes $\omega(5, N)$ as a comparison, which ceases to benefit from larger sample size when $N > {10}^{3}$ and contradicts the conventional 5-$\sigma$ rule for determining the statistical significance of an experimental result \cite{Statistical_Methods, Precisions_Physical_Measurements}. 

For a Uniform distribution, by definition $\hat{\kappa} = \sqrt{3}$ with $\varepsilon(\hat{\kappa}) = $.
The unbiased bounding range $\kappa$ of Uniform distribution is estimated as $1 + \frac{1}{N}$ times the minimal and maximal values in the sample \cite{Uniform_Distrobution_Estimator}.
Figure \ref{fig: Uniform_Function} shows that for the selected functions, $\varepsilon(\sqrt{3}, N) = \omega(\sqrt{3}, N)$, and both decrease linearly $N$.
The difference between Figures \ref{fig: Normal_Function} and \ref{fig: Uniform_Function} shows the effect of the input uncertainty distribution on the output deviation $\delta f$.

\subsection{Two-Dimensional Statistical Taylor Expansion}

\begin{align}
\label{eqn: bound moment 2d}
\zeta&(m, \kappa_x, n, \kappa_y) \equiv
	\frac{\int_{\varrho_x \sigma_x + \mu_x}^{\kappa_x \sigma_x + \mu_x} 
			\int_{\varrho_y \sigma_y + \mu_y}^{\kappa_y \sigma_y + \mu_y} 
			\rho(\tilde{x}, \tilde{y}) \big(\frac{\tilde{x} - \mu_x}{\sigma_x}\big)^m \big(\frac{\tilde{y} - \mu_y}{\sigma_y}\big)^n d \tilde{x} d \tilde{y}}
		   {\int_{\varrho_x \sigma_x + \mu_x}^{\kappa_x \sigma_x + \mu_x} 
			\int_{\varrho_y \sigma_y + \mu_y}^{\kappa_y \sigma_y + \mu_y}
			\rho(\tilde{x}, \tilde{y}) d \tilde{x} d \tilde{y}};\\
\label{eqn: mean reverting moment 2d}
\zeta&(1,\kappa_x,0,\kappa_y) = \zeta(0,\kappa_x,1,\kappa_y) = 0; \\
\label{eqn: Taylor 2d}
f&(x + \tilde{x}, y + \tilde{y}) = f(x, y) + \sum_{m, n \geq 0}^{m+n \geq 1} \frac{f^{(m,n)}_{(x,y)}}{m! n!} \tilde{x}^m \tilde{y}^n; \\
\label{eqn: Taylor 2d mean}
\overline{f(x,y)} &= f(x, y) + \sum_{m, n \geq 0}^{m+n \geq 2} (\delta x)^m (\delta y)^n \frac{f^{(m,n)}_{(x,y)}}{m!\;n!} \zeta(m, \kappa_x, n, \kappa_y);  \\
\label{eqn: Taylor 2d variance}
\delta^2 f(x, y) &= \sum_{m, n \geq 0}^{m+n \geq 2} (\delta x)^m (\delta y)^n \sum_{0 \leq i \leq m,\; 0 \leq j \leq n}^{1 \leq i + j \leq m + n - 1}
		\frac{f^{(i,j)}_{(x,y)}}{i!\;j!}\frac{f^{(m-i, n-j)}_{(x,y)}}{(m-i)!\;(n-j)!} \nu_{x,y}(i, m - i, j, n  -j); \\
\label{eqn: Taylor 2d covariance}
\nu_{x,y}&(i, m - i, j, n - j) \equiv \zeta(m, \kappa_x, n, \kappa_y) - \zeta(i, \kappa_x, j, \kappa_y) \zeta(m-i, \kappa_x, n-j, \kappa_y);
\end{align}
Formula \eqref{eqn: bound moment 2d} extends the bound moment to 2-dimensional using the compound normalized probability density function $\rho(\tilde{x}, \tilde{y})$, while Formula \eqref{eqn: mean reverting moment 2d} is the mean-reverting condition.
Formulas \eqref{eqn: Taylor 2d mean} and \eqref{eqn: Taylor 2d variance} compute the mean and variance of the Taylor expansion given in Formula \eqref{eqn: Taylor 2d}.
Although Formula \eqref{eqn: Taylor 2d variance} covers only the 2-dimensional case, it extends readily to any number of dimensions.

When $x \pm \delta x$ and $y \pm \delta y$ are independent of each other, $\zeta(m, \kappa_x, n, \kappa_y) = \zeta_{x}(m, \kappa_x) \zeta_{y}(n, \kappa_y)$:
\begin{align}
\label{eqn: addition mean}
\overline{x \pm y} &= x \pm y; \\
\label{eqn: addition variance}
\delta^2 (x \pm y) &= \zeta_x(2, \kappa_x) (\delta x)^2 + \zeta_y(2, \kappa_y) (\delta y)^2; \\
\label{eqn: sum leakage}
\varepsilon(x \pm y) &= \frac{\omega_x (\delta x)^2 + \omega_y (\delta y)^2}{(\delta x)^2 + (\delta y)^2};
\end{align}
Formulas \eqref{eqn: addition mean}, \eqref{eqn: addition variance}, and \eqref{eqn: sum leakage} show the mean, variance, and ideal leakage of $x \pm y$.
They become the convolution results for $x \pm y$ \cite{Probability_Statistics} when $N \rightarrow \infty$.
\begin{align}
\label{eqn: multiplication mean}
\overline{x y} &= xy; \\
\label{eqn: multiplication variance}
P(x y)^2 &= \zeta_x(2, \kappa_x) P(x)^2 + \zeta_y(2, \kappa_y) P(y)^2 + \zeta_x(2, \kappa_x) \zeta_y(2, \kappa_y) P(x)^2 P(y)^2; \\
\label{eqn: product leakage}
\varepsilon(x y) &= \frac{\omega_x P(x)^2 + \omega_y P(y)^2 +  (\omega_x + \omega_y - \omega_x \omega_y) P(x)^2 P(y)^2}{P(x)^2 + P(y)^2 + P(x)^2 P(y)^2}; 
\end{align}
Formulas \eqref{eqn: multiplication mean}, \eqref{eqn: multiplication variance}, and \eqref{eqn: product leakage} display the mean, variance, and ideal leakage of $x y$.
\begin{align}
a_{m,n} &\equiv \frac{x^m}{m!\,n!\,x^y}\,\frac{\partial^{m+n}x^y}{\partial x^m\,\partial y^n}; \\
\label{eqn: x^y mean}
\frac{\overline{x^y}}{x^y} &= 1 + \sum_{0 \le m,n}^{2 \le m+n} a_{m,n}\,P(x)^m(\delta y)^n\,\zeta_x(m,\kappa_x)\,\zeta_y(n,\kappa_y); \\
\label{eqn: x^y variance}
\frac{\delta^2 x^y}{(x^y)^2} &= \sum_{0 \le m,n}^{2 \le m+n} P(x)^m (\delta y)^n 
	\sum_{0 \le i \le m,\, 0 \le j \le n}^{1 \le i+j \le m+n-1} a_{i,j}\,a_{m-i,n-j} \\
	&\quad\quad \zeta_x(m, \kappa_x) \zeta_y(n, \kappa_y) - \zeta_x(i, \kappa_x) \zeta_y(j, \kappa_y) \zeta_x(m-i, \kappa_x) \zeta_y(n-j, \kappa_y); \nonumber
\end{align}
Formulas \eqref{eqn: x^y mean} and \eqref{eqn: x^y variance} display the mean, and variance of $x^y$, whose convergence range is $P(x)\,\kappa_x \lesssim 1$, with no limit on $\delta y$.

\subsection{Uncorrelated Uncertainty Condition}

\begin{figure}
\centering
\includegraphics[height=2.5in]{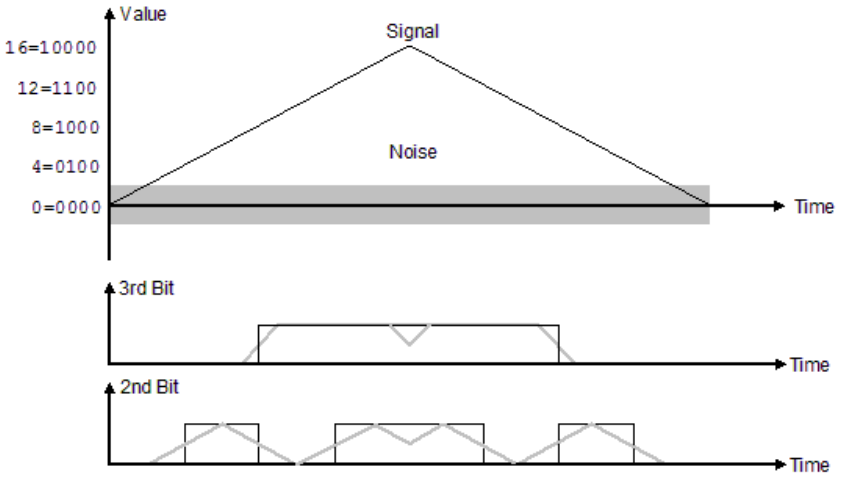} 
\captionof{figure}{
The digitization of a triangular wave, showing that noise attenuates the correlation to signal with each digit \cite{Prev_Precision_Arithmetic}.
}
\label{fig: Signal_and_Noise}
\end{figure}

\begin{figure}[p]
\centering
\includegraphics[height=2.25in]{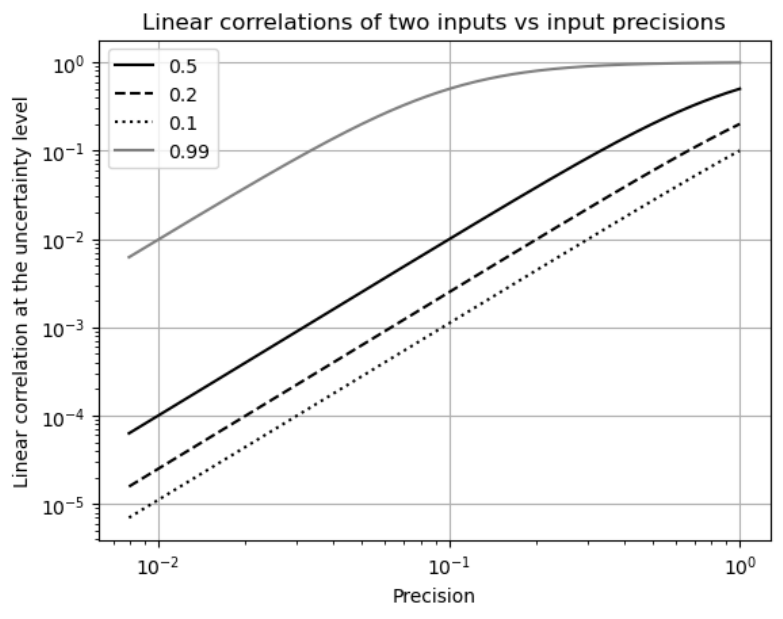} 
\captionof{figure}{
$\gamma_P^{1,1}$ versus $P$ for different $\gamma^{1,1}$ in Formula \eqref{eqn: uncertainty correlation (1,1)}.
The legend shows $\gamma^{1,1}$.
}
\label{fig: Independent_Uncertainty_Assumption}
\end{figure}

Formula \eqref{eqn: bound moment 2d} is applicable only when the two inputs $x$ and $y$ are measured together as a data set $\{(x_j, y_j), j=1,2 \dots N\}$.
When $x$ and $y$ are measured independently, their uncertainty can be approximated as independent of each other under the \emph{uncorrelated uncertainty condition} \cite{Prev_Precision_Arithmetic}.
The uncorrelated uncertainty condition can also apply to $x$ and $y$ which are measured together.

Figure \ref{fig: Signal_and_Noise} shows the effect of noise on the most significant two bits of a 4-bit measured signal.  
Its top chart shows a triangular waveform between 0 and 16 as a black line, and white noise between $-2$ and $+2$, using the grey area.  
The measured signal is the sum of the triangle waveform and the noise.  
The middle chart of Figure \ref{fig: Signal_and_Noise} shows the values of the 3rd bit of the true signal as a black line, and the mean values of the 3rd bit of the measurement as a grey line.  
The 3rd bit is affected by the noise during its transition between 0 and 1.  
For example, when the signal is slightly below 8, only a small positive noise can turn the 3rd bit from 0 to 1.  
The bottom chart of Figure \ref{fig: Signal_and_Noise} shows the values of the 2nd bit of the signal and the measurement as a black line and a grey line, respectively.  
Figure \ref{fig: Signal_and_Noise} clearly shows that the correlation between the measurement and the true signal is less at the 2nd bit than at the 3rd bit.  
The above conclusion agrees with the common experience that, below the noise level of measured signals, noise rather than true signals dominates each bit.  

Let two measured inputs be $x + y$ and $x + z$, in which $x$ is the part common to both inputs, while $y$ and $z$ are the parts independent of each other and of $x$, such as noises.
Let $\overline{x} = \overline{y} = \overline{z} = 0$, let $\sigma_x$, $\sigma_y$, and $\sigma_z$ be their distribution deviations, and let $\zeta_x(j) \equiv \overline{(x / \sigma_x)^j}$, $\zeta_y(j)$, and $\zeta_z(j)$ be their normalized moments.
For simplicity, let $\sigma_y = \sigma_z$, and let $\eta \equiv \sigma_y / \sigma_x$.
\begin{align}
\label{eqn: overal correlation}
\gamma^{m,n} &\equiv \frac{\overline{(x + y)^m (x + z)^n} - \overline{(x + y)^m}\;\overline{(x + z)^n}}
		{\big(\zeta_x(m+n) - \zeta_x(m) \zeta_x(n)\big) (\sigma_x^2 + \sigma_y^2)^{\frac{m}{2}} (\sigma_x^2 + \sigma_z^2)^{\frac{n}{2}}} \nonumber \\
	&= \frac{\sum_{j=0}^{m-1} \sum_{k=0}^{n-1} \binom{m}{j} \binom{n}{k}\, \eta^{j+k}\, \zeta_y(j)\, \zeta_z(k)\, \big(\zeta_x(m+n-j-k) - \zeta_x(m-j)\, \zeta_x(n-k)\big)}
		{\big(\zeta_x(m+n) - \zeta_x(m) \zeta_x(n)\big) (1 + \eta^2)^{\frac{m+n}{2}}}; \\
\label{eqn: overal correlation (1,1)}
\gamma^{1,1} &= \frac{1}{1 + \eta^2}; \\
\label{eqn: uncertainty correlation (1,1)}
\gamma_P^{1,1} &= \frac{1}{\Big(\frac{1}{\gamma^{1,1}} - 1\Big) \frac{1}{P^2} + 1}; \\
\label{eqn: uncertainty correlation}
\gamma_P^{m,n} &= \gamma^{m,n}\Big(\eta = \sqrt{\frac{1}{\gamma_P^{1,1}} - 1}\Big)
	= \frac{m\, n\, \zeta_y(m-1)\, \zeta_z(n-1)}{\zeta_x(m+n) - \zeta_x(m) \zeta_x(n)} \gamma_P^{1,1} + O\big((\gamma_P^{1,1})^\frac{3}{2}\big); \\
\label{eqn: uncertainty correlation (3,1)}
\gamma_P^{3,1} &= \frac{3}{\zeta_x(4)} \gamma_P^{1,1} + \Big(1 - \frac{3}{\zeta_x(4)}\Big) (\gamma_P^{1,1})^2;
\end{align}
Formula \eqref{eqn: overal correlation} calculates the correlation $\gamma^{m,n}$ between $(x + y)^m$ and $(x + z)^n$, which is $1$ when $\eta = 0$.
Formula \eqref{eqn: overal correlation (1,1)} shows that $\gamma^{1,1}$ is the correlation coefficient between the two inputs, so that $\eta^2 = 1/\gamma^{1,1} - 1$.
At the level of uncertainty with precision $P$, the deviation of the common part $x$ reduces to $P \sigma_x$ while those of $y$ and $z$ remain, so that $\eta$ becomes $\eta_P = \eta / P$, which results in Formula \eqref{eqn: uncertainty correlation (1,1)} for the correlation $\gamma_P^{1,1}$ at the level of uncertainty.
Substituting $\eta_P^2 = 1/\gamma_P^{1,1} - 1$ into Formula \eqref{eqn: overal correlation} turns each term into $(\gamma_P^{1,1})^{\frac{m+n-j-k}{2}} (1 - \gamma_P^{1,1})^{\frac{j+k}{2}}$, which results in Formula \eqref{eqn: uncertainty correlation}.
Thus, $\gamma_P^{m,n}$ is determined by $\gamma_P^{1,1}$ and the distributions of $x$, $y$, and $z$, and it is a polynomial of $\gamma_P^{1,1}$ when $x$, $y$, and $z$ are symmetric.
For example, Formula \eqref{eqn: uncertainty correlation (3,1)} shows $\gamma_P^{3,1}$ for any distribution of $y$ and $z$, and $\gamma_P^{3,1} = \gamma_P^{1,1}$ when $x$ is Gaussian.
Therefore, when $\gamma_P^{1,1}$ is close to zero, $\gamma_P^{m,n}$ of all orders are close to zero, and the two inputs can be regarded as independent of each other at the level of uncertainty.
When $\sigma_y \neq \sigma_z$, $\gamma^{1,1} = 1/\sqrt{(1 + (\sigma_y/\sigma_x)^2)(1 + (\sigma_z/\sigma_x)^2)} \equiv 1/(1 + \eta^2)$, and the same conclusion holds.

Figure \ref{fig: Independent_Uncertainty_Assumption} shows that $\gamma_P^{1,1}$ decreases from $\gamma^{1,1}$ at $P = 1$, and approaches $P^2 \gamma^{1,1}/(1 - \gamma^{1,1})$ when $P \ll 1$.
When $\gamma_P^{1,1}$ is below a threshold, the two inputs can be regarded as independent of each other.
For example, if such threshold is $10^{-2}$, $P \leq 0.3$ when $\gamma^{1,1} = 0.1$, or $P \leq 0.01$ when $\gamma^{1,1} = 0.5$.
Because digitization at the level of uncertainty further attenuates the correlation, as shown in Figure \ref{fig: Signal_and_Noise}, Figure \ref{fig: Independent_Uncertainty_Assumption} serves as an upper bound for the uncorrelated uncertainty condition.

\subsection{Dependency Tracing}

Formula \eqref{eqn: dependency expression} displays different algebraic forms, which has an identical Taylor expansion.
Because statistical Taylor expansion has deterministic and analytic closed form, it no longer has the conventional dependency problem \cite{Precise_Numerical_Methods, Algorithms_Accuracy, Interval_Analysis, Interval_Analysis_Theory_Applications}.

When all inputs satisfy the uncorrelated uncertainty condition, statistical Taylor expansion traces dependencies through the intermediate steps.
Some typical examples of dependency tracing:
\begin{itemize}
\item The dependency tracing of $\delta^2 (f + g)$ is illustrated by $\delta^2 (f - f) = 0$, and $\delta^2 (f(x) + g(y)) = \delta^2 f + \delta^2 g$, with the latter corresponding to Formula \eqref{eqn: addition variance}.   

\item The dependency tracing of $\delta^2 (f g)$ is illustrated by $\delta^2 (f/f) = 0$, $\delta^2 (f f) = \delta^2 (f^2)$, and $\delta^2 (f(x) g(y)) = \overline{f}^2 (\delta^2 g) + (\delta^2 f) \overline{g}^2 +  (\delta^2 f) (\delta^2 g)$, with the latter corresponding to Formula \eqref{eqn: multiplication variance}.  

\item The dependency tracing of $\delta^2 f(g(x))$ is demonstrated by $\delta^2 (f^{-1}(f(x))) = (\delta x)^2$.  
For a reversible transformation such as matrix inversion or FFT (Fast Fourier Transformation), after a \emph{round-trip transformation} which is a forward transformation followed by a reverse transformation, the original inputs should be restored.

\end{itemize}
The dependency tracing is more than cancellation in Taylor expansion.  
For example, Sections \ref{sec: matrix} and \ref{sec: FFT} will provide examples of round-trip transformations in which a forward transformation and its reverse transformation are executed as two separate steps.

\begin{align}
\delta^2 (f + g + h) &= \delta^2 f + \delta^2 g + \delta^2 h  \nonumber  \\ 
\label{eqn: sum variance}
	&+ 2 (\overline{fg} - \overline{f}\overline{g}) + 2 (\overline{fh} - \overline{f}\overline{h}) + 2 (\overline{gh} - \overline{g}\overline{h}); \\
\delta^2 \left( \frac{f^{(m)}_x}{m!} \tilde{x}^m + \frac{f^{(n)}_x}{n!} \tilde{x}^n \right) &=
    (\delta x)^{2m} (\frac{f^{(m)}_x }{m!})^2 \left(\zeta(2m) - \zeta(m)^2\right) 
    	+ (\delta x)^{2n} (\frac{f^{(n)}_x }{n!})^2 \left(\zeta(2n) - \zeta(n)^2 \right)  \nonumber  \\
\label{eqn: Taylor inner dependency}
	&+ 2 (\delta x)^{m+n} \frac{f^{(m)}_x f^{(n)}_x}{m! \;n!} \left(\zeta(m+n) - \zeta(m) \zeta(n) \right);
\end{align}
From the sum variance Formula \eqref{eqn: sum variance} \cite{Probability_Statistics}, Formula \eqref{eqn: Taylor inner dependency} decomposes and leads to Formula \eqref{eqn: Taylor 1d variance}.
Because Formula \eqref{eqn: sum variance} comes from the variance definition, Formula \eqref{eqn: Taylor 1d variance} does not contain the dependency problem, in contrast to Taylor expansion using interval arithmetic \cite{Interval_Analysis, Taylor_Models, Taylor_Forms_Limits}.
If $\delta x$ is replaced by $\tilde{x} - x$, the major difference between Formula \eqref{eqn: Taylor 1d variance} and the equation for interval Taylor expansion \cite{Taylor_Models, Taylor_Forms_Limits} is the distribution-specific factor $\zeta(m+n) - \zeta(m) \zeta(n)$.

Statistical Taylor expansion employs dependency tracing to ensure that the calculated mean and variance satisfy statistics rigorously.
Dependency tracing also implies that the results of statistical Taylor expansion must remain path-independent.
However, dependency tracing comes at a cost: variance calculations are generally more complex than value calculations and exhibit a narrower convergence range for input variables.

\subsection{Traditional Execution and Dependency Problem}

Dependency tracing requires applying statistical Taylor expansion to the complete analytic form of a function, not to its decomposed parts.
This requirement often conflicts with conventional numerical methods for analytic functions:
\begin{itemize}

\item 
Statistical Taylor expansion can explain the dependency problem. 
For example, the dependency problem in Formula \eqref{eqn: dependency expression} is due to false independence assumptions between $x^2$ and $x$, or between $x - 1$ and $x$, respectively.

\item
Large calculations are often divided into sequential steps, such as computing $f(g(x))$ in two steps: first $y = g(x)$, then $f(y)$.
This approach also introduces the dependency problem by ignoring dependency tracing within $g(x)$ affecting $f(g(x))$, such as $\overline{(\sqrt{x})^2} > \overline{\sqrt{x^2}} > \overline{x}$ and $\delta^2 (\sqrt{x})^2 > \delta^2 \sqrt{x^2} > \delta^2 x$.

\item
Conditional executions are often employed to optimize performance and minimize rounding errors, for example, using Gaussian elimination to minimize floating-point rounding errors in matrix inversion \cite{Linear_Algebra}.  
For dependency tracing, such conditional executions should instead be replaced by direct matrix inversion as described in Section \ref{sec: matrix}.

\item
Traditionally, intermediate variables are widely used in computations; however, this practice disrupts dependency tracing by obscuring the relationships among the original input variables.

\end{itemize}
Dependency tracing therefore removes nearly all flexibility from traditional numerical executions, effectively eliminating the associated dependency problems.
Consequently, conventional numerical algorithms must be reexamined, and many will need to be redesigned, to align with the principles of statistical Taylor expansion.

\section{Variance Arithmetic}
\label{sec: variance arithmetic}

Variance arithmetic implements statistical Taylor expansion.
It represents an imprecise value $x \pm \delta x$ as a pair of 64-bit floating-point numbers and performs all computation using standard floating-point arithmetic.

Because of the finite precision and limited range of the floating-point representation, $\zeta(n, \kappa)$ can be computed only to limited terms.
Consequently, the following numerical rules are introduced:
\begin{itemize}
\item \emph{finite}: The resulting value and variance must remain finite.

\item \emph{monotonic}: As a necessary condition for convergence, the last $20$ terms of the expansion must decrease monotonically in absolute value, ensuring that under a null model where each consecutive term is equally likely to increase or decrease, the probability of observing 20 spurious monotonic decreases by chance is at most $2^{-20} \simeq 9.53 \times 10^{-7}$.

\item \emph{stable}: To avoid truncation error \cite{Numerical_Recipes}, the absolute value of the last expansion term must be less than $\epsilon$ times both the result deviation and the result absolute value, where $\epsilon \simeq 5.73 \times 10^{-7}$ is the bounding leakage for Gaussian distribution with $\hat{\kappa} = 5$.
This rule ensures sufficiently fast convergence in the context of monotonic convergence.

\item \emph{positive}: At every expansion order, the expansion variance must be positive.

\item \emph{reliable}: At every order, the deviation of the variance must be less than $1/5$ times the value of the variance.

\end{itemize}

For simplicity, the Taylor coefficients in Formulas \eqref{eqn: Taylor 1d} and \eqref{eqn: Taylor 2d} are assumed to be precise.

\subsection{Monotonic}

\begin{figure}[p]
\centering
\includegraphics[height=2.5in]{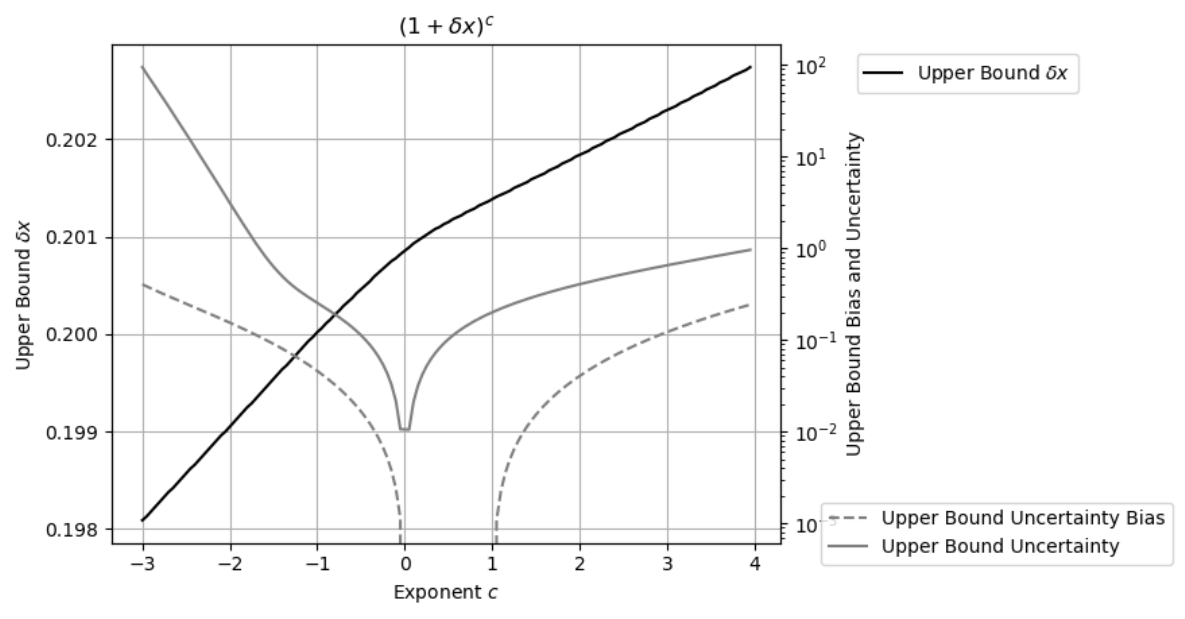} 
\captionof{figure}{
Measured upper bound $\delta x$ (left y-axis) for $(1 \pm \delta x)^c$ across different values of $c$ (x-axis) for Gaussian uncertainty.
The corresponding resulting uncertainty bias and deviation are also shown (right y-axis).
Natural-number values of $c$ are omitted because the Taylor series terminates at the $c$-th expansion term so that $\delta x$ has no upper bound.}
\label{fig: Pow_Conv_Edge}
\end{figure}

\ifdefined\Verbose

\begin{figure}[p]
\centering
\includegraphics[height=2.5in]{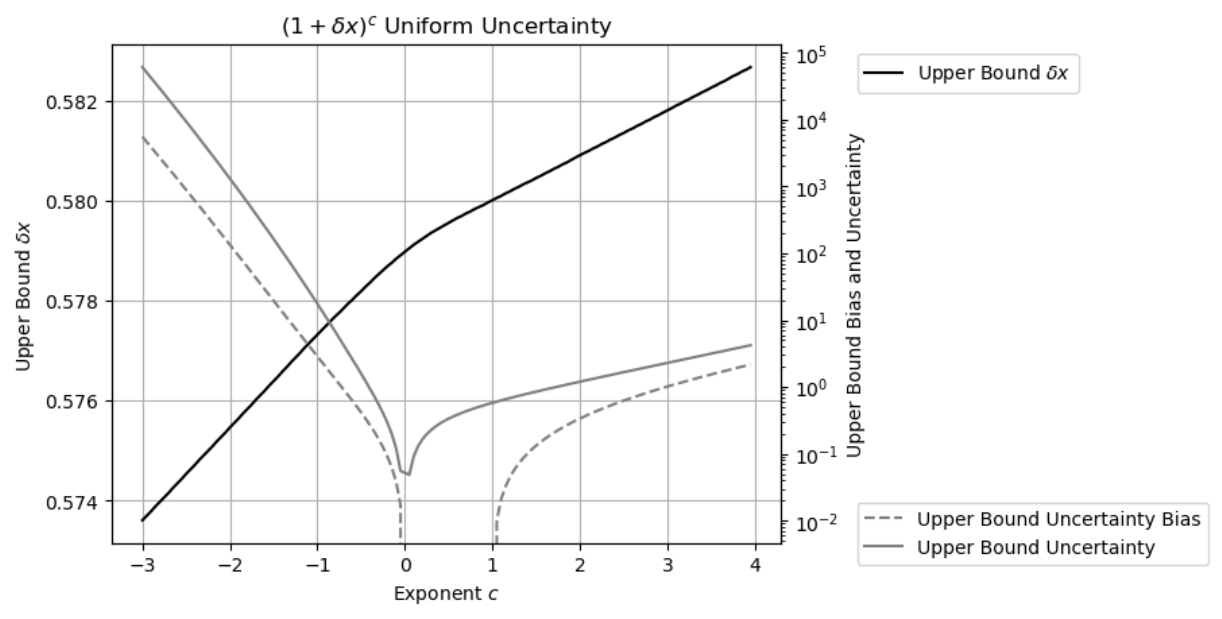} 
\captionof{figure}{
Measured upper bound $\delta x$ (left y-axis) for $(1 \pm \delta x)^c$ across different values of $c$ (x-axis) for uniform uncertainty.
The corresponding resulting uncertainty bias and deviation are also shown (right y-axis).
Natural-number values of $c$ are omitted because the Taylor series terminates at the $c$-th expansion term so that $\delta x$ has no upper bound.
For $c \in [0, 1]$, the uncertainty bias is negative and so does not appear on this log-scaled plot.
}
\label{fig: Pow_Conv_Edge_Uniform}
\end{figure}

\fi

Beyond an upper bound $\delta x$, the expansion is no longer monotonic for $e^{x \pm \delta x}$, $\log(x \pm \delta x)$, and $(x \pm \delta x)^c$.
For Gaussian input uncertainty with $\hat{\kappa}=5$, the numerical results of the monotonic requirement are:
\begin{itemize}
\item 
For $e^{x \pm \delta x}$, $\delta x \lesssim 19.864$ and $P(e^{x \pm \delta x}) \lesssim 1681.767$ regardless of $x$.
These limits follow directly from the relationship $\delta x \rightarrow P(e^x)$, as indicated in Formula \eqref{eqn: exp precision}.

\item 
For $\log(x \pm \delta x)$, $P(x) \lesssim 0.20086$ and $\delta \log(x \pm \delta x) \lesssim 0.213$ regardless of $x$.
These limits follow directly from the relationship $P(x) \rightarrow \delta \log(x)$, as indicated in Formula \eqref{eqn: log precision}.

\item
For $(x \pm \delta x)^c$, except when $c$ is a natural number, the upper bound $P(x)$ is close to $1/5$ but increasing with $c$. 
This trend is shown in Figure \ref{fig: Pow_Conv_Edge}.
\end{itemize}
\ifdefined\Verbose
Figure \ref{fig: Pow_Conv_Edge_Uniform} shows the upper bound $P(x)$ for $(x \pm \delta x)^c$ when the input uncertainty distribution is Uniform.
The Uniform-distribution trend of $P(x)$ upper bounds increasing with $c$ closely follows the Gaussian trend in Figure \ref{fig: Pow_Conv_Edge}, only with the y-axis scale set by $1/\sqrt{3}$ instead of $1/5$.
\else
A similar trend holds for Uniform input uncertainty, where $\hat{\kappa} = \sqrt{3}$.
\fi

\subsection{Positive}

\begin{figure}[p]
\centering
\includegraphics[height=2.5in]{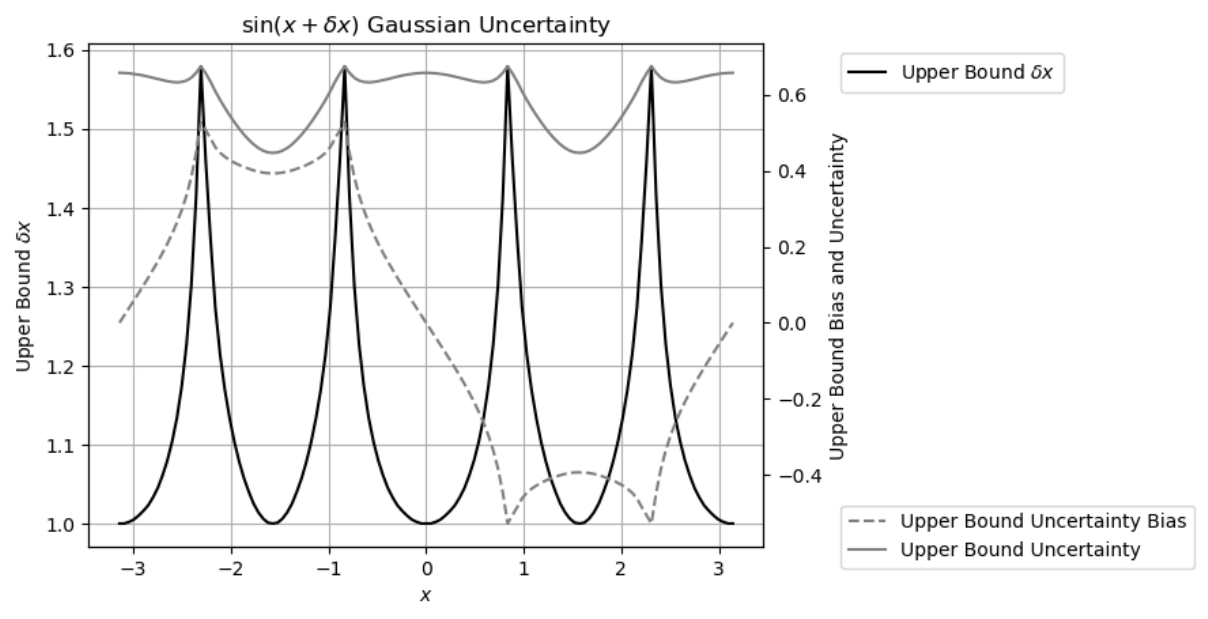} 
\captionof{figure}{
Measured upper bound $\delta x$ (left y-axis) for $\sin(x \pm \delta x)$ across different values of $x$ (x-axis) for Gaussian uncertainty.
The corresponding resulting uncertainty bias and deviation are also shown (right y-axis).
}
\label{fig: Sin_Conv_Edge}
\end{figure}

\ifdefined\Verbose
\begin{figure}[p]
\centering
\includegraphics[height=2.5in]{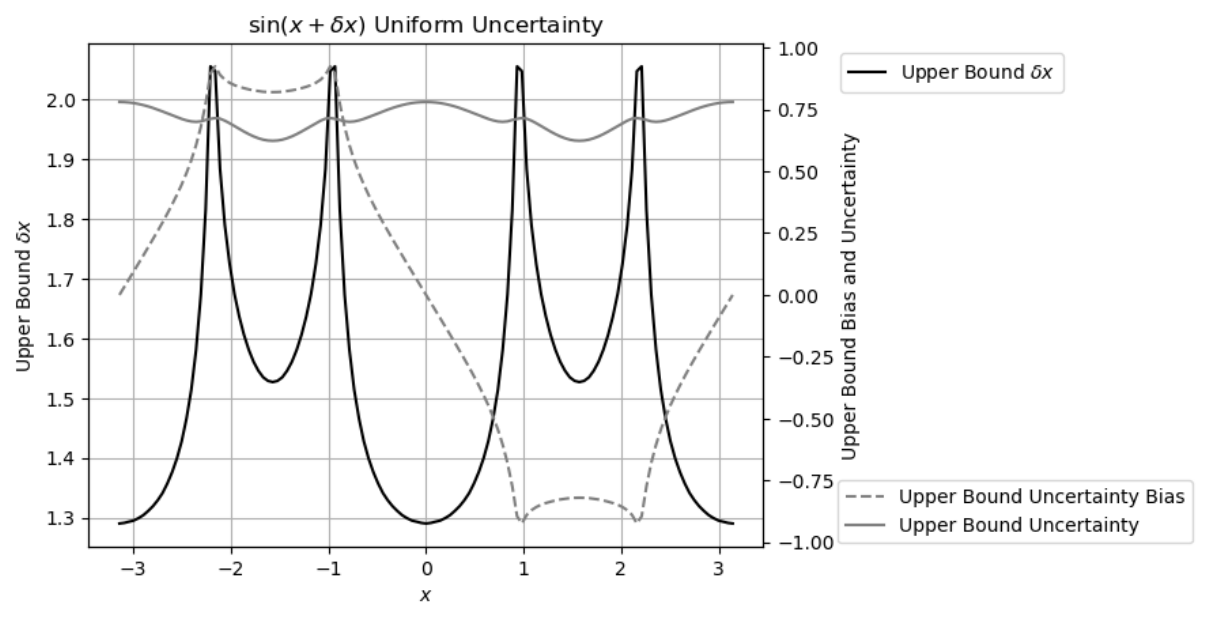} 
\captionof{figure}{
Measured upper bound $\delta x$ (left y-axis) for $\sin(x \pm \delta x)$ across different values of $x$ (x-axis) for Uniform uncertainty.
The corresponding resulting uncertainty bias and deviation are also shown (right y-axis).
}
\label{fig: Sin_Conv_Edge_Uniform}
\end{figure}
\fi

In addition to convergence, the variance expansion may yield negative results when the expansion is terminated at any finite order, as in Formula \eqref{eqn: sin precision} for $\sin(x \pm \delta x)$.
Therefore an upper bound for $\delta x$ is needed to guarantee non-negative variance.
Figure \ref{fig: Sin_Conv_Edge} shows that the upper bound of $\delta x$ for $\sin(x \pm \delta x)$ varies with period $\pi$ between $0.318 \pi$ and $0.416 \pi$ for Gaussian input uncertainty.
Because $\sin(x) \in [-1, +1]$, $\overline{\sin(x)} \pm \delta \sin(x)$ should also follow this range, as indicated in Figure \ref{fig: Sin_Conv_Edge}.
\ifdefined\Verbose
A similar trend holds when the input uncertainty is Uniform, but with alarger upper bound $\delta x$, as shown in Figure \ref{fig: Sin_Conv_Edge_Uniform}.
\else
A similar trend holds when the input uncertainty is Uniform, but with a larger upper bound $\delta x$.
\fi

\subsection{Floating-Point Rounding Errors}

Variance arithmetic incorporates floating-point rounding errors as $\delta x$ when converting a floating-point value $x$ into $x \pm \delta x$. 
Unless the least 20 bits of the significand of $x$ are all zero, $\delta x$ is assumed to be $1/\sqrt{3}$ times the ULP (Unit in the Last Place \cite{Floating_Point_Standard}) of $x$, because rounding errors are shown to be uniformly distributed over the range $[-\text{LSB}, +\text{LSB}]$ (Least significant bit) \cite{Prev_Precision_Arithmetic}.

\ifdefined\Verbose

\subsection{Comparison}

Two imprecise values can be compared statistically based on their difference:
\begin{itemize}
\item 
When the value difference is zero, the two imprecise values are considered equal.  
In statistics, such two values have a $50\%$ possibility of being either less than or greater than each other but zero probability of being exactly equal \cite{Probability_Statistics}.
In variance arithmetic, however, they are considered equal.

\item 
Otherwise, the standard z-statistic method \cite{Probability_Statistics} is applied to determine whether two imprecise values are statistically equal, less than, or greater than each other.
For example, the difference between $1.002 \pm 0.001$ and $1.000 \pm 0.002$ is $0.002 \pm 0.002236$, yielding $z = 0.002 / 0.002236$.
The probability that they are not equal is $\xi(|z|/\sqrt{2}) = 62.8\%$, in which $\xi(z)$ is the Normal error function \cite{Probability_Statistics}.
If the threshold probability for inequality is set at $50\%$, then $1.000 \pm 0.002 < 1.002 \pm 0.001$.
Because the result of comparison depends on the threshold probability, which is application-specific, comparison is not part of variance arithmetic.

\end{itemize}

\fi

\section{Validation}
\label{sec: validation}

Analytic functions or algorithms with precisely known results are used to evaluate the outputs of variance arithmetic based on the following statistical properties: 
\begin{itemize}

\item \emph{Value error}: the difference between the numerical result and the corresponding known precise analytic result.

\item \emph{Normalized error}: the ratio of a value error to the corresponding result deviation from statistical Taylor expansion.

\item \emph{Error deviation}: the standard deviation of a set of normalized errors.

\item \emph{Error distribution}: the histogram of a set of normalized errors.

\end{itemize}

Once input errors from all sources are accounted for precisely, \emph{ideal coverage} is achieved in either context:
\begin{itemize}
\item \emph{Distribution Test}:
When comparing the calculated mean and deviation with the result data set, the error deviation is exactly $1$ and the error distribution is Normal, regardless of input uncertainty distribution.
Such convergence to Gaussian distributions occurs rapidly \cite{Prev_Precision_Arithmetic} because of the central limit theorem \cite{Probability_Statistics} when a calculation contains a large amount of summation, subtraction, and multiplication (which is implemented as multiple additions digitally \cite{Floating_Point_Arithmetic}).

\item \emph{Value Test}: 
When comparing values one-by-one between a calculated data set and the corresponding result data set, the error deviation is much less than $1$ and the error distribution is Delta (Dirac-delta-like, concentrated at zero error).
For example, a round-trip test is a value test.

\end{itemize}

However, if the input uncertainty is known only to an order of magnitude, \emph{proper coverage} is achieved when the error deviations fall within the range $[0.1, 10]$.

When an input contains unspecified errors, such as numerical errors in library functions or floating-point rounding errors, Gaussian noise with progressively increasing deviations can be added to inputs, until ideal coverage is attained.
The minimal noise deviation required for the ideal coverage provides a good estimate of the magnitude of the unspecified input uncertainty deviations.
Achieving ideal coverage serves as a necessary verification step to ensure that statistical Taylor expansion has been applied correctly within the given context.
The input noise range that yields ideal coverage defines the ideal application range for the analytic function.
For the value test, the \emph{error slope}, which is defined as the slope of the linear regression of normalized error deviation on input noise, should be exactly $-1$.

\section{Polynomial}
\label{sec: polynomial}

Formula \eqref{eqn: polynomial Taylor} presents a polynomial Taylor expansion:
\begin{align}
\label{eqn: polynomial Taylor}
\sum_{j=0}^{N} c_j (x + \tilde{x})^j &= \sum_{j=0}^{N} \tilde{x}^{j} P_j, \quad
	P_j \equiv \sum_{k=0}^{N-j} x^k c_{j + k} \begin{pmatrix} j + k \\ j \end{pmatrix};
\end{align}

Because the variance computation for $N$-order polynomial requires moments up to order $2N$, $N$ can reach only half of the maximal expansion order of Formula \eqref{eqn: Taylor 1d variance}, for example, $N = 224 = 448/2$ for $\zeta(2n,5)$ of Gaussian input uncertainty.

\subsection{Residual Error}

\begin{figure}[p]
\centering
\includegraphics[height=2.5in]{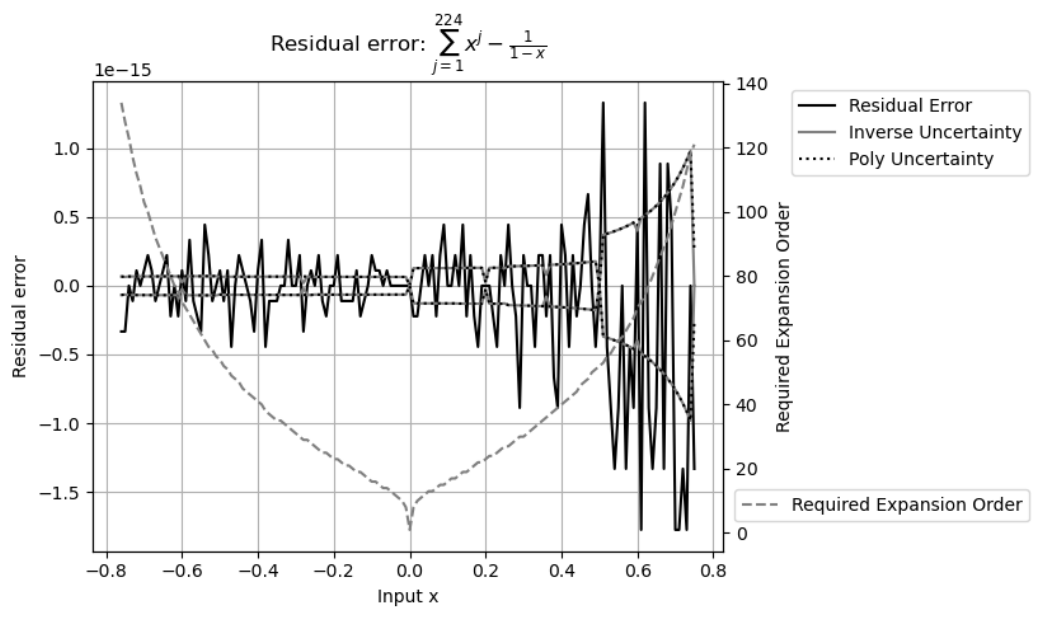}
\captionof{figure}{
Residual error of $\sum_{j=0}^{224} x^j - \frac{1}{1 - x}$ vs $x$ (x-axis).
The left y-axis shows both the value and the uncertainty of the residual errors.
The right y-axis indicates the expansion order needed to reach convergence for each $x$.
}
\label{fig: Poly_x}
\end{figure}

Figure \ref{fig: Poly_x} shows the residual error of $\sum_{j=0}^{224} x^j  - \frac{1}{1 - x}$.
The figure also displays the required expansion orders for $\frac{1}{1 - x}$, which are all less than $224$.
Therefore, the residual error reflects the rounding error between $\sum_{j=0}^{224} x^j$ and $\frac{1}{1 - x}$.
A detailed analysis indicates that the maximal residual error is four times the ULP of $\frac{1}{1 - x}$.
The calculated uncertainty bounds the residual error effectively for all $x$, with a $2.60$ error deviation when the expansion order is less than $224$.
Variance arithmetic can provide proper coverage for rounding errors.

\subsection{Continuity}

In variance arithmetic, the resulting mean, variance, and error distribution are generally continuous across parameter space.
For example, when $c$ is a natural number, $(x \pm \delta x)^c$ becomes a polynomial with no upper bound on $\delta x$, in contrast to when $c$ is not a natural number as shown in Figure \ref{fig: Pow_Conv_Edge}.
However, the resulting mean, variance and error distribution of $(x \pm \delta x)^c$ remain continuous across $c = n$.

\subsection{Distributional Pole}

\begin{figure}[p]
\centering
\includegraphics[height=2.5in]{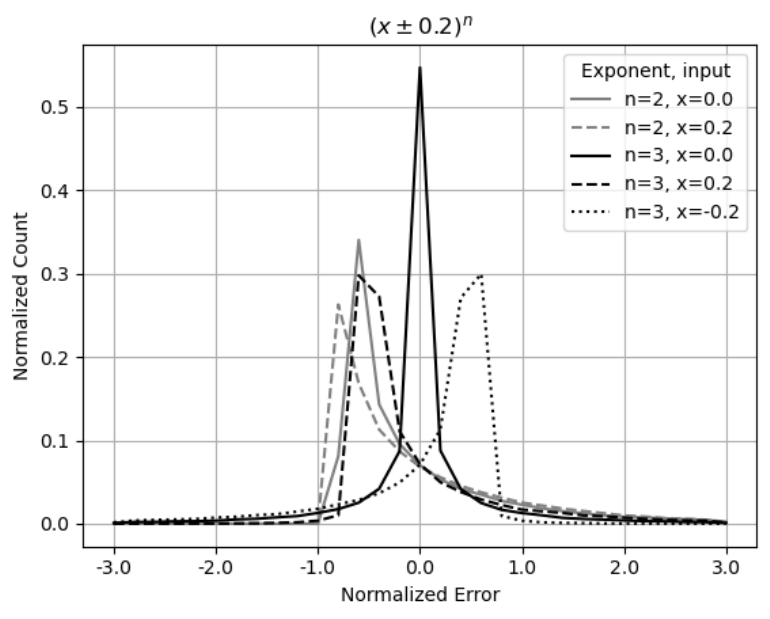} 
\captionof{figure}{
Error distributions for $(x \pm 0.2)^n$, with $x = 0, -0.2, +0.2$, and $n = 2, 3$, as indicated in the legend.
}
\label{fig: Poly_Continuity}
\end{figure}

A statistical bounding range in variance arithmetic can include a distributional pole, such as around $(0 \pm \delta x)^c, c > 1$.
The presence of such poles does not disrupt the continuity of the resulting mean, variance, or error distribution.
Figure \ref{fig: Poly_Continuity} illustrates the error distributions of $(x \pm 0.2)^n$ when $x = 0, -0.2, +0.2$ and $n = 2, 3$.
\begin{itemize}
\item When the second derivative is zero, the resulting distribution is symmetric, two-sided, and Delta-like, such as when $n = 3, x = 0$.

\item When the second derivative is positive, the resulting distribution is right-sided Delta-like, such as the distribution when $n = 2, x = 0$, or when $n = 2, x = \pm 0.2$, or when $n = 3, x = 0.2$.

\item When the second derivative is negative, the resulting distribution is left-sided Delta-like, such as when $n = 3, x = -0.2$, which is the mirror image of the distribution when $n = 3, x = 0.2$.

\end{itemize}

\section{Matrix Calculations}
\label{sec: matrix}

\begin{figure}[p]
\centering
\includegraphics[height=2.5in]{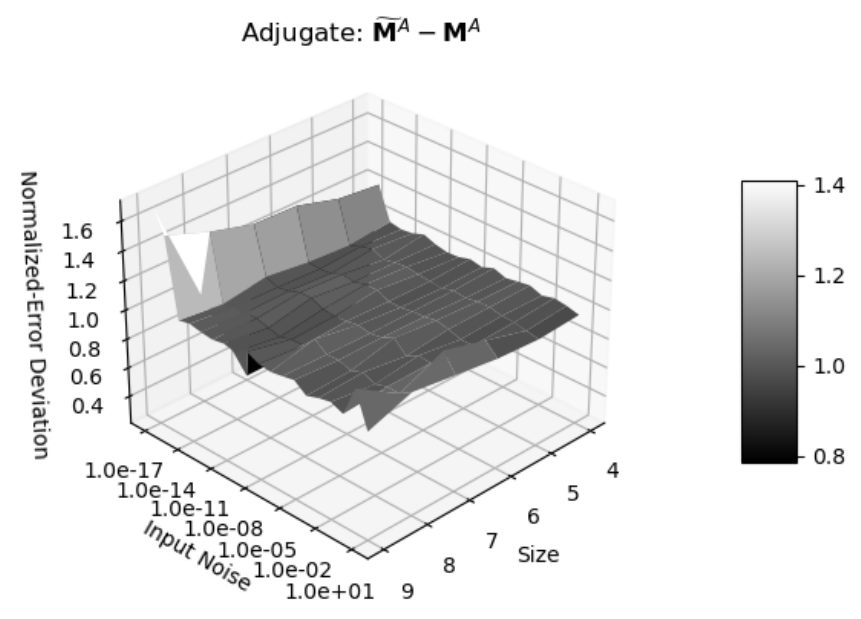} 
\captionof{figure}{
Error deviations (z-axis) of an adjugate matrix $\widetilde{\mathbf{M}}^A - \mathbf{M}^A$ as a function of input noise precision (x-axis) and matrix size (y-axis).
}
\label{fig: Adjugate_Error_vs_Size_Noise}
\end{figure}

\begin{figure}[p]
\centering
\includegraphics[height=2.5in]{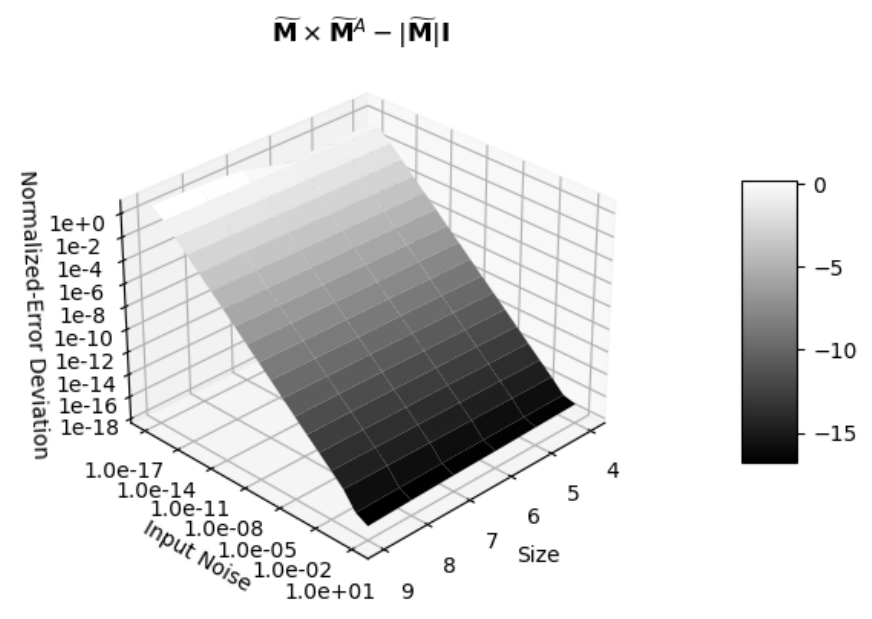} 
\captionof{figure}{
Error deviations (z-axis) of the difference between the two sides of Formula \eqref{eqn: adjugate matrix}, as a function of input noise precision (x-axis) and matrix size (y-axis).
}
\label{fig: Forward_Error_vs_Size_Noise}
\end{figure}

\begin{figure}[p]
\centering
\includegraphics[height=2.5in]{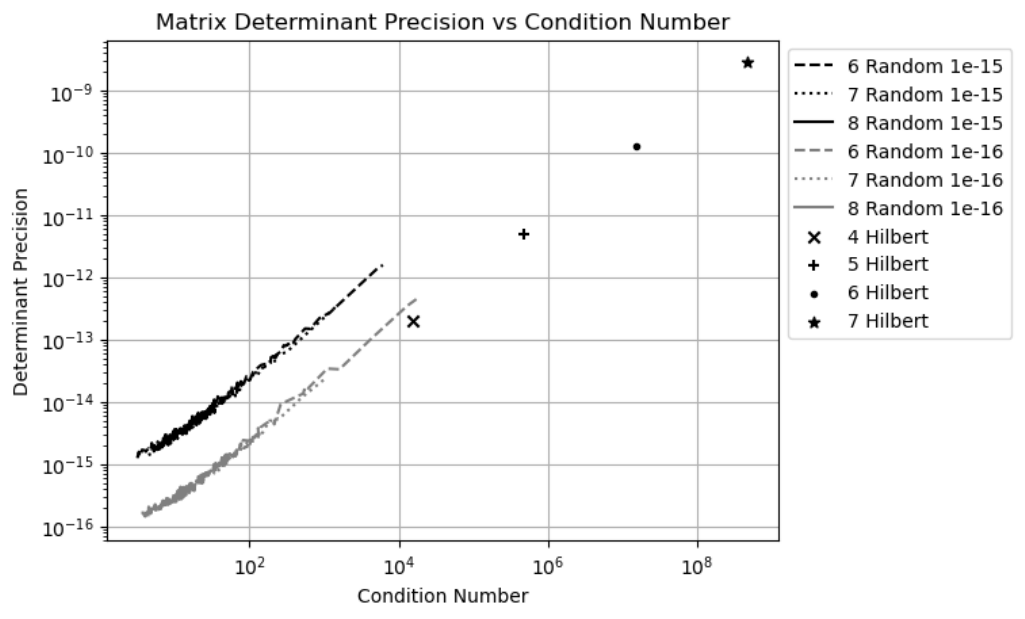} 
\captionof{figure}{
Linear correlation between the precision of a matrix determinant (y-axis) and its condition number (x-axis).
The legend shows the size of the matrix, the deviation of added noise, as well as the matrix type: \textit{Random} for randomly generated matrices, and \textit{Hilbert} for the Hilbert matrices.
The noise of a Hilbert matrix is the ULP of each of its elements, which is slightly less than $10^{-16}$ as expected.
}
\label{fig: Matrix_Determinant_Prec_vs_Condition}
\end{figure}

\begin{align}
\label{eqn: random determinant}
|\widetilde{\mathbf{M}}| &\equiv \sum_{[j_1\dots j_n]_n} (-1)^{[j_1\dots j_n]_n}  \prod_{i=1 \dots n} (x_{i,j_i} + \tilde{x}_{i,j_{i}}) \\
\label{eqn: determinant mean}
\overline{|\mathbf{M}|} &= |\mathbf{M}|; \\
\label{eqn: determinant variance}
\delta^2 |\mathbf{M}| &= \sum_{m=1}^{n} \sum_{<i_1 \dots i_m>_n} \sum_{[j_1 \dots j_m]_n}
  	|\mathbf{M}_{<i_1 \dots i_m>_n, <[j_1 \dots j_m]>_n}|^2 \prod _{k=1 \dots m} \zeta_{i_k,j_k}(2, \kappa_{i_k,j_k})(\delta x_{i_k, j_k})^2; 
\end{align}
Let $<j_1, j_{2} \dots j_m>_n$ denote a selection of $m$ numbers from the sequence of natural numbers $1,2 \dots n$ in sequence, which is a combination.
Let $[j_1, j_{2} \dots j_m]_n$ and $(-1)^{[j_1, j_{2} \dots j_m]_n}$ denote a permutation of $<j_1, j_{2} \dots j_m>_n$ and the corresponding sign \cite{Linear_Algebra}.
Let $\mathbf{M}$ be a square matrix of size $n$ with element $x_{i,j} \pm \delta x_{i,j}, i, j = 1,2\dots n$ at row index $i$ and column index $j$. 
Formula \eqref{eqn: random determinant} presents the Taylor expansion of the determinant for a matrix $\mathbf{M}$ when the uncertainties of matrix elements are all independent of each other, whereas Formulas \eqref{eqn: determinant mean} and \eqref{eqn: determinant variance} give the statistical Taylor expansion of the determinant $|\mathbf{M}|$.
In Formula \eqref{eqn: determinant variance}, $\mathbf{M}_{<i_1 \dots i_m>_n, [j_1 \dots j_m]_n}$ is a sub-matrix for $\mathbf{M}$, in which $<i_1 \dots i_m>_n$ contains the row indexes, $[j_1 \dots j_m]_n$ contains the column indexes \cite{Linear_Algebra}, and $<[j_1, j_{2} \dots j_m]>_n$ denotes the corresponding combination of $[j_1 \dots j_m]_n$. 

The square matrix whose element is $(-1)^{i+j}|M_{j,i}|$ is defined as the \emph{adjugate matrix} \cite{Linear_Algebra} $\mathbf{M}^A$ to the original square matrix $\mathbf{M}$. 
Each element of $\mathbf{M}$ is assigned a random integer from $[-2^8, +2^8]$ such that $\mathbf{M}^A$ can be computed exactly in floating-point arithmetic when $n \leq 7$.
By adding Gaussian noise $\tilde{x}_{i,j}$ of deviation $\delta x$ to each element of $\mathbf{M}$, $\widetilde{\mathbf{M}}$ is created as shown in Formula \eqref{eqn: random determinant}.
The value error of $\mathbf{M}^A$ is the difference between $\widetilde{\mathbf{M}^A}$ and $\mathbf{M}^A$, whereas the result deviation is calculated using Formula \eqref{eqn: determinant variance}, such that $\widetilde{\mathbf{M}^A} - \mathbf{M}^A$ is a distribution test.
Figure \ref{fig: Adjugate_Error_vs_Size_Noise} shows that the error deviations of $\widetilde{\mathbf{M}^A} - \mathbf{M}^A$ are very close to $1$ when $\delta x > 10^{-17}$, as $1.000 \pm 0.019$ on average.

Matrix calculation verifies both distribution test and value test of statistical Taylor expansion.

\begin{align}
\label{eqn: adjugate matrix}
\mathbf{M} \times \mathbf{M}^A &= \mathbf{M}^A \times \mathbf{M} = |\mathbf{M}| \mathbf{I}; \\
\label{eqn: inverse matrix}
\mathbf{M}^{-1} &\equiv\; \mathbf{M}^A / |\mathbf{M}|;
\end{align}
Let $\mathbf{I}$ be the identity matrix for $\mathbf{M}$ \cite{Linear_Algebra}.
Formula \eqref{eqn: adjugate matrix} shows the relationship between $\mathbf{M}^A$ and $\mathbf{M}$ which leads to the definition of inverse matrix $\mathbf{M}^{-1}$ in Formula \eqref{eqn: inverse matrix} \cite{Linear_Algebra}.
$\tilde{\mathbf{M}} \times \tilde{\mathbf{M}}^A - |\tilde{\mathbf{M}}| \mathbf{I}$ is a value test applied to each matrix element.
Figure \ref{fig: Forward_Error_vs_Size_Noise} shows the error deviation of a typical value test: the error deviation decreases linearly with increasing input uncertainty deviation $\delta x$, with error slopes very close to $-1$, as $-0.998 \pm 0.008$ on average.
The matrix size serves as the algorithm-specific dimension in Figure \ref{fig: Forward_Error_vs_Size_Noise}.

Matrix calculation verifies both distribution test and value test of statistical Taylor expansion.

\begin{align}
\label{eqn: inverse matrix 2 variance}
|\mathbf{M}|^4 &\delta^2 \left( \begin{matrix} x_{1,1}, x_{1,2} \\ x_{2,1}, x_{2,2} \end{matrix} \right)^{-1} \simeq\;  \\
	 &\left( \begin{matrix} x_{2,2}^4, &x_{1,2}^2 x_{2,2}^2 \\ x_{2,1}^2 x_{2,2}^2, &x_{1,2}^2 x_{2,1}^2 \end{matrix} \right) \zeta_{x_{1,1}}(2) (\delta x_{1,1})^2 
	 +\left( \begin{matrix} x_{2,1}^2 x_{2,2}^2, &x_{1,1}^2 x_{2,2}^2 \\ x_{2,1}^4, &x_{1,1}^2 x_{2,1}^2 \end{matrix} \right)\zeta_{x_{1,2}}(2)  (\delta x_{1,2})^2\nonumber \\
	&\left( \begin{matrix} x_{1,2}^2 x_{2,2}^2, &x_{1,2}^4 \\ x_{1,1}^2 x_{2,2}^2, &x_{1,1}^2 x_{1,2}^2 \end{matrix} \right) \zeta_{x_{2,1}}(2) (\delta x_{2,1})^2
	 +\left( \begin{matrix} x_{1,2}^2 x_{2,1}^2, &x_{1,1}^2 x_{1,2}^2 \\ x_{1,1}^2 x_{2,1}^2, &x_{1,1}^4 \end{matrix} \right) \zeta_{x_{2,2}}(2) (\delta x_{2,2})^2; \nonumber
\end{align}
Because an element of the original matrix $\mathbf{M}$ appears multiple times in Formula \eqref{eqn: inverse matrix}, the variance computed via Formula \eqref{eqn: Taylor 2d variance} is very complicated. 
Unlike a conventional numerical approach such Gaussian elimination \cite{Numerical_Recipes}, Formula \eqref{eqn: inverse matrix} is symmetric to each element.
$\delta^2 \mathbf{M}^{-1}$ increases exponentially with its matrix size $n$ when the uncertainties of matrix elements are all independent of each other.
Formula \eqref{eqn: inverse matrix 2 variance} shows the simplest case for Formula \eqref{eqn: inverse matrix}: the first-order approximation of a $2 \times 2$ matrix.

In Formula \eqref{eqn: inverse matrix}, $\mathbf{M}^{-1}$ is dominated by $1/|\mathbf{M}|$, suggesting that the precision of $\mathbf{M}^{-1}$ is largely determined by the precision of $|\mathbf{M}|$.
Figure \ref{fig: Matrix_Determinant_Prec_vs_Condition} shows that a strong linear correlation exists between condition numbers \cite{Linear_Algebra} and the corresponding determinant precision of the matrices.
As a reference, Figure \ref{fig: Matrix_Determinant_Prec_vs_Condition} presents the Hilbert matrix \cite{Linear_Algebra} (which is the most unstable matrix in theory) for each matrix size and shows that the Hilbert matrices also follow the linear relation between determinant precision and condition number.
In Figure \ref{fig: Matrix_Determinant_Prec_vs_Condition}, adding noise to a matrix does not change its condition number, while increasing the result precision, because conceptually a matrix with larger uncertainty is less stable.  
In fact, Formula \eqref{eqn: inverse matrix 2 variance} shows the stability of each matrix element, which may be a better characterization of matrix stability than the condition number.

\section{Mathematical Library Functions}
\label{sec: Math Library}

\begin{figure}[p]
\centering
\includegraphics[height=2.5in]{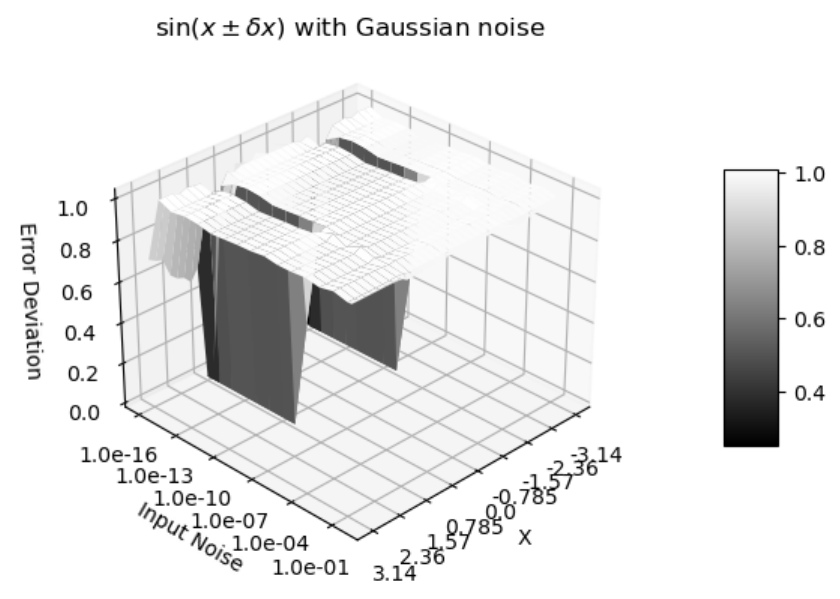} 
\captionof{figure}{
Error deviation for $\sin(x \pm \delta x)$ as a function of $x$ and $\delta x$.
The x-axis represents $x$ values between $-\pi$ and $+\pi$.
The y-axis represents $\delta x$ values between $10^{-16}$ and $1$.
The z-axis shows the corresponding error deviations. 
}
\label{fig: Sin_X_Dev}
\end{figure}

\begin{figure}[p]
\centering
\includegraphics[height=2.5in]{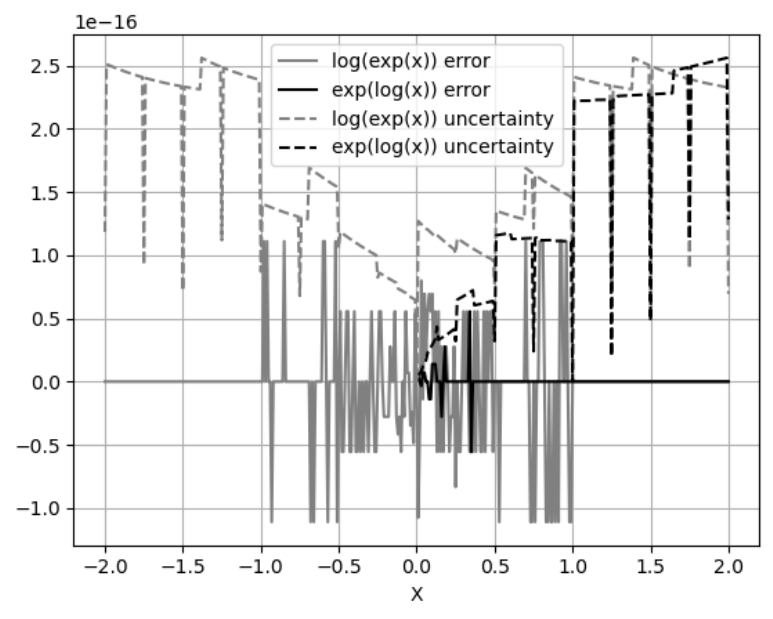} 
\captionof{figure}{
Values and uncertainties of $\log(e^x) - x$ and $e^{\log(x)} - x$ as functions of $x$, evaluated in increments of $0.1$.
When $x$ is a dyadic rational such as $1/2$ or $1$, the resulting uncertainties are significantly smaller because of floating-point representation.
}
\label{fig: ExpLog_Error}
\end{figure}

\begin{figure}[p]
\centering
\includegraphics[height=2.5in]{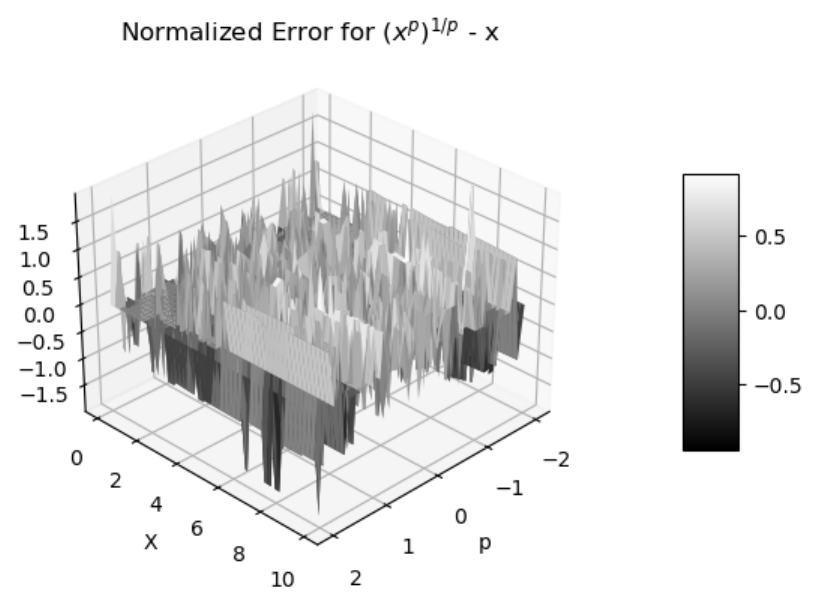} 
\captionof{figure}{
Normalized errors of $(x^p)^{\frac{1}{p}} - x$ as functions of $x$ and $p$.
}
\label{fig: Power_Error}
\end{figure}

\begin{table}
\centering
\resizebox{\textwidth}{!}{
\begin{tabular}{|c|c|c|c|c|} 
\hline 
Basic Function  & $e^{x \pm \delta x}$           & $\log(x \pm \delta x)$       & $(1 \pm \delta x)^c$         & $\sin(x \pm \delta x)$ \\ 
\hline  
Range             & $x \in [-100, +100]$            & $x \in [1/32, 32]$              & $c \in [-3, +3]$                 & $x \in [-\pi, +\pi]$     \\
\hline 
Uncertainty      & $\delta x \in [10^{-15}, 1]$ & $P(x) \in [10^{-15}, 0.2]$  & $P(x) \in [10^{-15}, 0.2]$ & $\delta x \in [10^{-15}, 1]$     \\
\hline 
Error Deviation & $1.000 \pm 0.010$              & $0.999 \pm 0.011$            & $1.000 \pm 0.007$            & $1.000 \pm 0.010$ \\
\hline 
\end{tabular}
}
\captionof{table}{
The result error deviations of selected basic functions with Gaussian input noise $x \pm \delta x$ when $\delta x > 10^{-15}$, obtained by variance arithmetic.
The error deviation for $(1 \pm \delta x)^c$ excludes the diverging regions.
The error deviation for $\sin(x \pm \delta x)$ excludes the pole regions.
}
\label{tbl: basic functions}
\end{table}

Table \ref{tbl: basic functions} shows that by sampling random inputs $\tilde{x}$ from Gaussian noise $x \pm \delta x$, Formulas \eqref{eqn: exp precision}, \eqref{eqn: log precision}, \eqref{eqn: sin precision}, and \eqref{eqn: power precision} provide nearly perfect characterization $f(x) \pm \delta f$ for the result distributions of $f(x+\tilde{x})$, where $f(x+\tilde{x})$ denotes the corresponding mathematical library functions.
The result error deviation is similar to that shown in Figure \ref{fig: Adjugate_Error_vs_Size_Noise}, but with the specific dimension as $x$ in $e^{x \pm \delta x}$, $\log(x \pm \delta x)$, and $\sin(x \pm \delta x)$ or $c$ in $(1 \pm \delta x)^c$.
The coverage is proper when $\delta x < 10^{-15}$.

Figure \ref{fig: Sin_X_Dev} shows that the error deviation for $\sin(x \pm \delta x)$ is $1.000 \pm 0.010$, except approaching $0$ when $x=\pm \pi/2$ and $\delta x < 10^{-8}$.
Near a distributional pole, the input uncertainty is suppressed, resulting in zero error deviation.
The numerical errors of the library functions $\sin(x)$ and $\cos(x)$ over a larger range of $x$ are examined in greater detail in Section \ref{sec: FFT}.

To test $f^{-1}(f(x)) - x = 0$ when $\delta x = 0$ for the library functions:
\begin{itemize}
\item Figure \ref{fig: ExpLog_Error} shows that the value errors in $e^{\log(x)} - x$ are much less than those in $\log(e^x) - x$. 
For $\log(e^x) - x$, the error deviation is $0.41$ when $|x| \leq 1$, or $0$ otherwise.

\item Figure \ref{fig: Power_Error} shows that the error deviation for $(x^p)^{1/p} - x$ is $0.56$, dependent on neither $x$ nor $p$.
\end{itemize}
The origin of this asymmetry --- why $(x^p)^{1/p} - x$ has larger value errors than $\log(e^x) - x$, while $e^{\log(x)} - x$ has nearly no value error --- remains unclear and may relate to floating-point representation details.

\section{Moving-Window Linear Regression}
\label{sec: Moving-Window Linear Regression}

An algorithm may need to be redesigned when uncertainty is itself a factor, particularly when its input data are tightly coupled across iterations, as in moving-window algorithms.

\subsection{Moving-Window Linear Regression Algorithm}

\begin{figure}[p]
\centering
\includegraphics[height=2.5in]{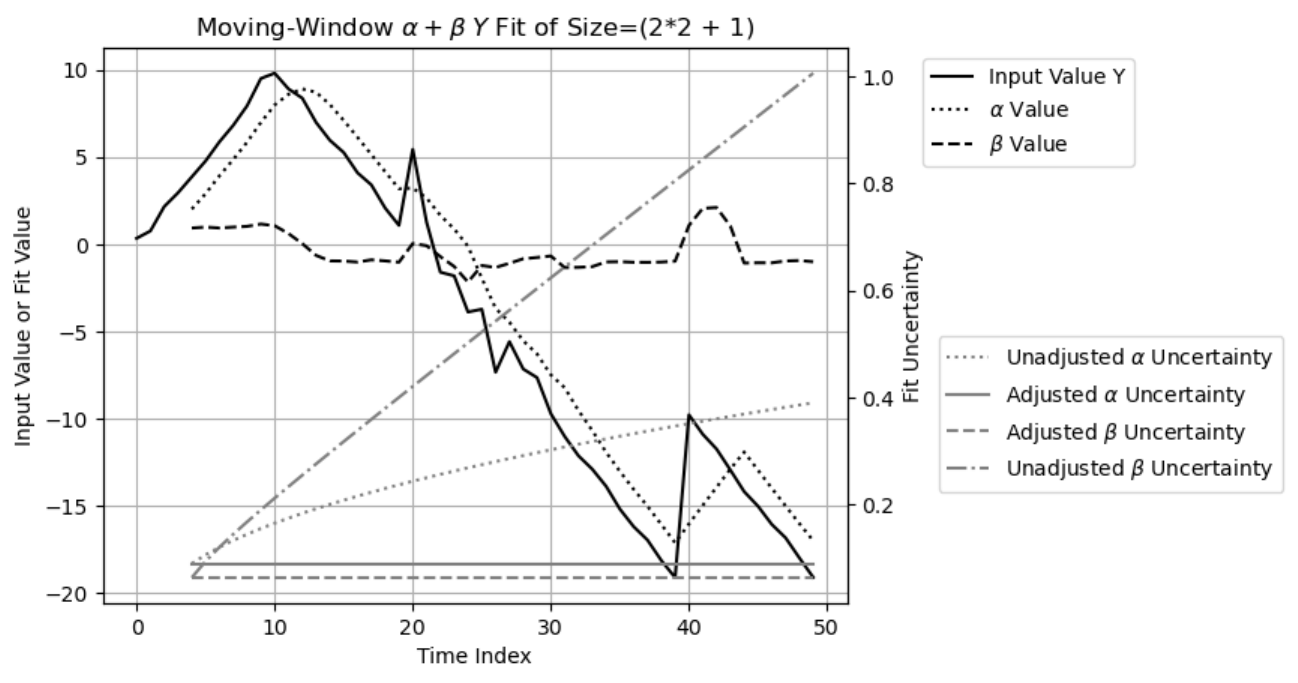} 
\captionof{figure}{ 
Result of fitting $Y = \alpha + \beta j$ to a time-series input $\{Y_j\}$ within a moving window of size $2 H + 1$ with $H = 2$.
The x-axis indicates the time index.
The y-axis on the left corresponds to the value of $\{Y_j\}$, $\alpha$, and $\beta$, while the y-axis on the right corresponds to the uncertainty of $\alpha$ and $\beta$.
The uncertainty for $Y$ is fixed at $0.2$.
In the legend, \textit{Unadjusted} refers to results obtained by directly applying Formulas \eqref{eqn: moving-window linear regression 0} and \eqref{eqn: moving-window linear regression 1} using variance arithmetic, whereas \textit{Adjusted} refers to using Formulas \eqref{eqn: moving-window linear regression 0} and \eqref{eqn: moving-window linear regression 1} for $\alpha$ and $\beta$ values, but Formulas \eqref{eqn: moving-window linear regression variance 0} and \eqref{eqn: moving-window linear regression variance 1} for their variances.
}
\label{fig: Moving_Linear_Fit_Value}
\end{figure}

\begin{figure}[p]
\centering
\includegraphics[height=2.5in]{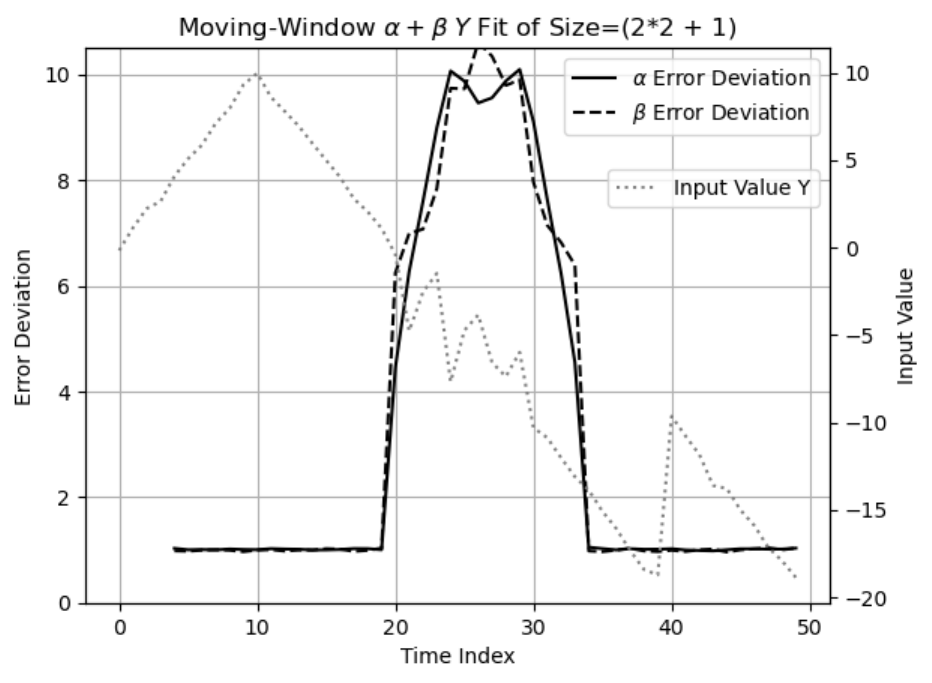} 
\captionof{figure}{ 
Error deviations of the $Y = \alpha + \beta j$ fit vs time index $j$. 
The x-axis represents the time index.
The y-axis on the left corresponds to the error deviation.
For reference, the input time-series signal $\{Y_j\}$ is also plotted, with its values indicated on the y-axis on the right.
}
\label{fig: Moving_Linear_Fit_Error}
\end{figure}

\begin{align}
\label{eqn: time-series linear regression 0}
\alpha _{j} &= \alpha \; 2 H = \sum_{X=-H+1}^{H} Y_{j-H+X}; \\
\label{eqn: time-series linear regression 1}
\beta _{j} &= \beta \; \frac{H (H+1)(2H+1)}{3} = \sum_{X=-H}^{H} X Y_{j-H+X}; 
\end{align}
In many applications, data set $\{Y_j\}$ denotes an input data stream where $j$ represents the time index or sequence index.
$\{Y_j\}$ is thus referred to as a time-series input.  
A moving window algorithm \cite{Numerical_Recipes} is applied within a small window centered on each $j$.  
For each calculation window, Formulas \eqref{eqn: time-series linear regression 0} and \eqref{eqn: time-series linear regression 1} present the least-squares line-fit of $Y = \alpha + \beta j$ for a window width $2H + 1$ of $\{Y_j\}$ \cite{Prev_Precision_Arithmetic}.

\begin{align}
\label{eqn: moving-window linear regression 0}
\beta _{j} &= \beta _{j-1} - \alpha _{j-1} + H \left(Y_{j-2H-1} + Y_{j} \right); \\
\label{eqn: moving-window linear regression 1}
\alpha_{j} &= \alpha _{j-1} - Y_{j-2H-1} + Y_{j};
\end{align}
The values of $(\alpha _{j}, \beta _{j})$ can be derived from the previous values $(\alpha _{j-1}, \beta _{j-1})$, allowing Formulas \eqref{eqn: time-series linear regression 0} and \eqref{eqn: time-series linear regression 1} to be reformulated into the progressive moving-window calculation given by Formulas \eqref{eqn: moving-window linear regression 0} and \eqref{eqn: moving-window linear regression 1}, respectively \cite{Prev_Precision_Arithmetic}.

Figure \ref{fig: Moving_Linear_Fit_Value} shows that the input signal $Y_j$ consists of the following components:
\begin{enumerate}
\item An increasing slope for $j = 0 \dots 9$.

\item A decreasing slope for $j = 10 \dots 39$.

\item A sudden jump of magnitude $+10$ at $j=40$.

\item A decreasing slope for $j = 41 \dots 49$.
\end{enumerate}
For each increment of $j$, the increasing and the decreasing rates are $+1$ and $-1$, respectively.
Gaussian noise with a deviation of $0.2$ is added to the slopes, except for the segment $j = 10 \dots 19$ where Normal noise with a deviation of $2$ is introduced, representing actual uncertainty $10$ times larger than the specified uncertainty.

In Figure \ref{fig: Moving_Linear_Fit_Value}, the fitted values of $\alpha$ and $\beta$ follow the expected behavior, exhibiting a characteristic delay of $H$ in $j$.

\subsection{Variance Adjustment}

\begin{align}
\label{eqn: moving-window linear regression variance 0}
\delta^2 \alpha_{j} &= \sum_{X=-H+1}^{H} (\delta Y_{j-H+X})^2 = \delta^2 \alpha_{j-1} - (\delta Y_{j-2H})^2 + (\delta Y_{j})^2; \\
\label{eqn: moving-window linear regression variance 1}
\delta^2 \beta_{j} &= \sum_{X=-H}^{H} X^2 (\delta Y_{j-H+X})^2;
\end{align}
In Figure \ref{fig: Moving_Linear_Fit_Value}, directly applying Formulas \eqref{eqn: moving-window linear regression 0} and \eqref{eqn: moving-window linear regression 1} results in quadratic growth for both $\delta \alpha$ and $\delta \beta$, because the multiple usage of the same input introduces the dependency problem.
Figure \ref{fig: Moving_Linear_Fit_Value} shows that the variance calculation has a much stronger dependency problem than the value calculation in this case.
Therefore, $\delta \alpha$ and $\delta \beta$ need to be calculated using Formulas \eqref{eqn: moving-window linear regression variance 0} and \eqref{eqn: moving-window linear regression variance 1}, in which Formula \eqref{eqn: moving-window linear regression variance 1} is no longer windowed because the windowed formula is more computationally expensive.
After this adjustment, $\delta \alpha$ and $\delta \beta$ both become nearly constant when $\delta Y$ is a constant.

\subsection{Unspecified Input Error}

To determine the error deviations of $\alpha$ and $\beta$, the fitting procedure is applied to multiple time-series data sets, each generated with independent noise realizations.
Figure \ref{fig: Moving_Linear_Fit_Error} illustrates the resulting error deviation as a function of the time index $j$, which remains close to $1$ except within the range $j = 10 \dots 19$ where the actual noise is ten times greater than the specified value.
This observation suggests that an error deviation exceeding 1 may indicate the presence of unspecified additional input errors beyond rounding errors, such as numerical errors in mathematical library functions.

\section{Fast Fourier Transformation (FFT)}
\label{sec: FFT}

\subsection{Discrete Fourier Transformation (DFT)}

\ifdefined\Verbose
\begin{figure}
\includegraphics[height=2.5in]{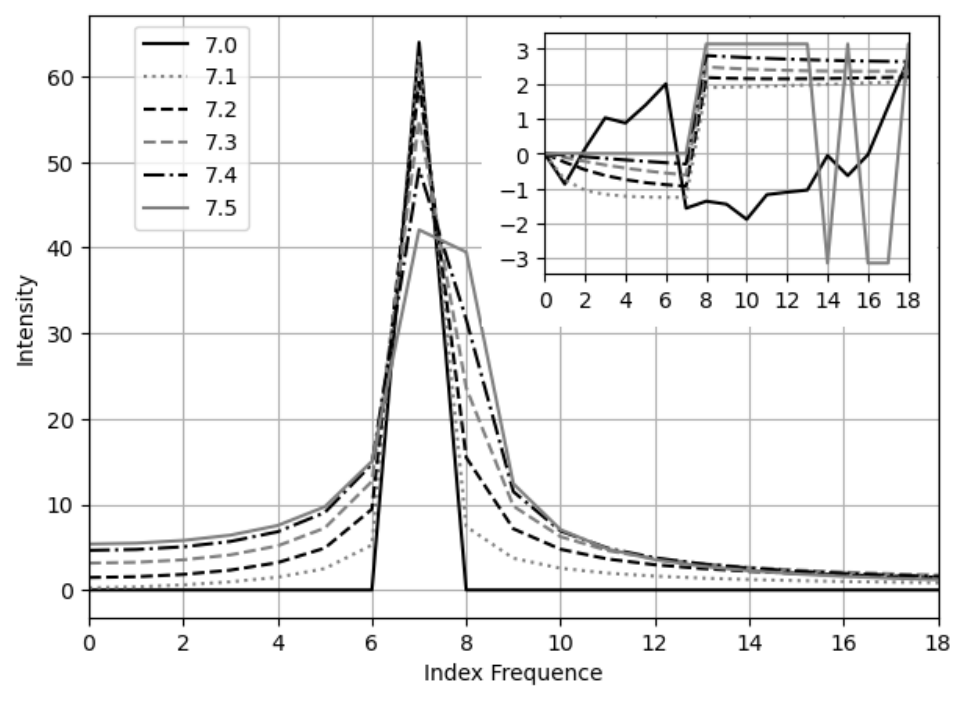} 
\captionof{figure}{
The DFT spectrum $H[n]$ of signal $h[k] = \sin(f \frac{2 \pi}{128} k), k \in [0, 127]$, as magnitude (y-axis) and phase (embedded y-axis) versus frequency index $n \in [0, 18]$ (x-axis and embedded x-axis) for different signal frequencies $f$ (legend).
This result agrees with both the theoretical formula \cite{Prev_Precision_Arithmetic} and the numerical computation from standard mathematical libraries such as \textit{SciPy}.
}
\label{fig: FFT_Unfaithful}
\end{figure}
\fi

\begin{align}
\label{eqn: Fourier forward}
H[n] &=\sum_{k=0}^{N-1} h[k] \; e^{\frac{-i 2\pi}{N} k n}; \\
\label{eqn: Fourier reverse}
h[k] &=\frac{1}{N} \sum_{n=0}^{N-1} H[n] \; e^{\frac{i 2\pi}{N} n k};
\end{align}

For each signal sequence $h[k]$, where $k = 0, 1, \dots, N-1$, and $N$ is a natural number, the discrete Fourier transform (DFT) $H[n]$, for $n = 0, 1, \dots, N-1$, along with its inverse transformation, is defined by Formulas \eqref{eqn: Fourier forward} and \eqref{eqn: Fourier reverse}, respectively \cite{Numerical_Recipes}.
As a convention, $k$ denotes the \emph{time index} for the waveform $h[k]$, whereas $n$ represents the \emph{frequency index} for the spectrum $H[n]$.

Although mathematically self-consistent, DFT is not a faithful digital implementation of the continuous Fourier transform (FT).
When the sampling window is moved from $k = 0,1,\dots,N-1$ to $k = 1,2,\dots,N$ and the new window is re-indexed to internal positions $0,\dots,N-1$, $H[n]$ changes to $e^{+i 2\pi n / N}(H[n] + h[N] - h[0])$, which matches the time-shifting property of FT \cite{Numerical_Recipes} only if $h[0] = h[N]$, or more generally $h[k] = h[k + N\,j]$ \cite{Prev_Precision_Arithmetic}.
This implied boundary condition causes the modeling errors of DFT except aliasing when DFT is viewed as the digital implementation of FT \cite{DFT_modeling_errors}.
\ifdefined\Verbose
For example, Figure \ref{fig: FFT_Unfaithful} shows the DFT spectra of the sine function with $f$ as frequency, which has a Dirac intensity of $N/2$ with phase $\pi/2$ only if $f$ is an integer, and violates FT in all other cases.
\fi
Similar modeling errors must exist for all digital implementations of infinite integration when the signal does not approach zero outside the digital integration window.

In this study, Formulas \eqref{eqn: Fourier forward} and \eqref{eqn: Fourier reverse} are taken as the canonical definitions of forward and reverse DFT, without claiming they faithfully implement FT.

\subsection{Fast Fourier Transformation (FFT)}

When $N = 2^{L}$, where $L$ is a natural number called the \emph{FFT order}, the generalized Danielson-Lanczos lemma can be applied to DFT to produce the FFT \cite{Numerical_Recipes}. 
\begin{itemize}

\item Within each output computation, each input contributes exactly once, so no dependency problem arises when decomposing the FFT into arithmetic operations such as Formulas \eqref{eqn: addition mean}, \eqref{eqn: addition variance}, \eqref{eqn: multiplication mean}, and \eqref{eqn: multiplication variance}.

\item When $L$ is large, the substantial volume of input and output data enables high-quality statistical analysis.

\item The  per-output computational complexity is proportional to $L$, because increasing $L$ by 1 adds an additional step involving a sum of two multiplications.

\item Each step in the forward transformation doubles the variance; hence the uncertainty deviation increases with the FFT order $L$ as $\sqrt{2}^L$.
Because the reverse transformation divides the result by $2^L$, its uncertainty deviation decreases with $L$ as $\sqrt{1/2}^L$.
Consequently, the uncertainty deviation for the round-trip transformation is $\sqrt{2}^L \times \sqrt{1/2}^L = 1$.

\end{itemize}

\subsection{Testing Signals}

The following signals are used for testing:
\begin{itemize}
\item \emph{Sin}: $h[k] = \sin(2\pi k f/N), f = 1, 2, \dots, \frac{N}{2} -1$.

\item \emph{Cos}: $h[k] = \cos(2\pi k f/N), f = 1, 2, \dots, \frac{N}{2} -1$.

\item \emph{Linear}: $h[k] = k$, whose DFT is given by Formula \eqref{eqn: Fourier spec for linear}.
\begin{align}
& y \equiv -i 2\pi \frac{n}{N}: \quad G(y) = \sum_{k=0}^{N-1}  e^{y k} = \frac{e^{N y} - 1}{e^y - 1}; \nonumber \\
\label{eqn: Fourier spec for linear}
H[n] &= \frac{d G}{d y} = \begin{cases} n = 0: \quad \frac{N (N-1)}{2} \\ n \neq 0: \quad
 - \frac{N}{2}(1 - i \frac{\cos(n \frac{\pi}{N})}{\sin(n \frac{\pi}{N})}) \end{cases};
\end{align}

\end{itemize}
The forward and reverse transformations differ only in the sign of the exponent and a $1/N$ normalization factor, implying that they are essentially the same algorithm; any observed difference arises mainly from the input data.
For Sin or Cos signals:
\begin{itemize}
\item The forward transformation converts a time-domain sine or cosine signal into a frequency-domain spectrum in which most values are zero, causing its uncertainties to grow more rapidly during mutual cancellation of input data. 

\item The reverse transformation spreads the frequency-domain spectrum, where most values are zero, into a time-domain sine or cosine signal, causing its uncertainties to grow more slowly. 
\end{itemize}
The question is whether variance arithmetic can work effectively in these two contrasting cases.

\subsection{Trigonometric Library Errors}

\begin{figure}[p]
\includegraphics[height=2.5in]{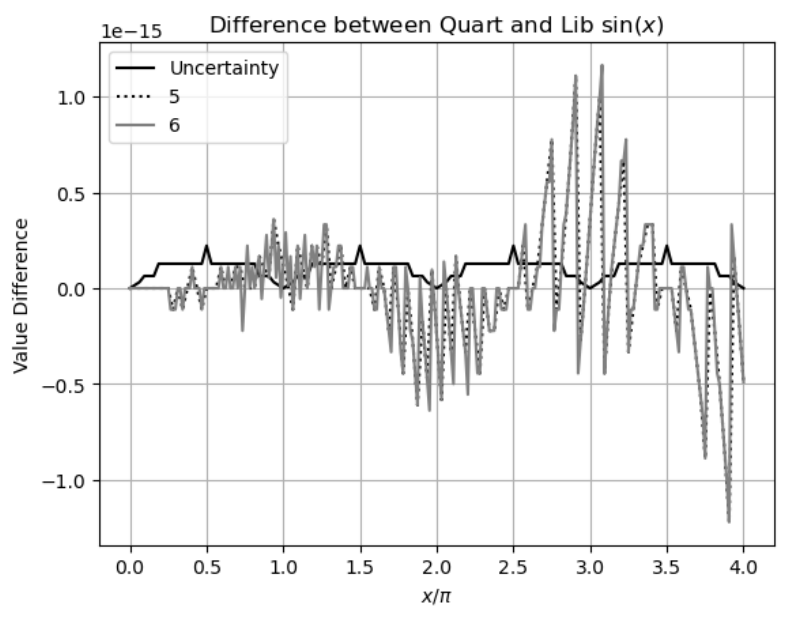} 
\captionof{figure}{
Difference between the Library and the Quart $\sin(x)$ (y-axis) for $x = 2\pi j /2^L, j =0, 1, \dots, 2^{L + 2}$ (x-axis), and $L = 5,6$ (legend).
The uncertainty of the Quart $\sin(x)$ is $\sin(x)$'s ULP, which shows a periodicity of $\pi$.
}
\label{fig: Sin_Diff}
\end{figure}

\begin{figure}[p]
\includegraphics[height=2.5in]{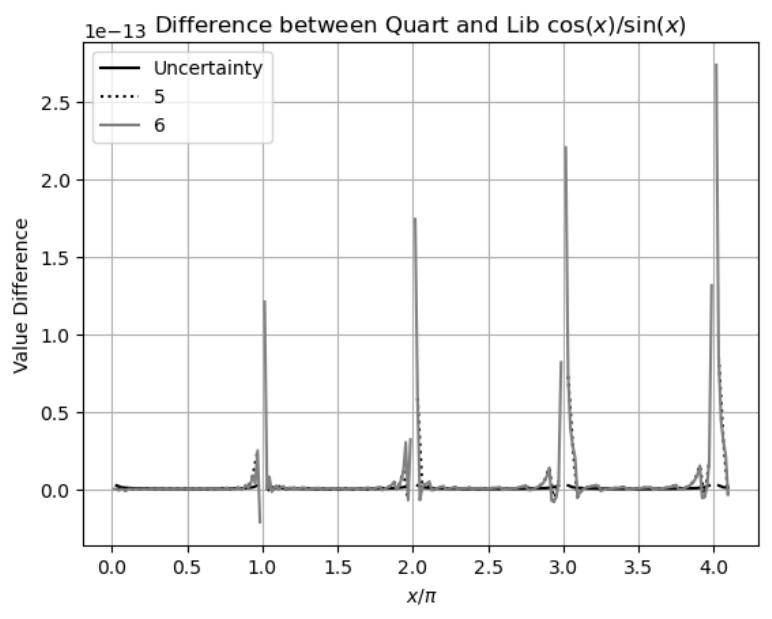} 
\captionof{figure}{
Difference between the Library and the Quart $\cos(x)/\sin(x)$ (y-axis) for $x = 2\pi j /2^L, j =0, 1, \dots, 2^{L + 2}$ (x-axis), and $L = 5,6$ (legend).
}
\label{fig: Cot_Diff}
\end{figure}

Formulas \eqref{eqn: Fourier forward} and \eqref{eqn: Fourier reverse} restrict the use of $\sin(x)$ and $\cos(x)$ to $x = 2\pi j /2^L$, where $L$ is the FFT order.
To minimize numerical errors in computing $\sin(x)$, the following \emph{Quart sine} can be used in place of standard \emph{Library Sine} functions:
\begin{enumerate}
\item Instead of a floating-point value $x$ as input for $\sin(x)$, an integer index $j$ defines the input as $\sin(\pi j/2^L)$, thereby eliminating the floating-point rounding error of $x$.

\item The values of $\sin(\pi j/2^L), j \in [0, 2^{L-2}]$ are the library sine directly, whereas the values of $\sin(\pi j/2^L), j \in [2^{L-2}, 2^{L-1}]$ are computed from the library $\cos(\pi (2^{L - 1} - j)/2^L), j \in [0, 2^{L-2}]$.

\item The values of $\sin(\pi j/2^L)$ are extended from $j \in [0, 2^{L-1}]$ to $j \in [0, 2^{L + 1}]$ by exploiting the symmetry of $\sin(\pi j/2^L)$.

\item The values of $\sin(\pi j/2^L)$ are extended to all integer values of $j$ by leveraging the periodicity of $\sin(\pi j/2^L)$ in $j$.

\end{enumerate}
Because the Quart sine function strictly preserves the symmetry and periodicity of the sine function, it provides better numerical accuracy than the Library sine function.
\begin{itemize}
\item Figure \ref{fig: Sin_Diff} shows that the value difference between the Library $\sin(x)$ and the Quart $\sin(x)$ increases approximately linearly with $|x|$.

\item Figure \ref{fig: Cot_Diff} shows that the value difference between the Quart and Library $\cos(x)/\sin(x)$ also increases roughly linearly with $|x|$, but is $10^2$ times larger than that observed for $\sin(x)$.
Therefore, the linear spectrum in Formula \eqref{eqn: Fourier spec for linear} contains significantly larger numerical errors when computed using the Library sine function.

\end{itemize}

\subsection{Using Quart Sine for Sin/Cos Signals}

\begin{figure}[p]
\centering
\includegraphics[height=2.5in]{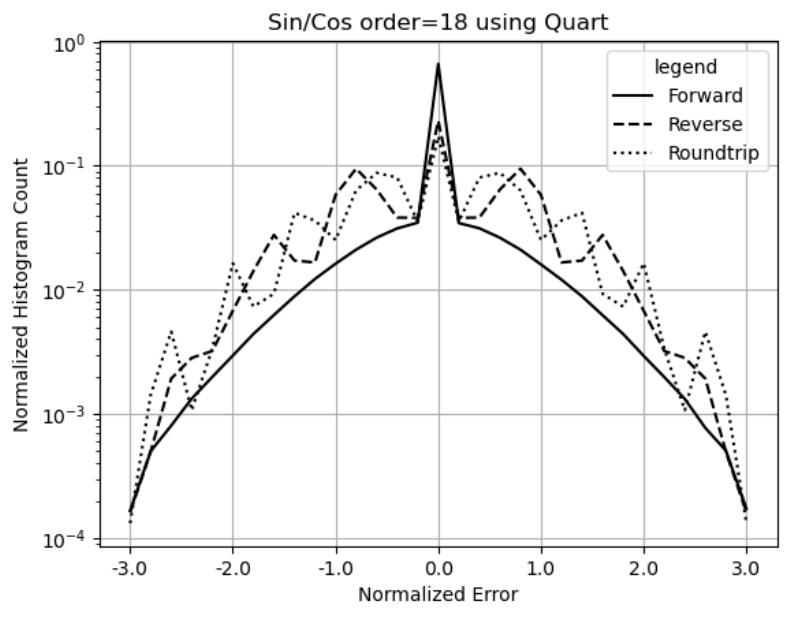} 
\captionof{figure}{
Error distributions of normalized errors of Sin/Cos signals without added input noise for forward, reverse, and round-trip transformations (legend) using the Quart sine function.
The FFT order is $18$.
}
\label{fig: FFT_SinCos_Clean_Histo_Indexed}
\end{figure}

With the FFT order as the specific dimension, the error deviations obtained using the Quart sine function for forward and reverse transformations resemble those shown in Figure \ref{fig: Adjugate_Error_vs_Size_Noise}, whereas those for round-trip transformation are nearly identical to those shown in Figure \ref{fig: Forward_Error_vs_Size_Noise}, independent of the signal frequency or whether a Sin or Cos signal is used.
Therefore, the results for Sin and Cos signals across all frequencies are pooled for statistical analysis, under the unified category \emph{Sin/Cos} signals.
When the FFT order $L$ is less than 8, the error deviations oscillate around $1$ because of an insufficient sample count of $2^L$.
Even though the data for forward and reverse transformations are substantially different, variance arithmetic works effectively in both cases.

When $L=18$ and $\delta x = 0$, Figure \ref{fig: FFT_SinCos_Clean_Histo_Indexed} shows that the error distributions of Sin/Cos signals resemble Normal distributions, with an additional Delta-like distribution at $\tilde{z} = 0$ for the forward transformation.
The error distribution of the reverse transformation shows additional structure on top of the Normal distribution, suggesting that the reverse transformation is more sensitive to numerical errors in the sine function.

\subsection{Using Library Sine for Sin/Cos Signals}

\begin{figure}[p]
\centering
\includegraphics[height=2.5in]{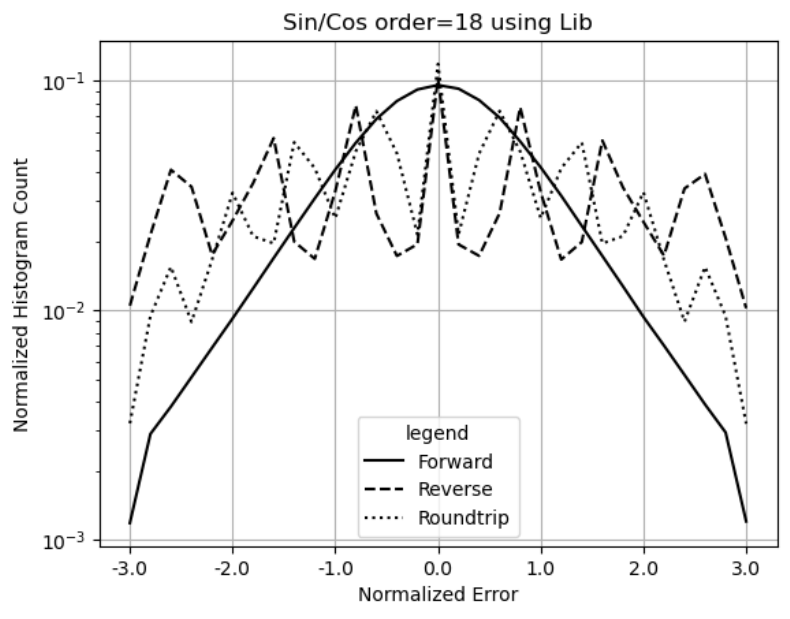} 
\captionof{figure}{
Error distributions of normalized errors of Sin/Cos signals without added input noise for forward, reverse, and round-trip transformations (legend) computed using the Library sine function.
The FFT order is $18$.
}
\label{fig: FFT_SinCos_Clean_Histo_Lib}
\end{figure}

\begin{figure}[p]
\includegraphics[height=2.5in]{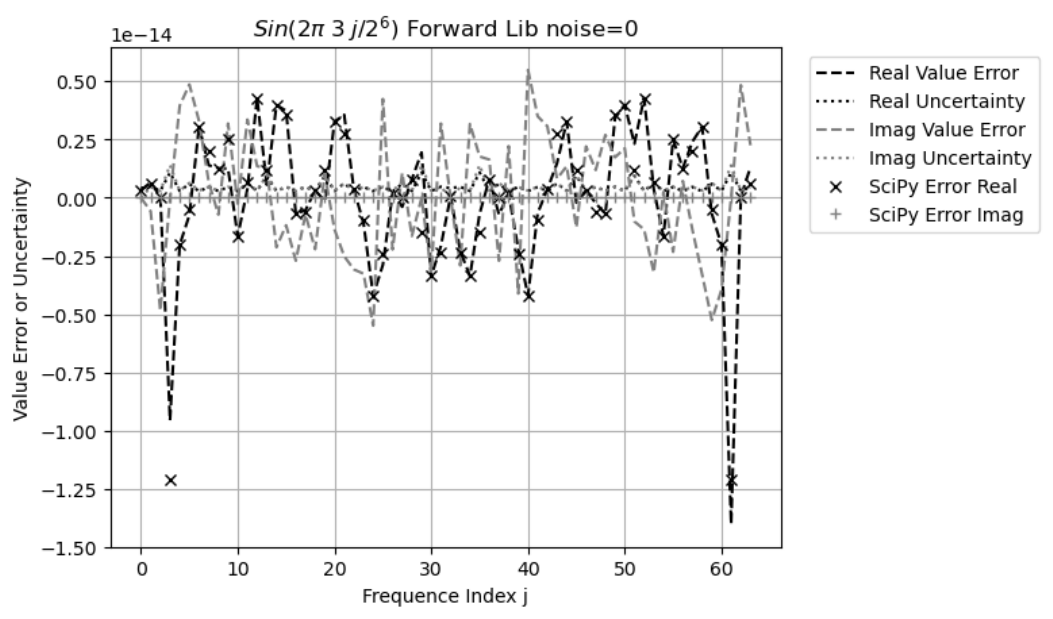} 
\captionof{figure}{
FFT value error spectrum of $\sin(3 \frac{2 \pi}{2^6} j)$ computed using either the Library sine function or \textit{SciPy} after the forward transformation.
The legend distinguishes between uncertainty and value error.
The x-axis represents the frequency index, and the y-axis represents both uncertainty and value error.
}
\label{fig: FFT_Sin_Clean_6_3_Spec_Lib}
\end{figure}

\begin{figure}[p]
\includegraphics[height=2.5in]{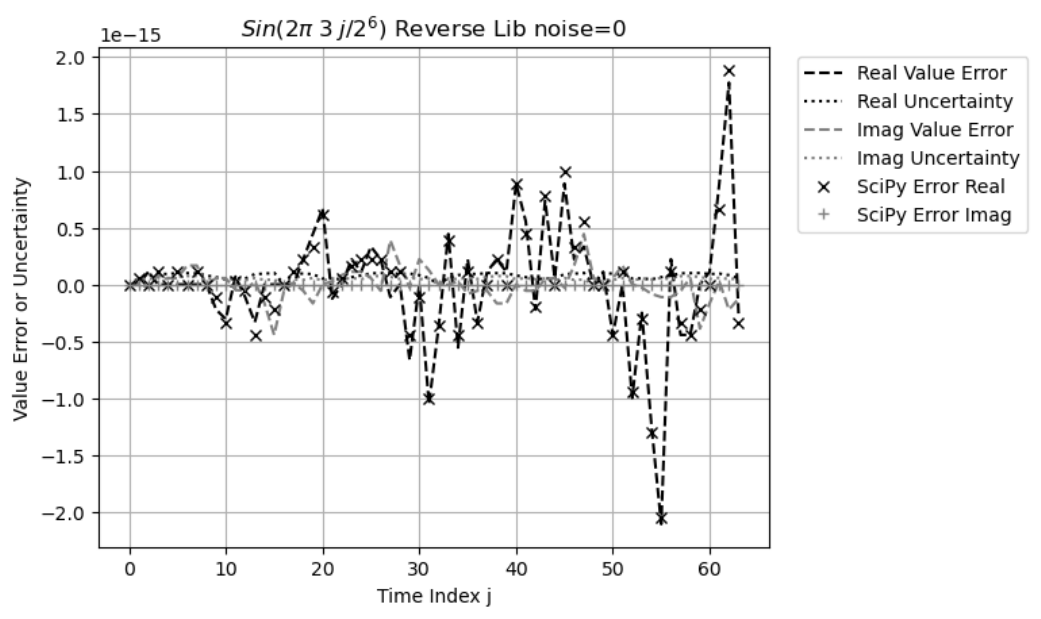} 
\captionof{figure}{
FFT value error waveform of $\sin(3 \frac{2 \pi}{2^6} j)$ computed using either the Library sine function or \textit{SciPy} after the reverse transformation.
The legend distinguishes between uncertainty and value error.
The x-axis represents the time index, and the y-axis represents both uncertainty and value error.
}
\label{fig: FFT_Sin_Clean_6_3_Wave_Lib}
\end{figure}

\begin{figure}[p]
\includegraphics[height=2.5in]{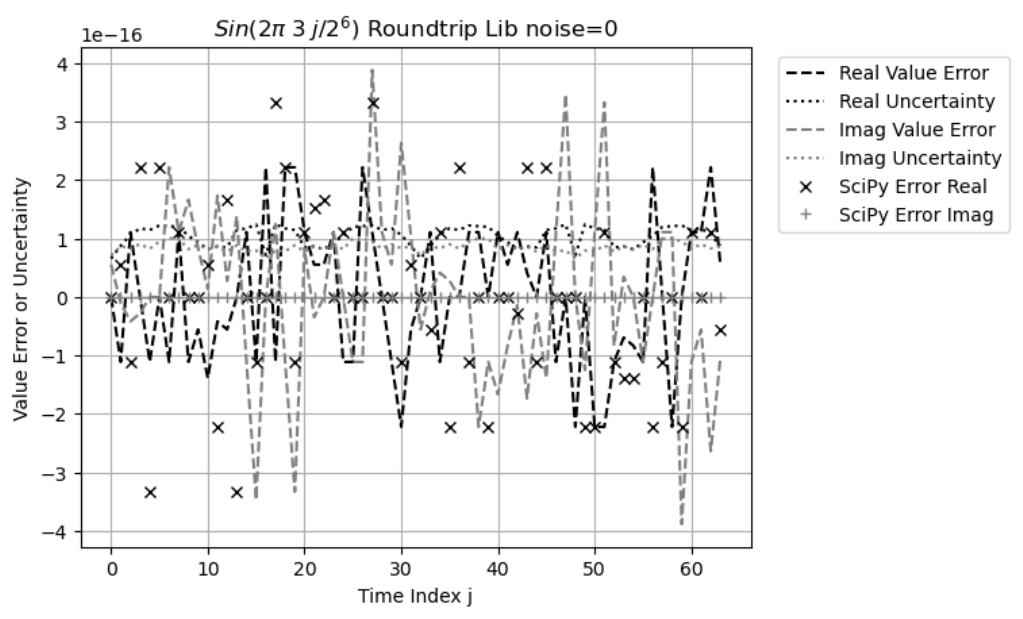} 
\captionof{figure}{
FFT value error waveform of $\sin(3 \frac{2 \pi}{2^6} j)$ computed using either the Library sine function or \textit{SciPy} after the round-trip transformation.
The legend distinguishes between uncertainty and value error.
The x-axis represents the time index, and the y-axis represents both uncertainty and value error.
}
\label{fig: FFT_Sin_Clean_6_3_Roundtrip_Lib}
\end{figure}

\begin{figure}[p]
\centering
\includegraphics[height=2.5in]{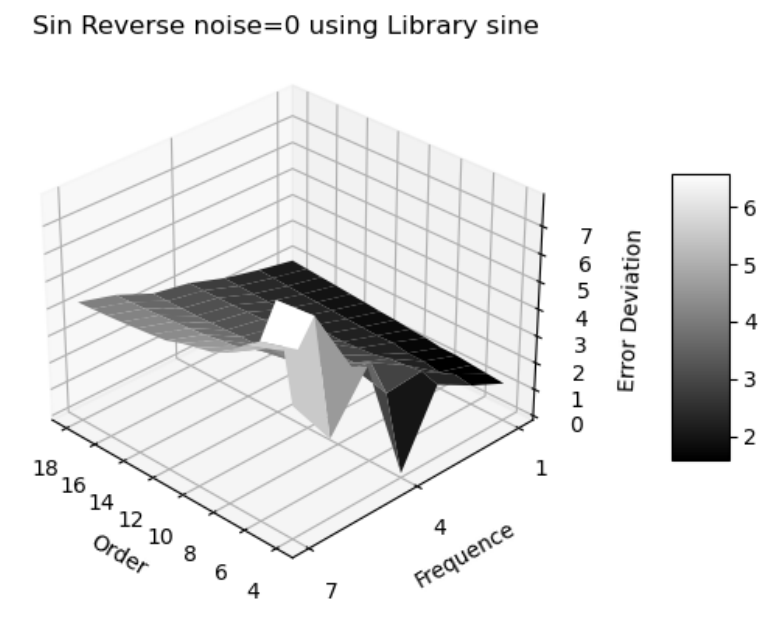} 
\captionof{figure}{
Error deviation (z-axis) of the FFT reverse transformation of $\sin(f \frac{2 \pi}{2^L} j)$ versus frequency $f$ (x-axis) and FFT order $L$ (y-axis).
}
\label{fig: Lib_Reverse_Error_vs_Freq_Order}
\end{figure}

With the FFT order as the specific dimension, the error deviations obtained using the Library sine function for the forward and reverse transformations resemble those shown in Figure \ref{fig: Adjugate_Error_vs_Size_Noise}, whereas those for the round-trip transformation are nearly identical to those shown in Figure \ref{fig: Forward_Error_vs_Size_Noise}.
In addition, the error deviations are larger than $1$ when $\delta x < 10^{-15}$ for the forward transformation and $\delta x < 10^{-14}$ for the reverse transformation.

When $\delta x = 0$, the error deviations for the reverse transformation increase with the FFT order, reaching $6.2$ at FFT order $18$.
As shown in Figure \ref{fig: Sin_Diff}, the Library sine function contains more numerical errors; as a result, the error distribution for the reverse transformation obtained using the Library sine function in Figure \ref{fig: FFT_SinCos_Clean_Histo_Lib} is more structured and broader than that shown in Figure \ref{fig: FFT_SinCos_Clean_Histo_Indexed} obtained using the Quart sine function, whereas the error distributions for the forward transformations are more similar.
This difference is consistent with a larger error deviation $6.2 > 1$ for the reverse transformation, compared with the comparable error deviation $1.1 > 1$ for the forward transformation when $\delta x = 0$.

Figures \ref{fig: FFT_Sin_Clean_6_3_Spec_Lib}, \ref{fig: FFT_Sin_Clean_6_3_Wave_Lib}, and \ref{fig: FFT_Sin_Clean_6_3_Roundtrip_Lib} present the value errors for the forward, reverse, and round-trip transformations, respectively, for a sine wave with a frequency of $3$ computed using the Library sine function with $\delta x = 0$.
In the reverse transformation, the value errors exhibit a clear trend of increasing with the time index. 
These large value errors appear systematic rather than random and visually resemble a resonant pattern.
Similar increases are observed at other frequencies and FFT orders, as well as in computational results obtained using mathematical libraries such as \textit{SciPy}.
In contrast, such resonance is absent from the round-trip transformation shown in Figure \ref{fig: FFT_Sin_Clean_6_3_Roundtrip_Lib}, as well as when using the Quart sine function.
Figure \ref{fig: Lib_Reverse_Error_vs_Freq_Order} demonstrates that the error deviations increase with sine or cosine frequency, regardless of FFT order $L$.
Figure \ref{fig: Sin_Diff} indicates that the numerical errors of $\sin(x)$ obtained using the Library sine function exhibit a periodicity of $\pi$, which can resonate with a signal whose periodicity is an integer multiple of $\pi$, producing the resonant pattern shown in Figure \ref{fig: FFT_Sin_Clean_6_3_Wave_Lib}.
At a higher frequency, the resonant beats between the signal and the numerical errors in the Library sine function become stronger.
To suppress this numerical error resonance, an input noise of $\delta x = 10^{-14}$ must be added to the sine or cosine signals.
Such \emph{resonance of numerical errors} can easily be mistaken for signals.

\subsection{Using Quart Sine for Linear Signals}

\begin{figure}[p]
\centering
\includegraphics[height=2.5in]{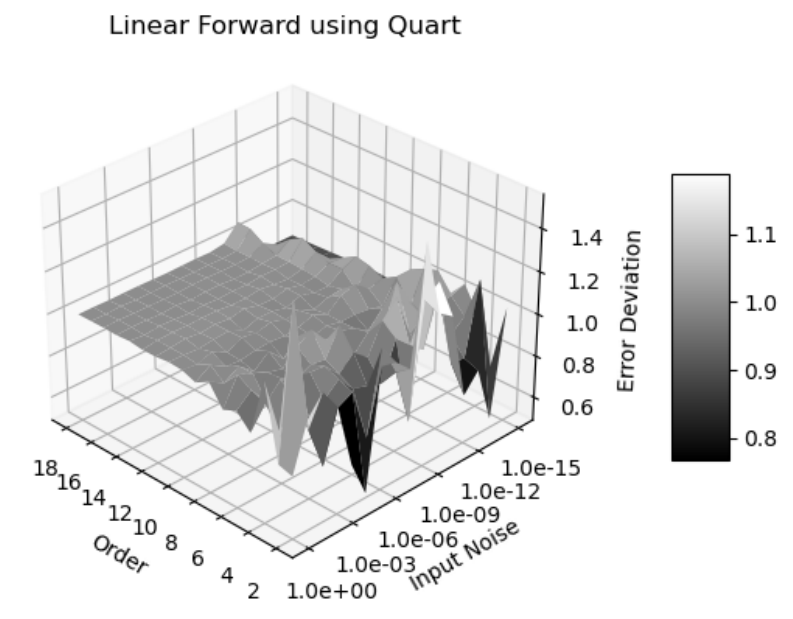} 
\captionof{figure}{
Error deviation (z-axis) versus input uncertainty (x-axis) and FFT order (y-axis) for the forward transformation of Linear signals computed using the Quart sine function.
}
\label{fig: FFT_Linear_Indexed_Forward_ErrorDev_vs_Noise_Order}
\end{figure}

\begin{figure}[p]
\centering
\includegraphics[height=2.5in]{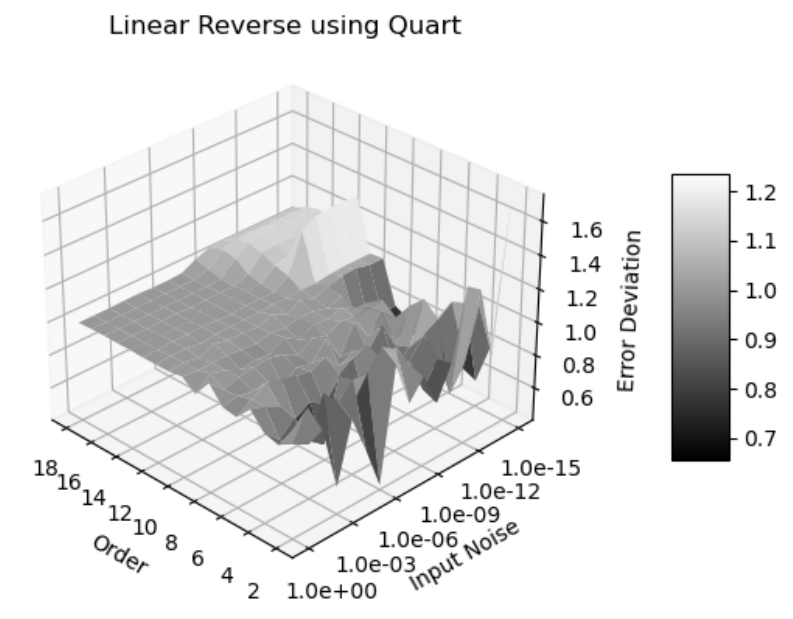} 
\captionof{figure}{
Error deviation (z-axis) versus input uncertainty (x-axis) and FFT order (y-axis) for the reverse transformation of Linear signals computed using the Quart sine function.
}
\label{fig: FFT_Linear_Indexed_Reverse_ErrorDev_vs_Noise_Order}
\end{figure}

\begin{figure}[p]
\centering
\includegraphics[height=2.5in]{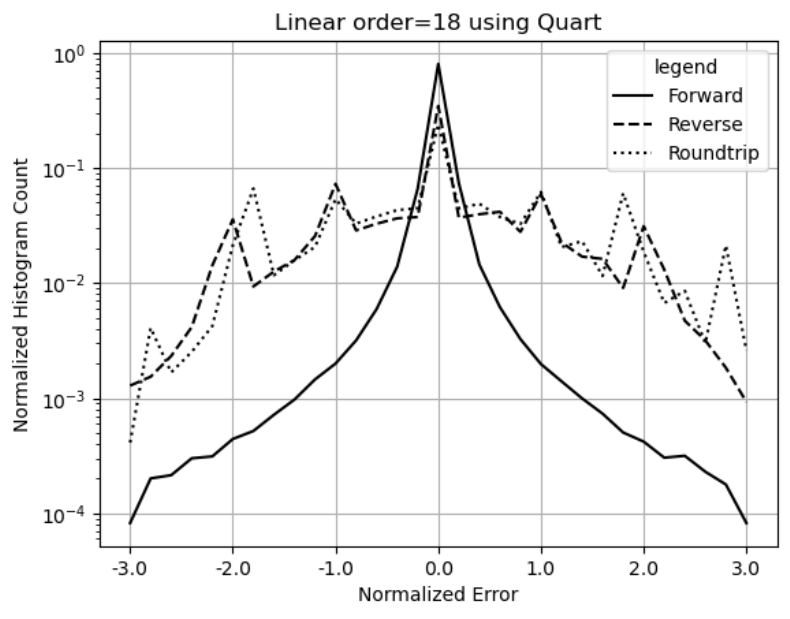} 
\captionof{figure}{
Error distributions of normalized errors of Linear signals for forward, reverse, and round-trip transformations (legend) computed using the Quart sine function.
The FFT order is $18$.
}
\label{fig: FFT_Linear_Clean_Histo_Indexed}
\end{figure}

Figures \ref{fig: FFT_Linear_Indexed_Forward_ErrorDev_vs_Noise_Order} and \ref{fig: FFT_Linear_Indexed_Reverse_ErrorDev_vs_Noise_Order} show the error deviations for the forward and the reverse transformations, respectively, both of which resemble Figure \ref{fig: Adjugate_Error_vs_Size_Noise}. 
The forward transformation exhibits a larger ideal coverage area than the reverse transformation: $\delta x > 10^{-12}$ for the forward transformation, and $\delta x > 10^{-8}$ for the reverse transformation.
In other areas, both transformations achieve proper coverage with error deviations around $1$.

When $L = 18$ and $\delta x = 0$, the error distribution of the reverse transformation in Figure \ref{fig: FFT_Linear_Clean_Histo_Indexed} is narrower than that shown in Figure \ref{fig: FFT_SinCos_Clean_Histo_Lib}.
The corresponding error deviations are $1.5 < 6.2$, respectively.

The error deviations for the round-trip transformation resemble those in Figure \ref{fig: Forward_Error_vs_Size_Noise}, but with the FFT order as the specific dimension.

\subsection{Using Library Sine for Linear Signals}

\begin{figure}[p]
\centering
\includegraphics[height=2.5in]{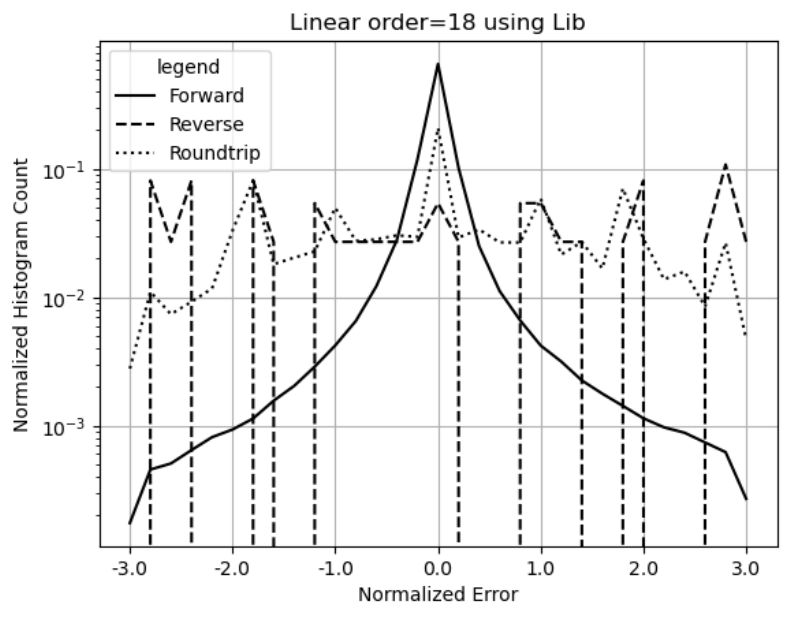} 
\captionof{figure}{
Error distributions of normalized errors of Linear signals for forward, reverse, and round-trip transformations (legend) computed using the Library sine function.
The FFT order is $18$.
}
\label{fig: FFT_Linear_Clean_Histo_Lib}
\end{figure}

\begin{figure}[p]
\centering
\includegraphics[height=2.5in]{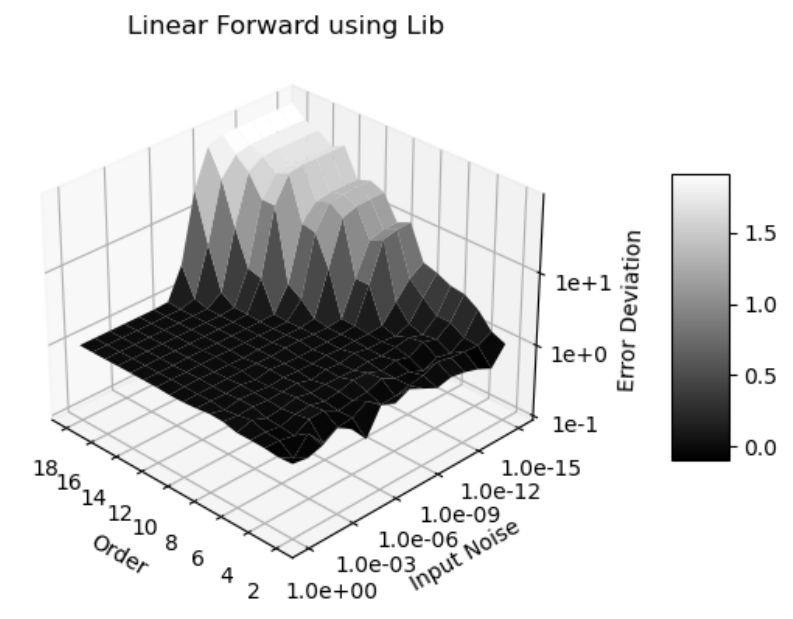} 
\captionof{figure}{
Error deviation (z-axis) versus input uncertainty (x-axis) and FFT order (y-axis) for the forward transformation of Linear signals computed using the Library sine function.
}
\label{fig: FFT_Linear_Lib_Forward_ErrorDev_vs_Noise_Order}
\end{figure}

\begin{figure}[p]
\centering
\includegraphics[height=2.5in]{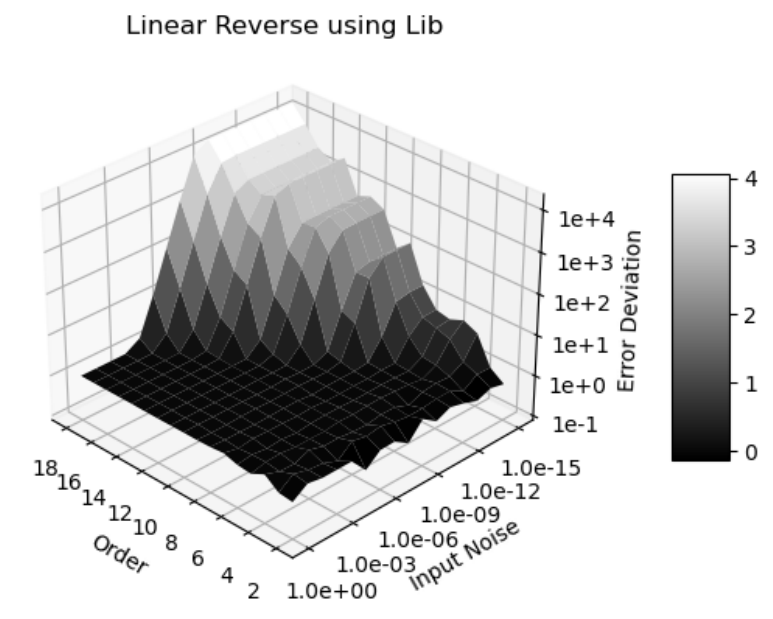} 
\captionof{figure}{
Error deviation (z-axis) versus input uncertainty (x-axis) and FFT order (y-axis) for the reverse transformation of Linear signals computed using the Library sine function.
}
\label{fig: FFT_Linear_Lib_Reverse_ErrorDev_vs_Noise_Order}
\end{figure}

\begin{figure}[p]
\centering
\includegraphics[height=2.5in]{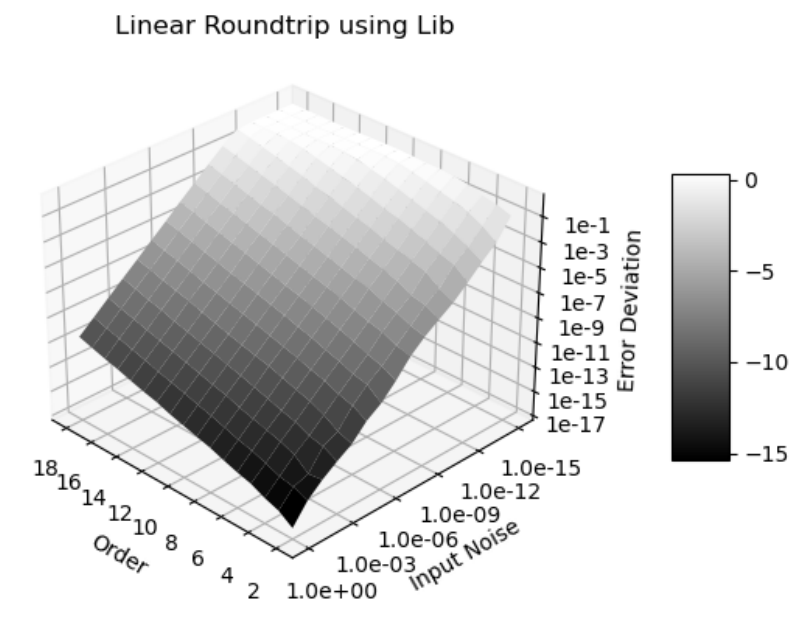} 
\captionof{figure}{
Error deviation (z-axis) versus input uncertainty (x-axis) and FFT order (y-axis) for the round-trip transformation of Linear signals computed using the Library sine function.
}
\label{fig: FFT_Linear_Lib_Roundtrip_ErrorDev_vs_Noise_Order}
\end{figure}

Figure \ref{fig: FFT_Linear_Clean_Histo_Lib} shows that the error distribution of the reverse transformation when $\delta x = 0$ is no longer bounded.
The difference between Figures \ref{fig: FFT_Linear_Clean_Histo_Lib} and \ref{fig: FFT_Linear_Clean_Histo_Indexed} is consistent with the large numerical errors demonstrated in Figure \ref{fig: Cot_Diff}.
Variance arithmetic fails because of the substantial and unspecified numerical errors from the Library sine function.

Figures \ref{fig: FFT_Linear_Lib_Forward_ErrorDev_vs_Noise_Order} and \ref{fig: FFT_Linear_Lib_Reverse_ErrorDev_vs_Noise_Order} show much smaller ideal coverage areas than those shown in Figures \ref{fig: FFT_Linear_Indexed_Forward_ErrorDev_vs_Noise_Order} and \ref{fig: FFT_Linear_Indexed_Reverse_ErrorDev_vs_Noise_Order}, respectively.
Because uncertainty deviations grow more slowly in the reverse transformation than in the forward transformation, the former exhibits a smaller ideal coverage region.
Outside of the ideal coverage region, proper coverage cannot be achieved for the reverse transformation.
Furthermore, the range of input noise that produces ideal coverage decreases with increasing FFT order.
At sufficiently high FFT orders (visually beyond FFT order $25$ for the reverse transformation), ideal coverage may no longer be achievable.
Although FFT is widely regarded as one of the most robust numerical algorithms \cite{Numerical_Recipes, Precise_Numerical_Methods} and is generally insensitive to input errors, it can still fail because extensive computation can amplify numerical errors in the Library sine function to obscure true signals.
Such deterioration of computed values is not easily detectable when using conventional floating-point arithmetic.

Figure \ref{fig: FFT_Linear_Lib_Roundtrip_ErrorDev_vs_Noise_Order} shows that, even when variance arithmetic can no longer effectively track the value errors for either the forward or the reverse transformations, it can still effectively track the value errors for the round-trip transformation, as shown by the plateau region of error deviations at high $L$ and low $\delta x$.
Such error cancellation arises from dependency tracing in statistical Taylor expansion.

\subsection{Ideal Coverage}

\begin{figure}[p]
\centering
\includegraphics[height=2.5in]{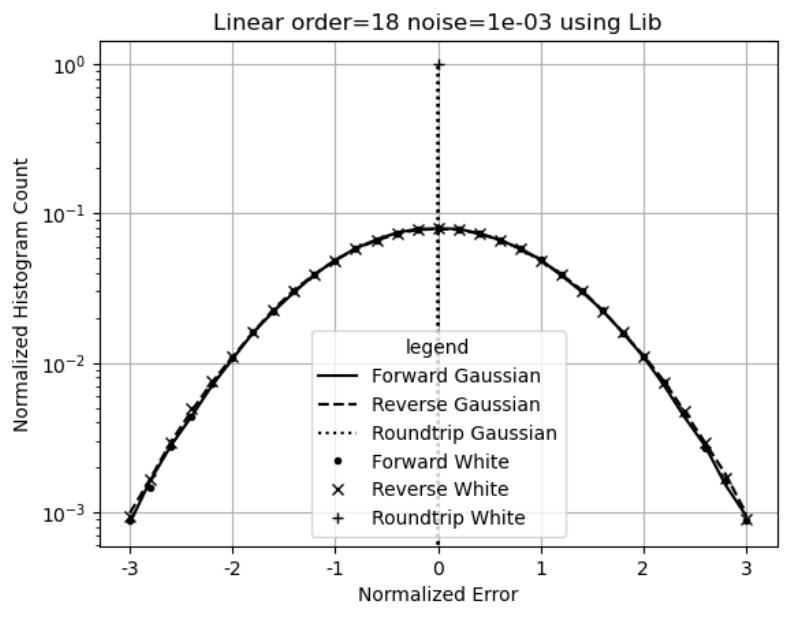} 
\captionof{figure}{
Error distributions for Linear signals with $10^{-3}$ input noise for forward, reverse, and round-trip transformations (legend) computed using the Library sine function.
The input noise is either Gaussian or white (legend).
The FFT order is $18$.
}
\label{fig: FFT_Linear_1e-3_Histo_Lib}
\end{figure}

\begin{figure}
\centering
\includegraphics[height=2.5in]{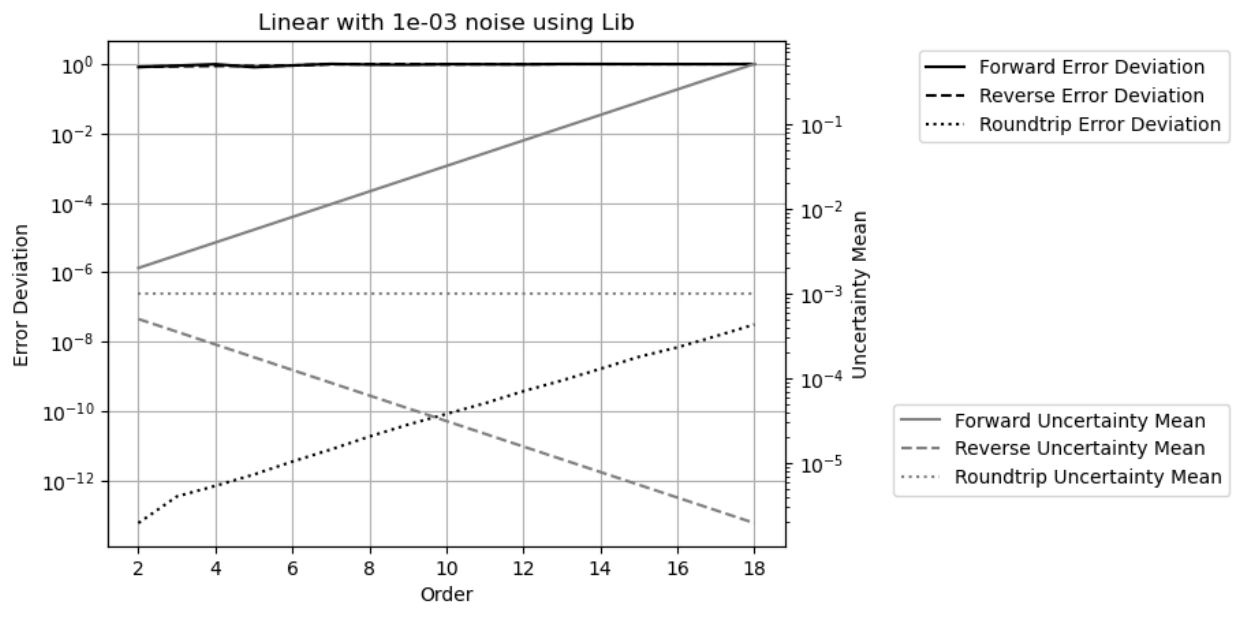} 
\captionof{figure}{
Error deviation (left y-axis) and uncertainty deviation (right y-axis) of Linear signal versus FFT order (x-axis) and transformation type (legend) computed using the Library sine function.
}
\label{fig: FFT_Linear_1e-3_vs_Order_Lib}
\end{figure}

\begin{table}
\centering
\begin{tabular}{|c|c|c|c|c|} 
\hline 
Signal     & Sine             & Forward       & Reverse       & Round-trip    \\ 
\hline 
Sin/Cos   & Quart           & $10^{-16}$  & $10^{-12}$ & $10^{-14}$ \\ 
\hline 
Sin/Cos   & Library         & $10^{-16}$  & $10^{-11}$ & $10^{-12}$ \\ 
\hline 
Linear     & Quart           & $10^{-11}$  & $10^{-7}$   & $10^{-8}$ \\
\hline 
Linear     & Library         & $10^{-11}$  & $10^{-3}$   & $10^{-8}$  \\
\hline 
\end{tabular}
\captionof{table}{
The measured minimal required noise to achieve ideal coverage for FFT transformations at the FFT order $18$ for different signals and sine functions.
}
\label{tbl: ideal coverage}
\end{table}

\begin{table}
\centering
\resizebox{\textwidth}{!}{
\begin{tabular}{|c|c|c|c|c|}
\hline
Signal   & Sine     & Forward Error Dev.       & Reverse Error Dev.         & Roundtrip Slope          \\
\hline
Sin/Cos & Quart   & $0.9997 \pm 0.0035$   & $1.0001 \pm 0.0041$     & $-0.9879 \pm 0.0012$     \\
\hline
Sin/Cos & Library & $0.9997 \pm 0.0037$   & $1.0001 \pm 0.0029$     & $-0.9888 \pm 0.0008$     \\
\hline
Linear   & Quart   & $1.000 \pm 0.017$      & $1.015 \pm 0.044$         & $-0.9326 \pm 0.0515$     \\
\hline
Linear   & Library & $6.53 \pm 17.63$        & $829.9 \pm 3055.8$       & $-0.9354 \pm 0.0507$     \\
\hline
\end{tabular}
}
\captionof{table}{
The measured average error deviations for forward and reverse transformations and the measured error slopes for the round-trip transformations for difference signals and sine functions.
The FFT order is larger than $7$ for statistical stability.
The input noise is larger than $10^{-15}$.
}
\label{tbl: fft error stats}
\end{table}

Adding noise to the input can dominate unspecified input errors.
After adding a Gaussian input noise of $\delta x = 10^{-3}$ to a Linear signal when using the Library sine function, the error distributions for both the forward and the reverse transformations become Normal, whereas the error distribution for the round-trip transformation becomes Delta, as shown in Figure \ref{fig: FFT_Linear_1e-3_Histo_Lib}.
Figure \ref{fig: FFT_Linear_1e-3_vs_Order_Lib} illustrates the corresponding error deviations and uncertainty deviations versus FFT order:
\begin{itemize}
\item As expected, the resulting uncertainty deviations for the forward transformations increase with FFT order $L$ as $\sqrt{2}^L$.

\item As expected, the resulting uncertainty deviations for the reverse transformations decrease with FFT order $L$ as $1/\sqrt{2}^L$.

\item As expected, the resulting uncertainty deviations for the round-trip transformations remain equal to the corresponding input uncertainties of $10^{-3}$.

\item As expected, the resulting error deviations for the forward and the reverse transformations remain constant at $1$.

\item As expected, the resulting error deviations for the round-trip transformations are far less than $1$ but increase exponentially with FFT order $L$ because increasing calculation amplifies rounding errors.
\end{itemize}

Table \ref{tbl: ideal coverage} shows the minimal required noise to achieve ideal coverage for FFT transformations at FFT order $L=18$ for different signals and sine functions, which is consistent with the corresponding error distributions shown in Figures \ref{fig: FFT_SinCos_Clean_Histo_Indexed}, \ref{fig: FFT_SinCos_Clean_Histo_Lib}, \ref{fig: FFT_Linear_Clean_Histo_Indexed}, and \ref{fig: FFT_Linear_Clean_Histo_Lib}.
The shape of an error distribution can indicate whether the input-uncertainty estimate provides ideal, proper, or no coverage.
Without knowing the precise result, a similar histogram can be constructed from the result data set for $f$ using the calculated mean $\overline{f}$ and deviation $\delta f$.
It is worth investigating if such an empirical histogram has similar power to reveal input uncertainty coverage.

Table \ref{tbl: fft error stats} shows the measured average error deviation for forward and reverse transformations and the measured error slope for round-trip transformations for different signals and sine functions.
The Linear/Library rows stand out with large reverse error deviations, consistent with the corresponding noise thresholds of $10^{-3}$ in Table \ref{tbl: ideal coverage}.
In other cases, the average error deviations are close to $1$ while the error slopes are close to $-1$.

\subsection{Comparison to Interval Arithmetic}

\begin{figure}[p]
\centering
\includegraphics[height=2.5in]{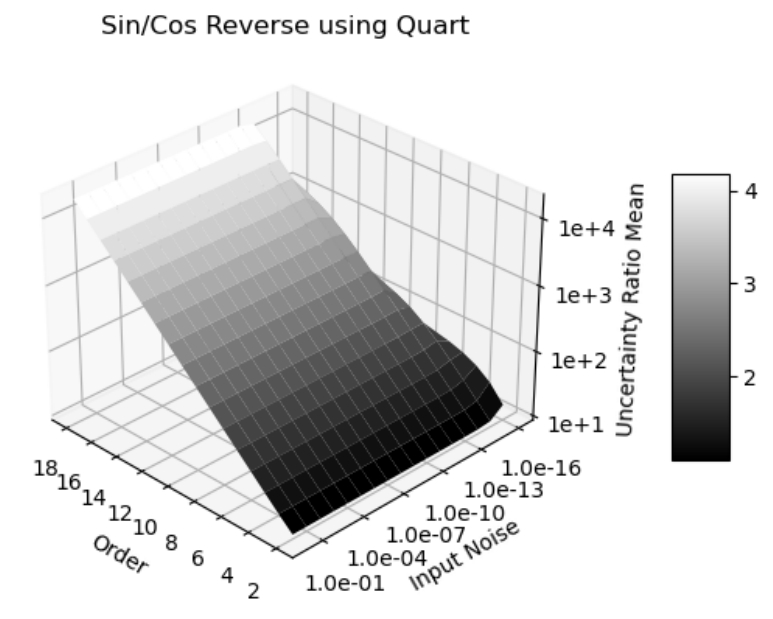}
\captionof{figure}{
The ratio of the interval range to the uncertainty deviation (z-axis) versus input uncertainty (x-axis) and FFT order (y-axis) for the reverse transformations of Sin/Cos signals computed using the Quart sine function.
}
\label{fig: FFT_Reverse_Uncertainty_Ratio}
\end{figure}

\begin{figure}[p]
\centering
\includegraphics[height=2.5in]{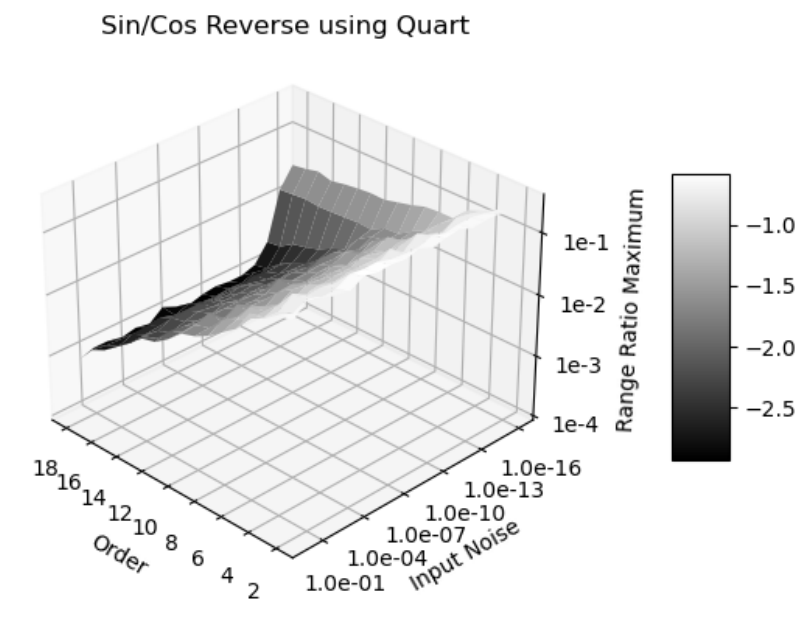}
\captionof{figure}{
The ratio of the maximal value error to the interval range (z-axis) versus input uncertainty (x-axis) and FFT order (y-axis) for the reverse transformations of Sin/Cos signals computed using the Quart sine function.
}
\label{fig: FFT_Reverse_Range_Ratio}
\end{figure}

It is worthwhile to compare statistical Taylor expansion with interval arithmetic, which is the current de facto standard for uncertainty tracking.
For this comparison, the input range is either $2\sqrt{3}$-fold of the deviation for Uniform input noise, or $10$-fold of the deviation for Gaussian input noise.
The above interval ranges do not account for rounding errors when the input noise deviation is less than $10^{-15}$.
Figure \ref{fig: FFT_Reverse_Uncertainty_Ratio} shows that the ratio of the interval range to the uncertainty deviation increases linearly with FFT order.
As a consequence, Figure \ref{fig: FFT_Reverse_Range_Ratio} shows that the ratio of the maximal value error to the interval range decreases linearly with FFT order.
This trend of uncertainty over-estimation is independent of input noise type (Gaussian versus Uniform), the choice of interval range (fixed multiples of the deviation or the actual input range), the transformation (forward, reverse, or round-trip), the signal under test (Sin, Cos, or Linear), and the sine function (Quart or Library).
While statistical Taylor expansion tracks the output value errors precisely with an error deviation of $1$ as shown in Table \ref{tbl: ideal coverage}, exactly the same calculation using interval arithmetic results in over-estimation of uncertainty, which is linearly proportional to the amount of calculation.
Because interval arithmetic tracks the worst-case of output range, it is unsuitable to track random uncertainty. 
Thus, statistical Taylor expansion has a much wider applicable range because the uncertainties in most applications are random in nature \cite{Statistical_Methods, Precisions_Physical_Measurements}.

\section{Recursive Generation of Sine}
\label{sec: recursion}

\begin{figure}
\centering
\includegraphics[height=2.5in]{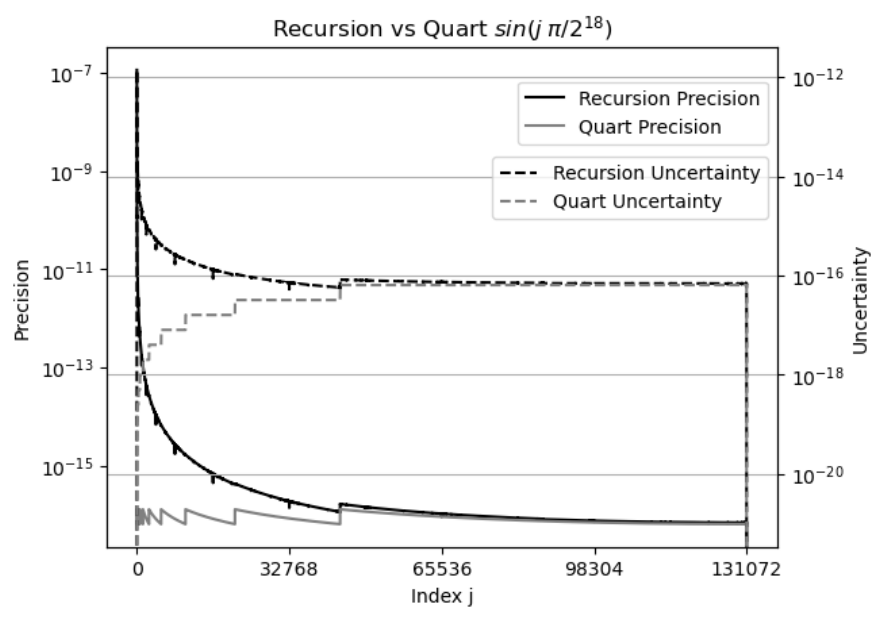}
\captionof{figure}{
The resulting precision (left y-axis) and uncertainty (right y-axis) of $\sin(\pi j/2^{18})$ versus index $j$ (x-axis) using either the Quart or the recursive sine function (legend).
}
\label{fig: Regression_Sin}
\end{figure}

\begin{align}
\label{eqn: phase boundary}
& \sin(0) = \cos(\frac{\pi}{2}) = 0;\quad \sin(\frac{\pi}{2}) = \cos(0) = 1; \\
\label{eqn: phase sin}
& \sin(\frac{\alpha + \beta}{2}) = \sqrt{\frac{1 - \cos(\alpha + \beta)}{2}} = \sqrt{\frac{1 - \cos(\alpha) \cos(\beta) + \sin(\alpha) \sin(\beta)}{2}}; \\
\label{eqn: phase cos}
& \cos(\frac{\alpha + \beta}{2}) = \sqrt{\frac{1 + \cos(\alpha + \beta)}{2}} = \sqrt{\frac{1 + \cos(\alpha) \cos(\beta) - \sin(\alpha) \sin(\beta)}{2}};
\end{align}
Formulas \eqref{eqn: phase sin} and \eqref{eqn: phase cos} compute $\sin(j \pi/2^L), \cos(j \pi/2^L), j = 0, 1, \dots, 2^{L - 2}$ recursively for recursion order $L = 0, 1, \dots, 17$ starting from Formula \eqref{eqn: phase boundary}.

Formula \eqref{eqn: phase sin} is not suitable for computing $\sin(x)$ as $x \rightarrow 0$ because it suffers from behavior analogous to catastrophic cancellation \cite{Rounding_Error, Precise_Numerical_Methods}.
As shown in Figure \ref{fig: Regression_Sin}, the Quart sine function exhibits a constant precision around $10^{-16}$, whereas the recursive sine function shows worsening precision up to $10^{-7}$ as $x \rightarrow 0$.
Unlike the hidden catastrophic cancellation in floating-point arithmetic, variance arithmetic uses coarser precision to demonstrate where and by how much the recursive algorithm becomes unfit to compute $\sin(x)$.

\section{Regression}
\label{sec: regression}

Statistical Taylor expansion brings not only numerical enhancement, but also conceptual extension of conventional mathematical concepts.
In particular, the value, deviation, and sample count of each variable should be treated jointly in analysis.
Conventional least-squares regression \cite{Statistical_Methods, Probability_Statistics, Numerical_Recipes} violates this principle and is therefore defective.

\subsection{Ordinary Regression}

\begin{align}
\label{eqn: ordinary polynomial residual}
R_P &\equiv \frac{1}{N} \sum_{n=1}^N (Y_n - c_0 - \sum_{p=1}^{P} c_p\, X_n^p)^2; \\
\label{eqn: ordinary polynomial equation}
&\sum_{p=0}^{P} c_p \overline{x^{p+i}} = \overline{x^i y}, \quad i = 0, 1, \dots, P;
\end{align}
An \emph{ordinary polynomial regression} \cite{Statistical_Methods, Probability_Statistics, Numerical_Recipes} fits a data set containing $N$ samples $\{(X_n, Y_n), n = 1, \dots, N\}$ to a polynomial by minimizing the residual $R_P$ defined in Formula \eqref{eqn: ordinary polynomial residual}, to result in Equation \eqref{eqn: ordinary polynomial equation} for $c_p,  p = 0, 1, \dots, P$.

\begin{align}
\label{eqn: ordinary multi residual}
R_D &\equiv \frac{1}{N} \sum_{n=1}^N \left(Y_n - c_0 - \sum_{d=1}^{D} c_d X_{n, d}\right)^2; \\
\label{eqn: ordinary multi slope}
&\boldsymbol{c} \equiv (c_0, \dots, c_D)^T = \mathbf{C}^{-1} \boldsymbol{\nu}; \\
\label{eqn: ordinary multi intercept}
&c_0 = \overline{y} - \sum_{d=1}^{D} c_d\, \overline{x_d}; \\
\label{eqn: ordinary multi slope cov}
&\nu(c_d, c_e) = (\mathbf{C}^{-1})_{d,e} \frac{(\delta y)^2 - \boldsymbol{\nu}^T \mathbf{C}^{-1} \boldsymbol{\nu}}{N} \quad d, e = 1, \dots, D;
\end{align}
When $P=1$, the regression is an \emph{ordinary linear regression}.
It can be extended to \emph{ordinary multi-dimensional linear regression} \cite{Statistical_Methods, Probability_Statistics, Numerical_Recipes}, which fits a data set containing $N$ samples $\{(X_{1, n}, \dots, X_{D, n}, Y_n)\}$ to a multi-dimensional linear formula by minimizing the residual $R_D$ defined in Formula \eqref{eqn: ordinary multi residual}, to result in the solution for the slope and intercept as Formulas \eqref{eqn: ordinary multi slope} and \eqref{eqn: ordinary multi intercept}, respectively, in which:
\begin{itemize}
\item $\nu(u, v) \equiv \overline{u v} - \overline{u}\, \overline{v}$ is the covariance between $u$ and $v$.
In the special case, $(\delta u)^2 = \nu(u, u)$.

\item $\mathbf{C}$ is the covariance matrix with $(\mathbf{C})_{d,e}$ as the covariance between dimension $d$ and $e$ of X,

\item  $\boldsymbol{\nu}$ is the covariance vector with $(\boldsymbol{\nu})_d$ as the covariance between the X dimension $d$ and Y.
\end{itemize}
Formula \eqref{eqn: ordinary multi slope cov} shows that slopes from different dimensions may be mutually dependent, which is a known weakness of ordinary multi-dimensional linear regression. 

The similarity between Formulas \eqref{eqn: ordinary polynomial residual} and \eqref{eqn: ordinary multi residual} suggests that the polynomial regression can also be applied as an ordinary multi-dimensional linear regression with $x^p$ for each dimension.
Conversely, Formula \eqref{eqn: ordinary multi slope} also yields Formula \eqref{eqn: ordinary polynomial equation}, confirming that these two methods are equivalent.

The dependency between $x^d$ and $x^e$ in ordinary polynomial regression is conceptually troublesome for statistical Taylor expansion: the uncorrelated uncertainty condition cannot hold between $x^d$ and $x^e$, so that the result of the polynomial fit has the dependency problem.
For example, when $(\delta x)/\overline{x} \rightarrow 0$, $\nu(x^d, x^e) \rightarrow (\delta x^d) (\delta x^e)$ making $\mathbf{C}$ near-singular with large $|(\mathbf{C}^{-1})_{d,e}|$ in an unstable $\mathbf{C}^{-1}$, so that this regression should be carried out in a coordinate with $\overline{x} = 0$.
These conditioning problems are well known and motivate the use of orthogonal polynomial bases (Chebyshev, Legendre, Hermite) \cite{Polynomial_Regression}, which shows that \emph{a dependency problem can be reduced by orthogonal transformations under restricted conditions}.

\subsection{Total Regression}

\begin{align}
\label{eqn: total regression distance}
0 &= (x_n - X_n) + \left( \sum_{p=0}^{P} c_p x_n^p -  Y_n \right) \sum_{p=1}^{P} c_p p x_n^{p - 1}; 
\end{align}
A total regression \cite{Linear_Regression} minimizes the squared distance between $(X_n, Y_n)$ and a point $(x_n, y_n)$ on the polynomial, to result in the distance Equation \eqref{eqn: total regression distance}. 
Total linear regression is invariant under swapping the roles of $x$ and $y$.
In contrast, ordinary linear regression minimizes the vertical squared-distance sum rather than the orthogonal distance to the line, and so always yields a lower $|c_0|$ and a larger residual than total linear regression when the fit is not perfect.
For total multi-dimensional linear regression \cite{Linear_Regression}, the solution for the regression residual and the dimensional fitting coefficients are the smallest eigenvalue and its corresponding eigenvector  $(-\mathbf{c},\ 1)^T$ for $\begin{pmatrix} \mathbf{C} & \boldsymbol{\nu} \\ \boldsymbol{\nu}^T & (\delta y)^2 \end{pmatrix}$, respectively.
From the perspective of statistical Taylor expansion, total multi-dimensional linear regression is superior to ordinary multi-dimensional linear regression because the former discovers independent dimensions without the dependency problem of the latter.
Moreover, as a principal component regression \cite{Linear_Algebra}, a total multi-dimensional linear regression can drop a dimension if the corresponding fitting coefficient has precision coarser than $1$.

However, because Equation \eqref{eqn: total regression distance} for total polynomial regression has no closed-form general solution when the polynomial degree $P$ is larger than $2$, Equation \eqref{eqn: ordinary polynomial equation} for ordinary polynomial regression has to be used.
This compromise brings back the dependency problem.

\subsection{Partition and Weight}

All regressions need the mean $\overline{u}$ and covariance $\nu(u, v)$ as inputs, in which $u$ and $v$ can be any of X or Y.

\begin{align}
\label{eqn: set mean}
\overline{u} &= \sum_{j} \frac{N_j}{N} \overline{u_j}; \\
\label{eqn: set covariance}
\nu(u, v) &= \sum_j \frac{N_j}{N} \left( \nu(u, v)_j + (\overline{u_j} - \overline{u})(\overline{v_j} - \overline{v}) \right);
\end{align}
Suppose the data set of $N$ samples is partitioned into $J$ subsets, where each $j$-th subset has samples $N_j$, mean $\overline{u_j}$, and covariance $\nu(u_j, v_j)$. 
Formulas \eqref{eqn: set mean} and \eqref{eqn: set covariance} give the set mean and covariance, respectively, in which $N_j$ is the weight of the $j$-th subset.
$\nu(u, v)$ contains both the within-subset covariance/variance $\nu(u, v)_j$ and the between-subset covariance/variance $(\overline{u_j} - \overline{u})(\overline{v_j} - \overline{v})$.

\begin{align}
\label{eqn: ordinary c_0}
c_0 &= \frac{\nu(x, y)}{(\delta x)^2} = \frac{\sum_j \frac{N_j}{N} \left( \nu(x, y)_j + (\overline{x_j} - \overline{x})(\overline{y_j} - \overline{y}) \right)}{\sum_j \frac{N_j}{N} \left( (\delta x_j)^2 + (\overline{x_j} - \overline{x})^2 \right)}; 
\end{align}
As the simplest case of Formula \eqref{eqn: ordinary multi slope}, Formula \eqref{eqn: ordinary c_0} shows an example of applying Formulas \eqref{eqn: set mean} and \eqref{eqn: set covariance} to ordinary linear fit of $y = c_0 + c_0\, x$, in which $\delta x_j$ is the deviation for $j$-th subset (but not the deviation $(\delta x_j)/\sqrt{N_j}$ for $j$-th subset mean $x_j$).
Formula \eqref{eqn: time-series linear regression 1} is a special case of the first part of Formula \eqref{eqn: ordinary c_0} with $X_j \in [-H,\dots,+H]$.

\begin{align}
\label{eqn: convention residual}
\chi^2(c_0, c_0) &= \sum_j N_j \left(\frac{\overline{y_j} - c_0 - c_0 \overline{x_j}}{\delta y_j} \right)^2; \\
\label{eqn: convention c_0}
c_0 &= \frac{\left(\sum_j \frac{N_j}{(\delta y_j)^2} \right) \left(\sum_j \frac{N_j}{(\delta y_j)^2} \overline{x_j} \overline{y_j}\right)
					- \left(\sum_j \frac{N_j}{(\delta y_j)^2} \overline{x_j} \right) \left(\sum_j \frac{N_j}{(\delta y_j)^2} \overline{y_j}\right)}
			{\left(\sum_j \frac{N_j}{(\delta y_j)^2} \right) \left(\sum_j \frac{N_j}{(\delta y_j)^2} \overline{x_j}^2\right)
				- \left(\sum_j \frac{N_j}{(\delta y_j)^2} \overline{x_j} \right)^2};
\end{align}
In contrast, common practice \cite{Statistical_Methods, Numerical_Recipes} assumes the input uncertainty to be Gaussian, constructs the residual $\chi^2$ as in Formula \eqref{eqn: convention residual}, and finds the extremum over $\{ c_0, c_0 \}$ to result in Formula \eqref{eqn: convention c_0}.
Formulas \eqref{eqn: convention residual} and \eqref{eqn: convention c_0} reproduce Formulas (15.2.2) and (15.2.6) from the reference \cite{Numerical_Recipes}, respectively, in which $c_0, c_0, \overline{x_j}, \overline{y_j}, \delta y_j/\sqrt{N_j}$ replace $a, b, x_i, y_i, \sigma_i$, respectively.
The residual difference between Formulas \eqref{eqn: ordinary polynomial residual} and \eqref{eqn: convention residual}, and the result incompatibility between Formulas \eqref{eqn: ordinary c_0} and \eqref{eqn: convention c_0}, suggest that this common practice differs from ordinary linear regression. 
Formula \eqref{eqn: ordinary c_0} is derived rigorously and directly from definitions with no assumption on the underlying input uncertainty distribution.
Conversely, Formula \eqref{eqn: convention c_0} can approximate Formula \eqref{eqn: ordinary c_0} under the following conditions:
\begin{itemize}
\item within-subset covariance/variance is ignored;
\item $\delta y_j \simeq \delta y$ when $N_j$ is sufficiently large;
\item $N_j$ is not known, but the subset-mean deviation $\delta y_j/\sqrt{N_j}$ is supplied, so that $N_j/(\delta y_j)^2$ replaces $N_j$ as the subset weight.
\end{itemize}
In a similar context, $N_j/(\delta x_j)^2$ can also be used as the weight, which means that Formula \eqref{eqn: convention c_0} applies in a broader context than that of Formula \eqref{eqn: convention residual}
.
When $\delta y$ is also known, a better approximation than Formula \eqref{eqn: convention c_0} is to apply the estimated $N_j$ directly to Formula \eqref{eqn: ordinary c_0} to include within-subset covariance/variance.
Compared with Formula \eqref{eqn: ordinary c_0}, Formula \eqref{eqn: convention c_0} has several drawbacks:
\begin{itemize}
\item ignoring within-subset covariance/variance,
\item introducing noise from approximating $\delta y_j$ as $\delta y$,
\item preventing mixing of precise and imprecise values as regression inputs.
\end{itemize}
In contrast, Formulas \eqref{eqn: set mean} and \eqref{eqn: set covariance} are much simpler, general, and well-suited for mixing of precise and imprecise values as regression inputs.

\ifdefined\Verbose
\subsection{A Practical Example: Hydrogen Atomic Spectra}

The interpretation of hydrogen atomic spectra gave rise to the mathematical formalism for quantum physics.
The Lamb shift, the $4.4\times10^{-6}\,\mathrm{eV}$ difference between $2S_{1/2}$ and $2P_{1/2}$, marks the start of quantum field theory \cite{Lamb_Shift}.

\begin{align}
\label{eqn: atomic hydrogen levels}
\frac{1}{\mu_e} = \frac{1}{m_e} + \frac{1}{m_p}:&\quad  
	E_n = -\frac{\alpha^2 \mu_e c^2}{2n^2}, \quad n=1,2,\dots; \\
\label{eqn: vacuum permittivity}
\varepsilon_0 \equiv \dfrac{e^2}{4 \pi \alpha \hbar c}:&\quad 
E_n =-\dfrac{\mu_e e^4}{2(4\pi\varepsilon_0)^2\hbar^2 n^2}, \quad n=1,2,\dots;
\end{align}
Formulas \eqref{eqn: atomic hydrogen levels} and \eqref{eqn: vacuum permittivity} both display the atomic hydrogen energy levels using quantum physics \cite{Fundamentals_of_Physics}, in which:
\begin{itemize}
\item $e$ is the electron electric charge, with $\delta e = 0$ as a definition.
\item $\hbar$ is the reduced Planck constant, with $\delta \hbar = 0$ as a definition.
\item $c$ is the light speed in vacuum, with $\delta c = 0$ as a definition.
\item $\mu_e$ is the effective electron mass, which is defined by Formula \eqref{eqn: atomic hydrogen levels}, with $P(\mu_e) = 3.0\times 10^{-10}$. In Formula \eqref{eqn: atomic hydrogen levels}, $m_e$ and $m_p$ are the mass of electron and proton, respectively.
\item $\alpha$ is the fine-structure constant, with $P(\alpha) = 1.5\times 10^{-10}$ using other methods \cite{CODATA_2018, Fine_Structure_Constant}.
\item $\varepsilon_0$ is the vacuum permittivity, which is defined by Formula \eqref{eqn: vacuum permittivity}.
\end{itemize}

\begin{table}
\centering
\begin{tabular}{|l|r|r|r|r|}
\hline
series & Lyman & Balmer & Paschen & Brackett \\
\hline
lower level $n$ & 1 & 2 & 3 & 4 \\
lines used & 11 & 35 & 28 & 7 \\
maximal level $m_{\max}$ & 12 & 37 & 40 & 16 \\
strongest intensity ($10^3$) & 840 & 500 & 32 & 11 \\
weakest intensity ($10^3$) & 5.6 & 0.70 & 0.09 & 0.80 \\
intensity weighted $\delta\lambda$ ($10^{-3}$ nm) & 0.53 & 3.05 & 6.63 & 6.15 \\
intensity weighted $\delta\lambda/\lambda$ ($10^{-6}$) & 5.42 & 5.78 & 6.17 & 2.47 \\
$S$-state QFT shift ($10^{-6}$) & 2.49 & 1.29 & 0.86 & 0.65 \\
\hline
precision ($10^{-6}$) & 4.25 & 3.57 & 3.55 & 3.14 \\
relative error ($10^{-6}$) & +5.81 & +5.48 & +2.58 & +1.78 \\
in-between variance percentage & $59\%$ & $35\%$ & $25\%$ & $85\%$ \\
\hline
\end{tabular}
\captionof{table}{
Determination of the fine-structure constant $\alpha$ from each hydrogen spectral series, using the measured vacuum wavelengths and their uncertainties from the NIST Atomic Spectra Database \cite{NIST_ASD}, with the tabulated relative intensity as the weight $N_j$ of Formulas \eqref{eqn: set mean} and \eqref{eqn: set covariance}.
The $S$-state QFT shift is the fraction of $E_n$ by which quantum field theory shifts the $S$ state of the lower level \cite{CODATA_2018}.
The relative error is $(\alpha - \alpha_0)/\alpha_0$ against the best available $\alpha_0 = 7.2973525693 \times 10^{-3}$ \cite{CODATA_2018}.
}
\label{tbl: hydrogen spectra}
\end{table}

Formula \eqref{eqn: atomic hydrogen levels} can be used to measure $\alpha$, which here serves only as a vehicle to expose the in-between variance of Formula \eqref{eqn: set covariance} on real experimental data.
Each observed vacuum wavelength $\lambda$ and its uncertainty $\delta \lambda$ are converted to the transition energy $E_m - E_n = hc/\lambda$ and its uncertainty $\delta (E_m - E_n)$, then normalized by $\left(\frac{1}{n^2} - \frac{1}{m^2}\right) \frac{\mu_e c^2}{2}$ to give $\alpha^2$.
Each spectral series is treated as one data set, whose $j$-th subset is a single line with mean $\overline{\alpha_j}$, within-subset deviation $\delta \alpha_j$, and weight $N_j$ given by the relative intensity.
Table \ref{tbl: hydrogen spectra} shows the result, one column per series.
The quantum field theory shift of the lowest energy level for each series is shown as the ``$S$-state QFT shift'' row, which is treated as a systematic error in this exercise, and which should decrease the ``intensity weighted $\delta\lambda$'' row.
Other systematic errors are due to relativistic effects \cite{Lamb_Shift}, which are much more than the ``$S$-state QFT shift'' row, and which should increase the ``intensity weighted $\delta\lambda$'' row instead.
The ``in-between variance percentage'' row decreases with the series, until suddenly rises to $85\%$ for the Brackett series.
This abnormality comes from an abnormally small "intensity weighted $\delta\lambda/\lambda$", which is due to its longer infrared wavelengths that shrink the within-subset variance. 
A single line at $m=12$ also contributes most of the in-between variance as a possible outlier.
The small ``in-between variance percentage'' suggests that the Paschen result can be improved by reducing $\delta \lambda$, but such improvement can not be achieved for the Brackett series.

The calculated $E_1$ bias offset is about $1.1\times10^{-10}\,\mathrm{eV}$, well below the Lamb shift.
However, the actual value of bias offset using Formula \eqref{eqn: vacuum permittivity} is about 3-fold of that using Formula \eqref{eqn: atomic hydrogen levels}.
Therefore, it is important to distinguish between the original input and the derived input when applying statistical Taylor expansion---the choice of inputs makes a difference.

\fi

\section{Solving Equations}
\label{sec: equations}

When an analytic equation $0 = g(f(x, \dots), \dots)$ has an analytic solution $f(x, \dots)$, $0 = g(f(x, \dots), \dots)_x^{(j)}, j = 0,1\dots$, which means $\forall \tilde{x}: 0 = g(f(x + \tilde{x}, \dots), \dots)$.
Thus, $x$ in $f(x, \dots)$ can be a random variable $x \pm \delta x$. 
A proposed \emph{statistical algebra}:
\begin{enumerate}
\item treats any of the variable, the coefficients, initial conditions, and boundary conditions in $f(x, \dots)$ as either precise or random as far as the term is treated as abstract without a precise range during the derivation of $f(x, \dots)$,

\item uses statistical Taylor expansion to provide the interpretation of the result solution $f(x, \dots)$ such as $\overline{f(x, \dots)}$ and $\delta^2 f(x, \dots)$.
\end{enumerate}
 However, some crucial fundamental questions remain for this new algebra.

\subsection{The Simplest Ordinary Differential Equation}

\begin{align}
\label{eqn: first-order ODE}
\frac{d}{d x} f(x) &= c_0; \\
\label{eqn: first-order ODE solution}
f(x) &= (x - x_0) c_0 + f(x_0); 
\end{align}
Equation \eqref{eqn: first-order ODE} shows the simplest ordinary differential equation, with its precise solution in Formula \eqref{eqn: first-order ODE solution} \cite{Numerical_Recipes}.

\begin{align}
\label{eqn: first-order random ODE solution}
f(x) &= (x - x_0) (c_0 \pm \delta c_0) + (f(x_0) \pm \delta f(x_0)); \\
\label{eqn: first-order ODE mean}
\overline{f(x)} &= (x - x_0) \overline{c_0} + \overline{f(x_0)}; \\
\label{eqn: first-order ODE variance}
\delta^2 f(x) &= (x - x_0)^2 (\delta c_0)^2 + (\delta f(x_0))^2;
\end{align}
Formula \eqref{eqn: first-order random ODE solution} replaces $c_0$ and $f(x_0)$ in Formula \eqref{eqn: first-order ODE solution} with $c_0 \pm \delta c_0$, and $f(x_0) \pm \delta f(x_0)$, respectively.
Formulas \eqref{eqn: first-order ODE mean} and \eqref{eqn: first-order ODE variance} give the mean and variance for the solution $f(x)$, respectively.

\begin{align}
\label{eqn: first-order ODE solution initial}
f(x) &= \big(x - (x_0 \pm \delta x_0)\big)\,(c_0 \pm \delta c_0) + \big(f(x_0) \pm \delta f(x_0)\big); \\
\label{eqn: first-order ODE mean initial}
\overline{f(x)} &= (x - \overline{x_0}) \overline{c_0} + \overline{f(x_0)}; \\
\label{eqn: first-order ODE variance initial}
\delta^2 f(x) &= (x - \overline{x_0})^2 (\delta c_0)^2 + (\delta f(x_0))^2 + \big(\overline{c_0}^2+(\delta c_0)^2\big)(\delta x_0)^2;
\end{align}
When the initial condition is at $x_0 \pm \delta x_0$ but with the same $\delta f(x_0)$ instead, Formula \eqref{eqn: first-order ODE solution initial} shows the solution after replacing $x_0$ with $x_0 \pm \delta x_0$ in Formula \eqref{eqn: first-order random ODE solution}.
Formulas \eqref{eqn: first-order ODE mean initial} and \eqref{eqn: first-order ODE variance initial} give the mean and variance for the solution $f(x)$, respectively.

\begin{align}
\label{eqn: first-order ODE solution imprecise}
f(x \pm \delta x) &= \big((x \pm \delta x) - (x_0 \pm \delta x_0)\big)\,(c_0 \pm \delta c_0) + \big(f(x_0) \pm \delta f(x_0)\big); \\
\label{eqn: first-order ODE mean imprecise}
\overline{f(x)} &= (\overline{x} - \overline{x_0}) \overline{c_0} + \overline{f(x_0)}; \\
\label{eqn: first-order ODE variance imprecise}
\delta^2 f(x \pm \delta x) &= (x - \overline{x_0})^2 (\delta c_0)^2 + (\delta f(x_0))^2
		 + \big(\overline{c_0}^2+(\delta c_0)^2\big) \big((\delta x)^2+(\delta x_0)^2 \big);
\end{align}
Formula \eqref{eqn: first-order ODE solution imprecise} evaluates the solution at the random point $x \pm \delta x$, in contrast to the deterministic evaluation point $x$ in Formula \eqref{eqn: first-order ODE solution initial}. 
Formulas \eqref{eqn: first-order ODE mean imprecise} and \eqref{eqn: first-order ODE variance imprecise} give the mean and variance for the solution $f(x)$, respectively.

\begin{align}
\label{eqn: first-order ODE two anchors solution}
f(x) &= \frac{x_1 - x}{x_1 - x_0} f(x_0) + \frac{x - x_0}{x_1 - x_0} f(x_1); 
\end{align}
Formula \eqref{eqn: first-order ODE two anchors solution} shows the precision solution with two initial conditions $f(x_0)$ and $f(x_1)$.
Likewise, it can easily turn into a random solution by replacing any precise term into random term.

The above solutions for the simplest differential equation show that equations with random variables provide much richer structure than their precise counterparts.
Although Formula \eqref{eqn: first-order random ODE solution} as a random solution of Equation \eqref{eqn: first-order ODE} can be obtained by another method \cite{Random_Differential_Equations}, the presentation of the equation, the method, the solution, the interpretation of the solution, and the extension to more complicated cases in this section are all much simpler.

\subsection{Second Order Constant-Coefficient Homogeneous Ordinary Differential Equation}

\begin{align}
\label{eqn: homogeneous ODE}
0 &= (c_0 \pm \delta c_0) x + (c_1 \pm \delta c_1) \frac{d x}{d t} + (c_2 \pm \delta c_2) \frac{d^2 x}{d t^2}; \\
\label{eqn: homogeneous damping}
\eta &\equiv \frac{1}{2} \frac{c_1 \pm \delta c_1}{c_2 \pm \delta c_2}; \\
\label{eqn: homogeneous delta}
\Delta &\equiv (c_1 \pm \delta c_1)^2 - 4 (c_2 \pm \delta c_2)(c_0 \pm \delta c_0); \\
\label{eqn: homogeneous ODE overdamped}
\hat{\kappa}\, \delta \Delta < \overline{\Delta}:&\quad
	x= e^{-\eta t} \big(A \cosh(\mu\, t) + B \sinh(\mu\, t)\big), \quad
		\mu \equiv \frac{1}{2} \frac{\sqrt{\Delta}}{c_2 \pm \delta c_2}; \\
\label{eqn: homogeneous ODE underdamped}
\hat{\kappa}\, \delta \Delta < - \overline{\Delta}:&\quad
	x = e^{-\eta t} \big(A \cos(\omega\, t) + B \sin(\omega\, t)\big), \quad
		\omega \equiv \frac{1}{2} \frac{\sqrt{-\Delta}}{c_2 \pm \delta c_2};
\end{align}
Equation \eqref{eqn: homogeneous ODE} is a random second-order constant-coefficient homogeneous ordinary differential equation, whose most familiar application is the natural damping of a displacement $x$ over time $t$.
Formula \eqref{eqn: homogeneous damping} describes the damping rate $\eta$.
The $\Delta$ in Formula \eqref{eqn: homogeneous delta} determines whether the solution is over-damping as in Formula \eqref{eqn: homogeneous ODE overdamped} or under-damping as in Formula \eqref{eqn: homogeneous ODE underdamped} \cite{Numerical_Recipes}.
When $\hat{\kappa}\, \delta \Delta > |\overline{\Delta}|$, neither solution is statistically stable, so it forms a separation band of the two solutions.

When $c_1 \pm \delta c_1 \neq 0$, over-damping and under-damping for a damped oscillator are only two shapes of decay toward the same rest position, the separation band has little observable consequence.

\begin{equation}
\label{eqn: quantum tunneling}
0 = \left( (E - V) \pm \delta (E - V) \right) \varphi(x) + \frac{\hbar^2}{2 m} \frac{d^2 \varphi(x)}{d x^2};
\end{equation}

When $c_1 \pm \delta c_1 = 0$, $\eta = 0$, such as the stationary Schrödinger equation \cite{Quantum_Mechanics}, the separation band becomes observable, as in the case of tunneling of a particle of mass $m$ at energy $E \pm \delta E$ through a region of constant potential $V \pm \delta V$, as in Equation \eqref{eqn: quantum tunneling}.
Formula \eqref{eqn: homogeneous ODE overdamped} is thus the evanescent solution of the classically forbidden region, with $\mu = \sqrt{2 m (V - E)}\,/\hbar$ as the tunneling decay rate, while Formula \eqref{eqn: homogeneous ODE underdamped} is the propagating solution of the classically allowed region, with $\omega = \sqrt{2 m (E - V)}\,/\hbar$ as the de Broglie wave number.
According to statistical algebra, the separation band near $E = V$ no longer has zero width due to $\delta V$: inside the separation band, it is a mixture of both cases of different probabilities, to transform smoothly from one case to anther.

\subsection{Second Order Constant-Coefficient Ordinary Differential Equation with $\sin(\omega\,t)$ Driving Forcing}

\begin{align}
\label{eqn: forced ODE}
A \sin(\omega\,t) &= (c_0 \pm \delta c_0) x + (c_1 \pm \delta c_1) \frac{d x}{d t} + (c_2 \pm \delta c_2) \frac{d^2 x}{d t^2}; \\
\label{eqn: forced ODE steady}
x_\infty &= A_\infty \sin(\omega\,t - \psi); \\
\label{eqn: forced ODE amplitude}
A_\infty &= \frac{A}{\sqrt{\big((c_0 \pm \delta c_0) - (c_2 \pm \delta c_2)\, \omega^2\big)^2 + \big((c_1 \pm \delta c_1)\, \omega\big)^2}}; \\
\label{eqn: forced ODE phase}
\tan\psi &= \frac{(c_1 \pm \delta c_1)\, \omega}{(c_0 \pm \delta c_0) - (c_2 \pm \delta c_2)\, \omega^2};
\end{align}
Equation \eqref{eqn: forced ODE} describes a forced oscillation of the displacement over time $x(t)$ \cite{Driven_Damped_Oscillator}, with its steady solution provided in Formula \eqref{eqn: forced ODE steady}, \eqref{eqn: forced ODE amplitude}, and \eqref{eqn: forced ODE phase}.

\begin{equation}
\label{eqn: wave packet}
A\sin\big((\omega\pm\delta\omega)t\big)  \equiv A\int \rho(\tilde{\omega}, \omega, \delta \omega) \sin(\tilde{\omega} t)\, d\tilde{\omega};
\end{equation}
In Equation \eqref{eqn: forced ODE}, if $A \sin(\omega\,t)$ is replaced by a wave packet $A \sin((\omega \pm \delta \omega)\,t)$ as defined in Formula \eqref{eqn: wave packet}, in which $\rho(\tilde{\omega}, \omega, \delta \omega)$ is a probability density function, then $\omega$ in Formula \eqref{eqn: forced ODE steady} can be replaced by $\omega \pm \delta \omega$ because Equation \eqref{eqn: forced ODE} is linear.
Such replacement also seems to apply to Formulas \eqref{eqn: forced ODE amplitude} and \eqref{eqn: forced ODE phase} after applying the same integration over $\rho(\tilde{\omega}, \omega, \delta \omega)$ as in Formula \eqref{eqn: wave packet}.
While its notation is more concise and powerful, the proposed statistical algebra needs a solid mathematical ground for wider applicability.
For example, such application of $\omega \rightarrow \omega \pm \delta \omega$ is faulty when Equation \eqref{eqn: forced ODE} is a Duffing equation  \cite{Driven_Damped_Oscillator}.

\subsection{A Comment on Numerical Solution}

When available, an analytic solution is preferable to a numerical one because the latter must sample over the distributional range of coefficients, initial conditions, and boundary conditions.
For example, a numerical solution for Equation \eqref{eqn: first-order ODE} has to sample from the two distributions in Formula \eqref{eqn: first-order ODE solution}.

\section{Conclusion and Discussion}
\label{sec: conclusion and discussion}

\subsection{Summary}

Statistical Taylor expansion incorporates the sample counts and uncertainty distributions of each input to calculate the mean, deviation, and reliability of an analytic expression. 
It tracks the variable dependencies in intermediate steps and rejects invalid calculations.
Although statistical Taylor expansion eliminates the dependency problem, it also reduces execution flexibility.

When all input uncertainties are specified, statistical Taylor expansion achieves ideal coverage:
\begin{itemize}
\item For a distribution test, the normalized error distribution is Normal, with error deviation $1$.
\item For a value test, the normalized error distribution is Delta, with error slope $-1$.
\end{itemize}
Such precise quantification allows validation of each algorithm.

Variance arithmetic simplifies statistical Taylor expansion by introducing numerical rules that eliminate invalid results: divergent, negative-variance, unstable, infinite, or unreliable.
It also provides proper coverage for floating-point rounding errors.
The applicability of variance arithmetic has been demonstrated across a wide range of computational scenarios.

In statistical algebra, if any term in an analytic expression is derived abstractly without precise range, the term can be either precise or random.

The code and analysis framework for variance arithmetic are available as an open-source project at \url{https://github.com/Chengpu0707/VarianceArithmetic}.
A more detailed description of this study is available at \url{https://arxiv.org/abs/2410.01223}.

\subsection{Improvements Needed}

This study presents statistical Taylor expansion and variance arithmetic, which are still in early stages of development.
Accordingly, several important questions remain.

Mathematical library functions should be recalculated using variance arithmetic to ensure that each output value is accompanied by its corresponding uncertainty.
Without this refinement, the value errors in the library functions can produce unpredictable and potentially significant result errors.
The attempt to cast an 80-bit sine library into a 64-bit sine library and use the difference as the uncertainty deviation results in slightly worse error deviations.

The bound moment $\zeta(n, \kappa)$ should be extended to all probability distributions.
The choice of ideal bounding range $\hat{\kappa}$ should be extended to other distributions.
The procedure for determining the bound range $\kappa$ from sample count $N$ should be developed for discrete distributions.

The measured error slope rates so far are slightly but significantly less than $-1$.
This needs further investigation.

The performance of variance arithmetic must be improved for broader practical adoption.
The fundamental formulas of statistical Taylor expansion, Formulas \eqref{eqn: Taylor 1d mean}, \eqref{eqn: Taylor 1d variance}, \eqref{eqn: Taylor 2d mean}, and \eqref{eqn: Taylor 2d variance}, contain many independent summations, making them excellent candidates for parallel processing.
Moreover, the inherently procedural nature of these formulas allows statistical Taylor expansion to be implemented efficiently at the hardware level.

A key open question is more accurately quantification of rounding errors, because many theoretical calculations lack explicit input uncertainties. 
The current approach of variance arithmetic is only a rough estimation to achieve proper coverage only. 
A better quantification of rounding errors needs to be real-time and digital. 

In variance arithmetic, deviations are comparable to values; however, variances are used in calculation.
This approach effectively limits the range of deviations to the square root of that of the values.
If the sign bit of the floating-point type can be repurposed as an exponent bit in a new unsigned floating-point representation, the range of the deviations will be identical to that of the values.

When an analytic expression undergoes statistical Taylor expansion, the resulting expression can become highly complex, as in the case of matrix inversion.
Modern symbolic computation tools such as \textit{SymPy} and \textit{Mathematica} can significantly facilitate such calculations.
This observation suggests that it may be time to shift from purely numerical programming toward analytic programming, particularly for problems that possess inherently analytic formulations.

As an enhancement to dependency tracing, source tracing identifies each input's contribution to the overall result uncertainty.
This capability enables engineers to pinpoint the primary sources of measurement inaccuracy and to make targeted improvements in data acquisition and processing strategies.
For example, Formulas \eqref{eqn: sum leakage} and \eqref{eqn: product leakage} can provide guidance on improving the ideal leakage of $x \pm y$ and $xy$, respectively.

Figure \ref{fig: Normal_Bounding_Leakage} shows that linear leakage converges toward $0$, with faster convergence at larger $\kappa_s$.
It indicates that the ideal bound range $\hat{\kappa}$ should be as large as possible just to ensure convergence.
However, such an approach is valid only after the statistical meaning of ideal leakage --- and how it should inform the choice of ideal bounding --- is clarified.
For example, why do different choices of $\hat{\kappa}$ result in different $\delta^2 f$ in Figure \ref{fig: Normal_Bounding_Leakage}?

In variance arithmetic, $\delta^2 f$ in $f \pm \delta f$ is also calculated as an imprecise value.
The implication of such uncertainty-of-uncertainty requires further clarification.

Because conventional numerical approaches are based on floating-point arithmetic and are path-dependent in general, they must be reexamined or even reinvented within the framework of variance arithmetic.
Conventional numerical algorithms aim to identify optimal computational paths, whereas statistical Taylor expansion conceptually rejects all path-dependent calculations.
Reconciling these two paradigms may present a significant and ongoing challenge.

Establishing a theoretical foundation for applying statistical Taylor expansion in the absence of a closed-form analytic solution, or when only limited low-order numerical derivatives are available, as in solving differential equations, remains an important direction for future research.

The convergence of $\delta^2 f$ depends on $\kappa$, while $\kappa$ can be changed in a measurement by an introduced limit on the space-time of the wave function.
Such change of convergence may explain the collapse of waveform during the measurement.
The coexistence of two solutions with predefined probabilities within a separation band suggests similarity to Born interpretation \cite{Quantum_Mechanics}.
As a mathematics to describe uncertainties in analytic expressions, the correlation between statistical algebra and quantum physics is quite natural and waits to be investigated.

\ifdefined\NoAuthor
\else
\section{Statements and Declarations}

\subsection{Acknowledgments}

As an independent researcher without institutional affiliation, the author expresses sincere gratitude to Dr. Zhong Zhong (Brookhaven National Laboratory) and Prof. Weigang Qiu (Hunter College) for their encouragement and valuable discussions.
The author also gratefully acknowledges Prof. Dongfeng Wu (University of Louisville) for her guidance on statistical topics.
Unrelated to this work, the author is very grateful to Prof. Lizhi Fang (University of Arizona) and Prof. Paul Hough (Brookhaven National Laboratory) for their life-long leadership in scientific adventures, especially their encouragement to work on fundamental and novel research.
Special thanks are extended to the organizers of \emph{AMCS 2005}, particularly Prof. Hamid R. Arabnia (University of Georgia), and to the organizers of the \emph{NKS Mathematica Forum 2007}.
Finally, heartfelt appreciation is extended to the editors and reviewers of \emph{Reliable Computing} for their substantial assistance in shaping and accepting an earlier version of this work, with special recognition to Managing Editor Prof. Ralph Baker Kearfott.

\subsection{Data Availability Statement}

All data sets used in this study are generated by the open-source project at \url{https://github.com/Chengpu0707/VarianceArithmetic}.
Assistance with running and understanding the code is available from the author upon request.

\subsection{Competing Interests}

The author has no competing interests to declare that are relevant to the content of this article.

\subsection{Funding}

No funding was received from any organization or agency in support of this research.
\fi

\ifdefined\ManualReference

\else
\bibliographystyle{unsrt}
\bibliography{VarianceArithmetic}
\fi

\end{document}